\documentclass[nolongbibliography,%
 reprint,
 amsmath,amssymb,
 aps,
pra,
]{revtex4-2}
\usepackage{graphicx}% Include figure files
\usepackage{dcolumn}% Align table columns on decimal point
\usepackage{bm}% bold math
\usepackage{siunitx}
\usepackage{amsmath}
\usepackage[english]{babel}
\usepackage{svg}
\usepackage{booktabs}
\usepackage{xcolor}

\begin{document}
\preprint{APS/123-QED}
\twocolumngrid

\title{Resonant multi-harmonic acousto-optics for programmable frequency control of visible light in a CMOS platform}

% Alternative title: "Acousto-optic frequency-comb synthesis with
% harmonically engineered mechanical modes"

\author{Jacob M. Freedman$^{1}$}
\author{Matthew J. Storey$^{2}$}
\author{Daniel Dominguez$^{2}$}
\author{Andrew Leenheer$^{2}$}
\author{Nils T. Otterstrom$^{2,3}$}
\author{Matt Eichenfield$^{1,2}$}
\email{matt.eichenfield@colorado.edu}
\affiliation{\phantom{$^1$ James C. Wyant College of Optical Sciences, University of Arizona, Tucson, Arizona, USA} \\ $^1$ Department of Electrical, Computer and Energy Engineering, University of Colorado Boulder, Colorado, USA \\ $^2$ Microsystems Engineering, Science, and Applications, Sandia National Laboratories, Albuquerque, New Mexico, USA \\ $^3$ Present address: Manzano Systems Inc., 2100 Louisiana Blvd Suite 480 NE, Albuquerque, New Mexico, USA
} 

\begin{abstract}
Scaling quantum control for atoms, ions, and solid-state emitters requires gigahertz-frequency spectral control of high-power visible light in a volume-manufacturable platform. Silicon nitride photonics provides high power handling and CMOS-foundry compatibility but has no intrinsic mechanism for high-speed modulation. Integration with piezoelectric materials enables acousto-optic phase modulation, and mechanical resonant enhancement has made it efficient at gigahertz frequencies. However, a single resonance restricts the modulation waveform to a single tone, imposing Bessel-function sideband amplitudes and limiting frequency-shifting efficiency to $\SI{33.9}{\percent}$. Here we engineer a silicon nitride acousto-optic microstructure to support harmonically spaced resonances at $\SI{1.14}{\giga\hertz}$ and $\SI{2.28}{\giga\hertz}$, each strongly optomechanically coupled to a $\SI{730}{\nano\metre}$ guided optical mode, so that tailored non-sinusoidal modulation waveforms can be resonantly synthesized. By piezoelectrically controlling the two mechanical amplitudes and their relative phase, we demonstrate $\SI{50}{\percent}$ conversion to one sideband ($1.5\times$ the single-tone theoretical maximum), a flat seven-line comb, and a frequency shift with $\SI{60}{\decibel}$ carrier and $\SI{53}{\decibel}$ image suppression---to our knowledge the highest reported for an integrated modulator. The devices are fabricated in a $\SI{200}{\milli\metre}$ CMOS foundry, and we measure $\SI{91.7}{\percent}$ yield without post-fabrication tuning across 36 devices from three wafers. We also show how the technique can be straightforwardly scaled to three or more harmonics. This result overcomes the trade-off between resonant enhancement and spectral programmability, with important consequences including improved single-qubit gate efficiency for hyperfine qubits.  
\end{abstract}

\maketitle

Visible and near-visible laser light is the interface through which many of today's most advanced quantum technologies are controlled and measured \cite{park_technologies_2024, bruzewicz_trapped-ion_2019, menssen_strategic_2026}. It provides cooling, initialization, trapping, gates, and readout for trapped ions \cite{ransford_98-qubit_2026}, neutral atoms \cite{bluvstein_fault-tolerant_2026}, and solid-state emitters \cite{li_heterogeneous_2024}. Many of these functions require programming the spectral content of a continuous-wave laser at the gigahertz scale to tailor it into prescribed multi-tone optical fields \cite{levine_dispersive_2022, doi:10.1126/sciadv.ade4454}. As the number of physical qubits in these systems grows from tens toward millions \cite{manetsch_tweezer_2025, chiu_continuous_2025, holman_trapping_2026}, the control must scale to a comparable number of individually addressable channels, for which much of the optical power demultiplexing needs to be performed on-chip \cite{hall_innovations_2025}. Thus, the material platform in which these control systems are implemented must combine efficient modulation at gigahertz frequencies, a small footprint, high visible-wavelength power handling, and volume manufacturability \cite{bruzewicz_trapped-ion_2019, menssen_strategic_2026}.

Silicon nitride (SiN) photonic platforms are well-suited for this application space because SiN is transparent from $\SI{400}{\nano\metre}$ to $\SI{2350}{\nano\metre}$ \cite{blumenthal_silicon_2018}, handles high optical power at visible wavelengths \cite{freedman_gigahertz-frequency_2025, zhao_integrated_2025, mehta_integrated_2020}, and is compatible with complementary metal oxide semiconductor (CMOS) foundry processes \cite{blumenthal_silicon_2018, zimmermann_monolithic_2026}. Building on this foundation, an assortment of device capabilities has been realized with SiN integrated photonics \cite{xiang_silicon_2022}, ranging from microcomb generation \cite{shen_integrated_2020} and Brillouin lasing \cite{chauhan_visible_2021, klaver_surface_2026, gundavarapu_sub-hertz_2019} to ultra-low-loss waveguides \cite{morin_cmos-foundry-based_2021, mishra_ultra-low_2026, jin_hertz-linewidth_2021} and beam delivery systems for ion traps \cite{mehta_integrated_2020, niffenegger_integrated_2020}.

\begin{figure*}
    \includegraphics[width=173mm]{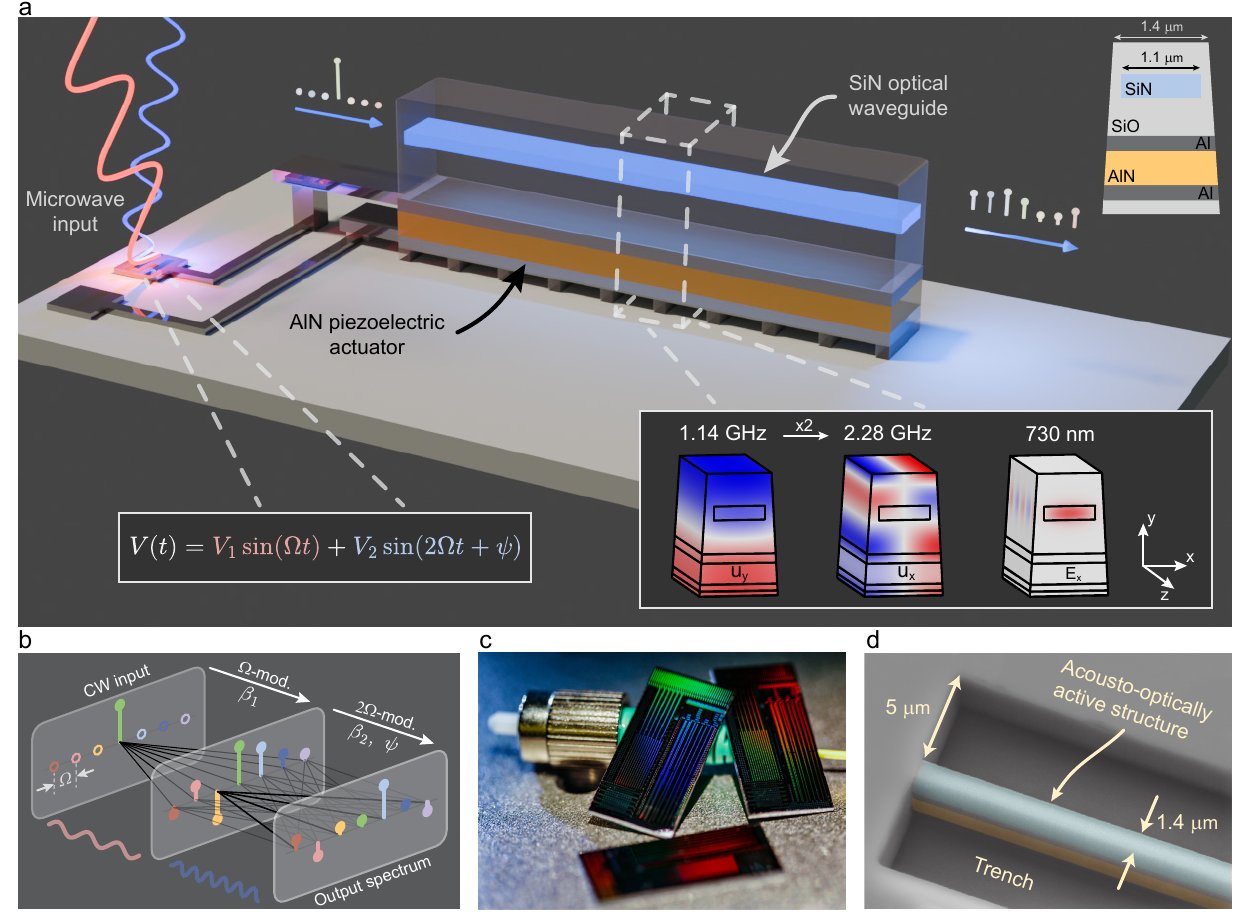}
    \caption{{\bf Design and working principle of multi-harmonic acousto-optic phase modulators.} {\bf a}  Rendering of the modulator depicting a released microstructure that supports piezoelectrically excitable breathing-mode mechanical resonances and propagating optical modes. The right inset shows two simulated displacement profiles of harmonically spaced mechanical resonances and the transverse-electric optical mode to which they are coupled. The diagram in the top right depicts the layer stack and geometry of the device's cross-section. {\bf b} Schematic showing the working principle of two-tone harmonic phase modulation. Modulation at $\Omega$ produces a Bessel-weighted comb, then each of its comblines acts as a carrier for the modulation at $2\Omega$, which produces combs whose combined interference forms the output spectrum. {\bf c} Photograph of the die taken from three different wafers used in this work with a standard fiber patch cord for scale. {\bf d} False-colored scanning electron micrograph of the $\SI{1.4}{\micro\metre}$-wide released structure between the $\SI{5}{\micro\metre}$-wide trenches on either side that mechanically separate it from the rest of the die.
    \label{fig:device}}
\end{figure*}

Because SiN is amorphous, it supports neither a Pockels nor a piezoelectric effect, meaning electrically excitable acousto-optic and electro-optic modulation are not intrinsically available. Instead, high-speed modulation is typically enabled by monolithic integration with a piezoelectric material such as aluminum nitride (AlN) or scandium-doped AlN \cite{tian_piezoelectric_2024}. Strain actuators comprising these materials can then drive photoelastic and moving-boundary interactions in a SiN waveguide without degrading its excellent passive optical properties \cite{hosseini_stress-optic_2015, erdil_wideband_2025, bian_demonstration_2024, stanfield_cmos-compatible_2019}. This approach has yielded quasi-static control of microcombs \cite{liu_monolithic_2020}, ring resonators \cite{jin_piezoelectrically_2018, stanfield_cmos-compatible_2019, menssen_scalable_2023}, Mach-Zehnder interferometers \cite{dong_high-speed_2022}, and beam scanners \cite{saha_nanophotonic_2026}.

Achieving efficient modulation at or near gigahertz frequencies has required either a mechanical resonance to enhance the acousto-optic interaction \cite{tian_magnetic-free_2021, freedman_gigahertz-frequency_2025, tian_hybrid_2020} or a waveguide routed through several passes of a traveling acoustic wave \cite{kenning_broadband_2025}. Both of these mechanisms restrict modulation to the close vicinity of a single microwave frequency, and any modulation waveform with richer spectral content is filtered to a single tone. This imposes Bessel-function sideband amplitudes, forbids asymmetric output spectra, and limits conversion to any shifted frequency to the single-tone bound of $\SI{33.9}{\percent}$.

Phase modulation can overcome these spectral limitations when the sidebands of one modulation serve as carriers for another, allowing the resulting combs to overlap and interfere. Frequency-encoded quantum processors achieve this by arranging a pulse shaper between two single-tone modulators, so that sidebands generated by the first modulator recombine in the second with programmed relative phases \cite{lukens_frequency-encoded_2017, lu_electro-optic_2018}, but pulse shapers are challenging to implement and control in integrated platforms \cite{cohen_silicon_2024}. Alternatively, a non-sinusoidal periodic modulation waveform produces the overlapping combs within a single modulator (Fig. \ref{fig:device}b), and a mechanical resonator with modes at integer multiples of the waveform's fundamental frequency could provide the necessary resonant enhancement at each of its Fourier components. Mechanical resonances with integer frequency ratios have been engineered in micromechanical systems for frequency stabilization and phononic frequency-comb generation \cite{antonio_frequency_2012, chen_direct_2017, shoshani_resonant_2021, ganesan_phononic_2017}, but they have not been incorporated into a photonic platform to provide a strong modulation response at each harmonic of an externally synthesized drive, which would remove the trade-off between resonant enhancement and spectral programmability.

Here we realize such a device by engineering the mechanical spectrum of a microstructure in a SiN-AlN CMOS platform to contain two harmonically spaced gigahertz-frequency mechanical breathing modes that are each optomechanically coupled to a guided visible-wavelength optical mode (Fig. \ref{fig:device}a). We achieve this by lithographically controlling the structure's width and the waveguide core's width, which provide parameters that tune the mechanical resonance frequencies and optomechanical coupling strengths. We embed an AlN piezoelectric actuator within the structure that allows for electromechanical control of the two mechanical amplitudes and their relative phase. Using these three control parameters, we can tailor output spectra covering a space far beyond that achievable with single-tone modulation.

We design and fabricate devices with a structure width $w_\text{struct}=\SI{1.4}{\micro\metre}$ and SiN core width $w_\text{core}=\SI{1.1}{\micro\metre}$ that have harmonically spaced resonances at $\SI{1.14}{\giga\hertz}$ and $\SI{2.28}{\giga\hertz}$ with measured half-wave voltages of $V_{\pi,1}=\SI{2.87}{\volt}$ and $V_{\pi,2}=\SI{1.49}{\volt}$, respectively. Operating on $\SI{730}{\nano\metre}$ light, we use these devices to synthesize programmed spectra including a frequency shift with $\SI{50}{\percent}$ conversion efficiency (a $1.5\times$ improvement over the single-tone bound), a flat seven-line comb, and a single-sideband frequency shift with $\SI{60}{\decibel}$ carrier suppression and $\SI{53}{\decibel}$ image (negative sideband) suppression at $\SI{40}{\percent}$ conversion efficiency, to our knowledge the highest simultaneous carrier and image suppression reported for an integrated frequency shifter (see comparison table in Supplementary Information Section 1) \cite{doi:10.1126/sciadv.ade4454}. Despite relying on the precise alignment of two mechanical resonance frequencies, we demonstrate $\SI{91.7}{\percent}$ yield across 36 devices from three different wafers with identically designed cross-sections (Fig. \ref{fig:device}c). From these device statistics, we find that the average modulation bandwidths for the $\SI{1.14}{\giga\hertz}$ and $\SI{2.28}{\giga\hertz}$ modes are $\SI{16.5}{\mega\hertz}$ and $\SI{24.4}{\mega\hertz}$, respectively. The same foundry process already supports watt-level optical power in single-mode visible waveguides \cite{freedman_gigahertz-frequency_2025, zhao_integrated_2025} and monolithic integration with CMOS drive electronics \cite{zimmermann_monolithic_2026}, so the devices are built in a platform that meets the power-handling and manufacturing requirements set out above.

We also measure an optomechanically coupled third-harmonic resonance at $\SI{3.39}{\giga\hertz}$ in a device with $w_\text{struct}=\SI{900}{\nano\metre}$ and $w_\text{core}=\SI{600}{\nano\metre}$ and show that concatenating modulators hosting individually designed resonances enables the use of more harmonics in a straightforward and scalable manner. Finally, we discuss advantages for quantum technologies offered by multi-harmonic phase modulation. Crucially, we calculate that driving at two sub-harmonics of a hyperfine qubit transition frequency can generate Rabi rates as or more efficiently than driving at the transition frequency itself. As an example of this, we note that many of the devices presented in this work operate at one sixth and one third of the $^{87}$Rb hyperfine splitting ($\SI{6.83}{\giga\hertz}$), and we show that using these two sub-harmonics can achieve \SI{97}{\percent} of the gate efficiency of a direct drive using the same mechanism of single-qubit gate generation implemented in state-of-the-art quantum information processors \cite{bluvstein_fault-tolerant_2026, manetsch_tweezer_2025}. The platform can therefore condition light for qubit control in leading quantum computing systems such as those based on rubidium, cesium, barium, and ytterbium without supporting higher-frequency mechanical resonances directly at their transitions, which range from $\SI{6.8}{\giga\hertz}$ to $\SI{12.6}{\giga\hertz}$ \cite{bluvstein_fault-tolerant_2026, graham_multi-qubit_2022, ransford_98-qubit_2026, moses_race-track_2023}.

\section*{Results}

\begin{figure*}
    \includegraphics[width=173mm]{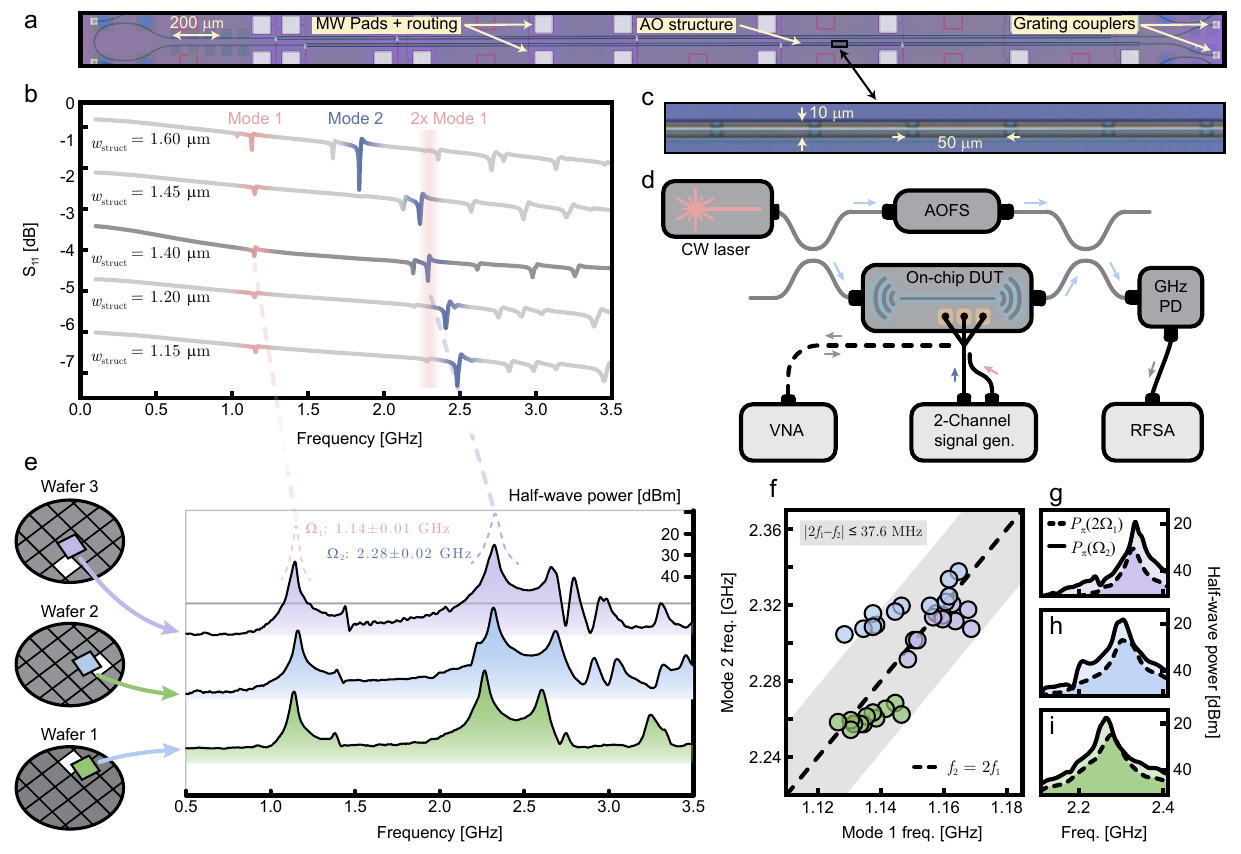}
    \caption{\textbf{Multi-wafer production of dual-harmonic resonantly enhanced modulators.}
    \textbf{a} Optical micrograph of the chip showing a grating-coupled waveguide meandering through several devices. The grating couplers, microwave pads, microwave routing, and an acousto-optic structure are indicated. {\bf b} Measured microwave reflection $S_{11}$ versus frequency for devices with structure widths ranging from $\SI{1.15}{\micro\metre}$ to $\SI{1.60}{\micro\meter}$ (vertically offset for clarity). Breathing-mode resonances appear as sharp dips, and a pair of optomechanically coupled modes is shown in red and blue. The red band roughly indicates where the second mode must exist to enable dual-harmonic modulation. {\bf c} Optical micrograph of a portion of single device showing the trenches and release structure. \textbf{d} Experimental setup used to characterize the frequency-control device. A VNA probes the electromechanical properties of the on-chip structures, and a self-heterodyne interferometer is used to measure the power in each sideband produced by the modulation. \textbf{e} Half-wave power versus drive frequency for $\SI{1.4}{\micro\metre}$-wide devices from three different wafers. Resonant peaks correspond to efficient modulation. Dashed curves on the back panel represent the average across the different devices, with statistics of the resonance frequency shown. {\bf f} Measured Mode 2 frequency versus Mode 1 frequency for all characterized devices. The dashed line marks exactly where $\Omega_2=2\Omega_1$, and the shaded band indicates the region within which harmonic alignment is achieved according to equation \eqref{eq:alignment}. {\bf g-i} Half-wave power versus frequency near Mode 2 for three representative devices from each of the wafers. Mode 1's half-wave power is overlaid on a doubled frequency axis to display the degree of harmonic alignment.
    \label{fig:char}}
\end{figure*}

\subsection*{Device fabrication and design}

\begin{figure*}
    \includegraphics[width=173mm]{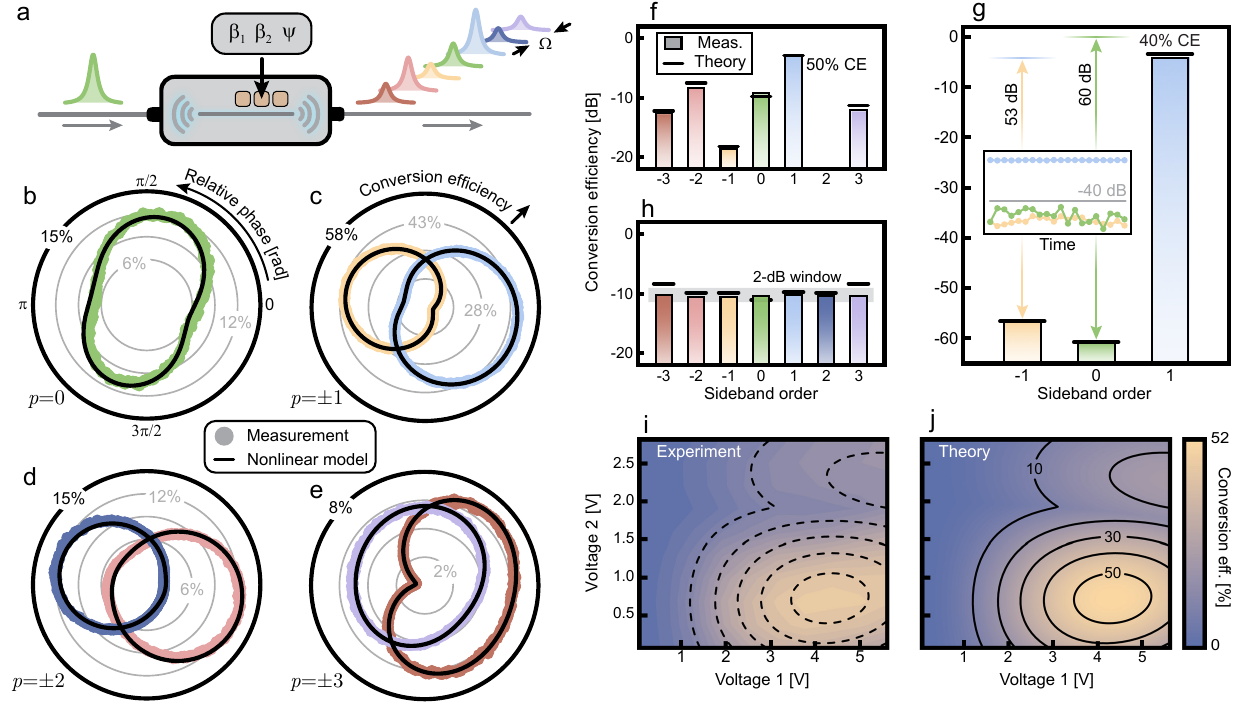}
    \caption{{\bf Integrated programmable gigahertz-scale frequency control of visible light.} {\bf a} Schematic of the two-tone modulation: the modulation depths $\beta_1$ and $\beta_2$ together with the relative phase $\psi$ determine the output spectrum. {\bf b--e} Conversion efficiency to the zeroth, $\pm1$, $\pm2$, and $\pm3$ order sidebands versus the relative phase between the drive tones, with modulation depths fixed at the values that maximize $+1$ conversion. Markers are measurements, and solid curves are a model accounting for nonlinear mechanical second-harmonic generation. {\bf f} Measured conversion efficiency for orders $-3$ to $3$ with the drive optimized for $+1$ conversion. {\bf g} Measured conversion efficiency for the carrier and $\pm 1$ sidebands with the drive optimized for single-sideband modulation with simultaneous carrier and $-1$ sideband suppression. Inset: free-running stability over $\SI{1}{\minute}$. {\bf h} As in {\bf f}, with the drive optimized for a flat spectrum. The shaded band indicates a $\SI{2}{\decibel}$ spread in conversion efficiency. {\bf i--j} Measured and modeled conversion efficiency to the $+1$ sideband versus the drive voltages at $\Omega$ and $2\Omega$, with $\psi$ optimized at each point.
    \label{fig:comb}}
\end{figure*}

The devices, depicted in Fig. \ref{fig:device}a, are fabricated on $\SI{200}{\milli\metre}$ wafers in a CMOS foundry process at Sandia National Laboratories. As described in previous work \cite{freedman_gigahertz-frequency_2025, dong_high-speed_2022, stanfield_cmos-compatible_2019}, the layer stack consists of a silicon nitride optical waveguide situated atop an aluminum nitride piezoelectric actuator. We release a patterned layer of amorphous silicon buried beneath the actuator to form the periodically supported suspended structure depicted in Fig. \ref{fig:device}a. The etch that accesses this buried layer from the upper oxide cladding of the waveguide also defines the width of the device (Fig. \ref{fig:device}d). The resulting structure hosts a spectrum of mechanical breathing modes and guided optical modes (insets of Fig. \ref{fig:device}a). The frequencies at which these resonances occur are functions of the mechanical properties of the materials within the layer stack, the thicknesses of the layers, and the width of the structure. The lattermost of these parameters provides a lithographically tunable knob that controls the mechanical spectrum (see Supplementary Information Section 2.1).

To enable resonantly enhanced multi-harmonic phase modulation, the structure must support at least two harmonically spaced mechanical modes that are each strongly optomechanically coupled to a propagating optical mode of the waveguide. Here, strongly coupled means that order-$\pi$ modulation depths are reached at microwave powers that lie below the damage threshold of the device and within the range of the driving electronics available for a given application. Because the coupling is governed by the overlap integral of the photoelastically weighted strain field with the optical intensity distribution, this condition amounts to localizing the mechanical mode energy at the SiN core and ensuring the strain in this region is free of spatial oscillations on the scale of the optical mode, which would otherwise cancel the overlap (see Supplementary Information Section 2.2 for details of the optomechanical coupling model).

With this, a two-harmonic modulator must satisfy the following condition, which we call harmonic alignment: 
\begin{align}
|2\Omega_1 - \Omega_2| \leq \gamma_1\sqrt{\left(\frac{\pi V_\text{max}}{\beta_1 V_{\pi,1}}\right)^2-1}+\frac{\gamma_2}{2}\sqrt{\left(\frac{\pi V_\text{max}}{\beta_2 V_{\pi,2}}\right)^2-1},
\label{eq:alignment}
\end{align}
where $\beta_1$ and $\beta_2$ are the maximum modulation depths respectively required from Mode 1 and Mode 2, $V_\text{max}$ is the maximum voltage that can be applied to the device, and both modes are assumed to have Lorentzian power spectra with center angular frequencies of $\Omega_1$ and $\Omega_2$, full widths at half maximum of $\gamma_1$ and $\gamma_2$, and corresponding half-wave voltages as a function of frequency of $V_{\pi,1}(\Omega)$ and $V_{\pi,2}(\Omega)$, respectively. Equation \eqref{eq:alignment} states that the two resonance frequencies must be spaced by a factor of two to within a tolerance set by their linewidths, scaled by a factor that depends on the available drive voltage and the modulation depths demanded by the target spectrum. 

For example, if the maximum voltage that can be applied to the device is large, then the required alignment is eased. Conversely, if large modulation depths are required to achieve desired output spectra, the alignment requirements are more stringent. In cases where the required modulation depth cannot be reached despite applying the maximum voltage on resonance, the argument of the square root is negative, and harmonic alignment cannot be achieved even if $\Omega_2=2\Omega_1$. For the devices and applications of this work, these parameters amount to modifying factors (the square-root terms) of order unity, so aligning the resonance frequencies to within the span of their linewidths is a useful rule of thumb (see Supplementary Information Section 3.2 for details on the derivation of this condition and its analysis).

\subsection*{Characterization of acousto-optic modulation}

Considering the above-mentioned condition of harmonic alignment, we design and fabricate an array of devices with swept $w_\text{struct}$ and $w_\text{core}$ (Fig. \ref{fig:char}a and c). We probe the electromechanical properties of these devices with a vector network analyzer (VNA) by extracting their microwave reflection coefficients $S_{11}$ as a function of frequency. We then record the electromechanically coupled subset of the mechanical spectrum of a device by identifying the frequencies at which dips in the reflection occur. The results of this analysis are shown for devices with five different widths in Fig. \ref{fig:char}b. In general, there is a mode, which we call Mode 1, whose resonance frequency $\Omega_1/2\pi$ is approximately $\SI{1.14}{\giga\hertz}$ and is relatively insensitive to the structure width. There is also a mode, which we call Mode 2, whose frequency $\Omega_2/2\pi$ exists within the range of $2.0$-$2.5$ $\SI{}{\giga\hertz}$ and is sensitive to the structure width. For a structure width $w_\text{struct}=\SI{1.4}{\micro\metre}$ and core width $w_\text{core}=\SI{1.1}{\micro\metre}$, these two modes have approximately harmonically spaced resonance frequencies. Simulations corroborate this dependence of frequency on width and predict that both of these modes have strong optomechanical coupling (Supplementary Information Section 2.2). 

Next, we measure the optical phase modulation produced by the two modes for this device geometry using the experimental setup schematically represented in Fig. \ref{fig:char}d. To do this, we drive the device with a microwave signal at a set power and frequency and measure the optical power in two distinct sidebands generated by the resulting on-chip phase modulation process. We obtain these optical powers through heterodyne interferometric readout (Supplementary Information Section 4). Then, by comparing the ratio of these powers, we extract the modulation depth and from there calculate a modulation efficiency metric for the device at that frequency. In Fig. \ref{fig:char}e we report the microwave power $P_\pi$ required to produce a modulation depth of $\pi$ for three select devices each chosen from separate wafers and a different die location within that wafer. For one of these devices we perform measurements of $V_\pi$ for which the delivered voltage is calibrated with a VNA, and we find that Mode 1 has $V_{\pi,1}=\SI{2.87}{\volt}$ and Mode 2 has $V_{\pi,2}=\SI{1.49}{\volt}$. We further characterize 36 devices from three different wafers that each have the same cross section and find that the average modulation bandwidths for Mode 1 and Mode 2 are $\SI{16.5}{\mega\hertz}$ and $\SI{24.4}{\mega\hertz}$, respectively (Supplementary Information Section 5). 

Using these values to calculate $\gamma_1$ and $\gamma_2$ in equation \eqref{eq:alignment}, assuming $V_\text{max} = \SI{2.5}{\volt}$, $\beta_1 = \SI{2.46}{\radian}$, and $\beta_2 = \SI{1.10}{\radian}$, we calculate $|2\Omega_1 - \Omega_2|$ must not exceed $2\pi \times \SI{37.6}{\mega\hertz}$. The choice of $V_\text{max} = \SI{2.5}{\volt}$ ensures operation below the threshold at which devices are damaged and maintains compatibility with CMOS amplifier drive voltages given the recent demonstration of this platform's monolithic integration with CMOS driving electronics \cite{zimmermann_monolithic_2026, wang_integrated_2018}. The choices for the required modulation depths allow for all spectra generated in this work to be produced. Using this tolerance, we evaluate that $\SI{91.7}{\percent}$ of the tested devices satisfy harmonic alignment (Fig. \ref{fig:char}f-i). We observe that each of the wafers produces a distinct cluster of resonance frequencies in Fig. \ref{fig:char}f, which is likely due to wafer-to-wafer thickness variations in the silicon dioxide cladding above and below the SiN. These thicknesses are defined by a chemical-mechanical-polishing process whose polish time is optimized once at the start of a multi-wafer fabrication run and subsequently used for all wafers. The wafer-to-wafer variability could be minimized by recalibrating the polish time after the condition of the polishing pad changes or by utilizing in-situ monitoring of the film thicknesses during polishing. 

We also infer the time-domain response of a representative device from frequency response data and find that Mode 1's $10$-to-$\SI{90}{\percent}$ rise time is $\SI{102}{\nano\second}$, and Mode 2's is $\SI{95}{\nano\second}$ (Supplementary Information Section 6). These rise times are limited by the mechanical quality factors, and they could be reduced in future work by modifying the design of the device's periodic supports. This, however, would come at the cost of reduced resonant enhancement and commensurately higher $V_\pi$. We also note here that the total insertion loss of one representative device is $\SI{17.7}{\decibel}$. Since the expected propagation loss is well under $\SI{1}{\decibel}$ given measurements on test structures fabricated in the same process and with the same layer stack \cite{dong_high-speed_2022, stanfield_cmos-compatible_2019, freedman_gigahertz-frequency_2025}, this quantity is very likely dominated by grating coupler loss and can be significantly improved in future work.

\begin{figure*}
    \includegraphics[width=173mm]{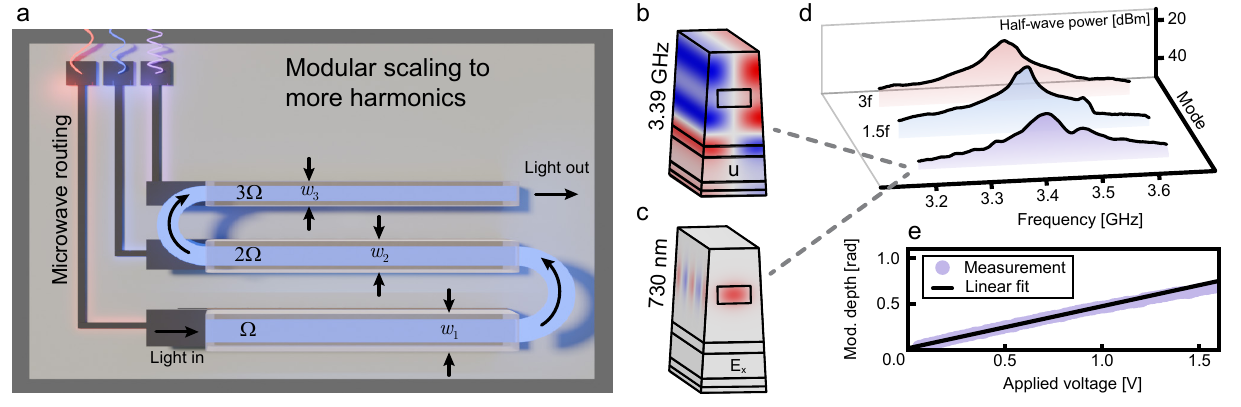}
    \caption{{\bf Concatenating devices for a scalable path to more harmonics.} {\bf a} Schematic showing how each harmonic can be generated in its own acousto-optic structure, with a single optical waveguide routed through them in series. The labels $w_1$, $w_2$, and $w_3$, indicate that each structure has a unique width designed for modulation at $\Omega$, $2\Omega$, and $3\Omega$, respectively. {\bf b} Simulated strain profile of the \SI{3.39}{\giga\hertz} breathing mode in the modified geometry with a structure width of $\SI{900}{\nano\metre}$ and a core width of $\SI{600}{\nano\metre}$. {\bf c} Simulated transverse-electric optical mode at $\SI{730}{\nano\metre}$ in the same geometry. {\bf d} Experimentally measured half-wave power versus frequency for the third-harmonic mode, together with Modes 1 and 2 of a device on the same die. The frequency axes for Mode 1 and Mode 2 are respectively scaled by $3$ and $1.5$ so that all three are referred to the third-harmonic frequency. {\bf e} Modulation depth versus applied voltage for the third-harmonic mode with a linear fit.
    \label{fig:more_harmonics}}
\end{figure*}

\subsection*{Demonstration of programmable frequency control}

Once these device-level conditions are met, the modulator can be treated abstractly as generating an optical signal $U(t)=U_0\exp\left\{{i\left[-\omega t + \phi(t)\right]}\right\}$ where the phase modulation $\phi(t)$ has the form
\begin{align}
    \phi(t) = \beta_1\sin(\Omega t) + \beta_2\sin(2\Omega t + \psi),
\end{align}
where $\omega$ is the optical angular frequency, $\beta_1$ is the modulation depth produced at the angular frequency $\Omega$, $\beta_2$ is the modulation depth produced at the angular frequency $2\Omega$, and $\psi$ is the phase difference between the two modulation functions. We omit any explicit phase from the modulation at $\Omega$ because its only effect is a global time translation of the signal that can be absorbed into $\psi$. From this expression, one can show (Supplementary Information Section 3.1) that the field amplitude, $\tilde{U}_p$, of the optical signal oscillating at $\omega -p\Omega$ is
\begin{align}
    \tilde{U}_p = U_0\sum_{k=-\infty}^\infty J_{2k+p}(\beta_1)J_{-k}(\beta_2)e^{-ik\psi}.
    \label{eq:up}
\end{align}
Because the two phase modulations commute, one can consider the two-harmonic operation as acting in stages in which the modulation at $2\Omega$ takes place after that at $\Omega$. From this point of view, each sideband produced by the $\Omega$ modulation acts as a carrier for the $2\Omega$ modulation, generating a family of secondary combs. Each term in equation \eqref{eq:up} is the amplitude that one secondary comb contributes at order $p$, and the sum is their coherent interference. This interpretation is pictorially represented in Fig. \ref{fig:device}b. For example, since the $2\Omega$ modulation produces sidebands at every other comb site from the $\Omega$ modulation, every odd-ordered sideband from the primary comb will generate power at the $p=1$ order in its secondary comb.

Given equation \eqref{eq:up}, a target spectrum can be specified and the three free parameters $\beta_1$, $\beta_2$, and $\psi$ optimized to approximate it. To demonstrate the spectra made accessible by two-tone driving, we identify and generate three qualitatively distinct optical combs that are impossible to achieve with Bessel-function-weighted sidebands. First, we maximize conversion efficiency to the positive first-order sideband. To do this, we run a simple optimization over the two modulation depths and the phase of the $2\Omega$ signal relative to the $\Omega$ signal (see Methods). In accordance with previous results \cite{zhu_generation_2026}, we identify that a conversion efficiency of $\SI{55}{\percent}$ is achieved when $\beta_1=\SI{1.94}{\radian}$, $\beta_2 = \SI{0.88}{\radian}$, and $\psi = \SI{180}{\degree}$. We then develop a protocol with which to experimentally replicate these theoretically identified operating points, which we describe in Methods.

The measured spectrum of the comb corresponding to maximizing the first-order power is shown in Fig. \ref{fig:comb}f. We measure a conversion efficiency to the $+1$ sideband of $\SI{50}{\percent}$, which is five percentage points below the two-tone optimum. We attribute this shortfall to the weakly frequency-dependent transmission of the characterization chain, which is not calibrated out from these measured spectra (Supplementary Information Section 4). Since we consistently measure a higher transmission coefficient for the carrier ($\SI{125}{\mega\hertz}$ in the microwave domain) than the $+1$ sideband (approximately $\SI{1}{\giga\hertz}$ in the microwave domain), this measured conversion efficiency is a slight underestimate (Supplementary Table 5). In Fig. \ref{fig:comb}i and j, we plot the experimentally observed conversion efficiency to the first-order sideband as a function of the driving voltage for each mode. At each pair of voltages, the relative phase is chosen to maximize the conversion efficiency. We note that for these two plots, the axes are based on voltages applied before the driving signal passes through a microwave directional coupler (to combine the two driving frequencies) and thus do not reflect the voltages actually incident upon the device itself.

Under the operating settings of maximum first-order conversion, we also measure the power in each sideband up to the third order as a function of the relative phase between the two driving signals. The results are shown in Fig. \ref{fig:comb}a-e, in which we observe close agreement with a fitted model that accounts for mechanical second-harmonic generation, which causes the measured phase-dependences of the sideband powers to deviate from their expected form (Supplementary Information Section 7). These data indicate the theoretical fact (Supplementary Information Section 7.3) that rotating $\psi$ by $\SI{180}{\degree}$ reflects any generated power spectrum about the carrier frequency, meaning, for instance, that the conversion efficiency achieved to the $+1$ sideband can also be achieved to the $-1$ sideband. 

Next, we identify a flat, seven-line comb produced by $\beta_1=\SI{1.36}{\radian}$, $\beta_2=\SI{1.41}{\radian}$, and $\psi = \SI{90}{\degree}$, which contains $\SI{92}{\percent}$ of the optical power and a peak-to-peak variation of $\SI{2}{\decibel}$ across the seven lines. The results of the measured comb are shown in Fig. \ref{fig:comb}h. Finally, we find a parameter configuration that produces single-sideband frequency shifting to the first order with a nine-percentage-point reduction from the two-tone optimum in exchange for simultaneous nulling of the carrier and the $-1$ sideband. The parameter values are $\beta_1=\SI{2.46}{\radian}$, $\beta_2=\SI{1.10}{\radian}$, and $\psi=\SI{180}{\degree}$. The experimentally observed comb is displayed in Fig. \ref{fig:comb}g with an inset showing its free-running stability over the span of one minute over which the carrier and image remain suppressed by more than $\SI{40}{\decibel}$. At the theoretically identified operating point, the carrier and image are exactly nulled, so the measured suppressions of $\SI{60}{\decibel}$ and $\SI{53}{\decibel}$ (calculated relative to the total input power and output $+1$ power, respectively) are set by how precisely the operating point is reached and how stable it remains within the integration time of the measurement. In future work, feedback control of the driving signals could potentially achieve even higher suppression and maintain it over a longer duration. Additionally, the higher-order sidebands, such as the $-2$ order containing $\SI{11}{\percent}$ of the input power, are not strongly suppressed. Future work can use additional harmonics to null a greater number of sidebands (Supplementary Information Section 8.2).

\subsection*{Scaling to more harmonics}

Inclusion of higher modulation harmonics expands the range of possible output spectra and improves key performance metrics. For example, with three harmonics, a conversion efficiency of $\SI{66}{\percent}$ can be achieved in shifting to either of the first-order sidebands. The lithographic parameters $w_\text{struct}$ and $w_\text{core}$ suffice to place two resonances in harmonic alignment but generically cannot enforce a third. Rather than seek additional degrees of freedom within one structure, we envision separating each harmonic into its own optimized structure (Fig. \ref{fig:more_harmonics}a). These structures can be designed for a particular harmonic by optimizing its optomechanical coupling, electromechanical coupling, microwave impedance matching, and mechanical quality factor. Then, a single optical waveguide can travel through them in series, which is mathematically equivalent to simultaneous modulation at each frequency within the same structure. Each structure occupies a $\SI{5}{\micro\metre}\times\SI{2}{\milli\metre}$ area, so concatenating, for instance, three of them as in Fig. \ref{fig:more_harmonics}a only marginally increases the overall footprint to approximately $\SI{15}{\micro\metre}\times\SI{2}{\milli\metre}$.

To experimentally support the feasibility of this approach, we characterize a phase modulator in a modified device geometry ($w_\text{struct}=\SI{900}{\nano\metre}$ and $w_\text{core}=\SI{600}{\nano\metre}$) that has an optomechanically coupled resonance at $\SI{3.39}{\giga\hertz}$, which is at the third harmonic of Mode 1 for the die from which it originates. The structure's simulated mechanical mode and transverse-electric optical mode are shown in Fig. \ref{fig:more_harmonics}b and c, and its half-wave power as a function of driving frequency is shown along with those of Mode 1 and Mode 2 from a device on the same die with their frequency axes scaled to show the degree of harmonic alignment (Fig. \ref{fig:more_harmonics}d). We also measure the modulation depth as a function of applied voltage on resonance (Fig. \ref{fig:more_harmonics}e). At a drive voltage of $\SI{1.5}{\volt}$, the modulation depth exceeds $\SI{0.6}{\radian}$, above the $\beta_3=\SI{0.58}{\radian}$ required for optimal three-harmonic frequency-shifting conversion efficiency (Fig. \ref{fig:advantages}a). This experimentally confirms that three-harmonic modulation is possible in this platform with no additional design complications besides routing an optical waveguide through both the $w_\text{struct}=\SI{1.4}{\micro\metre}$ device and the $w_\text{struct}=\SI{900}{\nano\metre}$ device.

\section*{Discussion}

Resonant enhancement enables efficient gigahertz-frequency acousto-optic modulation in a SiN platform, but it has so far restricted operation to a single drive tone and thus severely limited which spectra can be generated. Here we unlock a much larger space of output spectra by engineering the mechanical response of an acousto-optic microstructure to contain harmonically spaced resonances, which we utilize to perform multi-tone phase modulation. 

\begin{figure}
   \includegraphics[width=85mm]{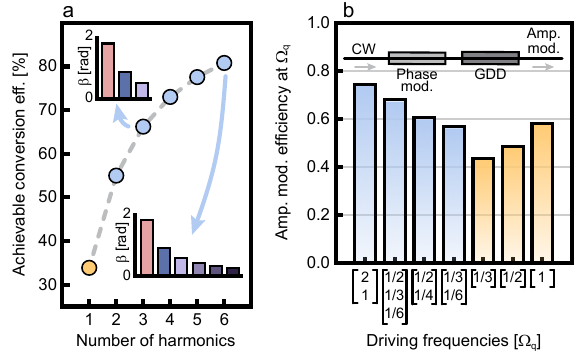}
   \caption{{\bf Application-specific advantages of multi-harmonic modulation.} {\bf a} Maximum achievable single-sideband conversion efficiency versus the number of harmonically related drive tones, obtained by numerical optimization over the modulation depths $\{\beta_n\}$ and relative phases $\{\psi_n\}$. The largest single increase, from $\SI{33.9}{\percent}$ to $\SI{55}{\percent}$, comes from adding the second harmonic. Insets show the modulation depths $\beta_n$ required at the optima for three and six harmonics. {\bf b} Calculated amplitude modulation generated at a target frequency $\Omega_\text{q}$ after a group-delay-dispersion element for drives composed of the indicated sub-harmonics and harmonics of $\Omega$ (blue) compared to driving at single tones (orange). Inset: the multi-harmonic phase modulator followed by a group-delay-dispersion element that converts phase to amplitude modulation.}
   \label{fig:advantages}
\end{figure}

The three spectra we identify and generate in this work illustrate the extent of this expanded design space. Maximizing conversion to the $+1$ sideband transfers $\SI{50}{\percent}$ of the optical power to a single sideband, a $1.5\times$ improvement over the $\SI{33.9}{\percent}$ ceiling of single-tone modulation. Figure \ref{fig:advantages}a shows how this maximum further improves with an increasing number of drive tones, which can be straightforwardly achieved in future work following the architecture presented in Fig. \ref{fig:more_harmonics}a (Supplementary Information Section 8.3). Slightly relaxing that maximum allows for an operating point at which the carrier and the $-1$ sideband null simultaneously. In this configuration, we measure $\SI{60}{\decibel}$ carrier suppression and $\SI{53}{\decibel}$ unwanted-sideband suppression at $\SI{40}{\percent}$ conversion efficiency. To our knowledge, these are the largest simultaneous suppressions reported for an integrated frequency shifter \cite{doi:10.1126/sciadv.ade4454}. A third setting produces a seven-line comb with a peak-to-peak power variation within $\SI{2}{\decibel}$ that contains greater than $\SI{90}{\percent}$ of the input power.

Access to multiple harmonics can improve the efficiency with which single-qubit rotations are generated. Hyperfine qubits underpin several leading quantum computing platforms, including those based on $^{171}$Yb$^{+}$ at \SI{12.643}{\giga\hertz}, $^{133}$Cs at \SI{9.193}{\giga\hertz}, $^{137}$Ba$^{+}$ at \SI{8.037}{\giga\hertz}, and $^{87}$Rb  at \SI{6.835}{\giga\hertz} \cite{moses_race-track_2023, graham_multi-qubit_2022, ransford_98-qubit_2026, bluvstein_fault-tolerant_2026}. In these systems, single-qubit gates are driven by phase modulating a continuous-wave laser and converting that phase modulation to amplitude modulation at the qubit transition frequency $\Omega_\text{q}$. The strength of this amplitude modulation is directly proportional to the gate's Rabi frequency, and conversion can be achieved by notch-filtering a comb line, interferometric filtering, or imparting a frequency-dependent phase with a dispersive element \cite{levine_dispersive_2022}. The last of these, implemented with a group-delay-dispersion (GDD) element, performs the conversion more efficiently than the other two and is used in state-of-the-art quantum information processors \cite{bluvstein_fault-tolerant_2026, manetsch_tweezer_2025}. 

A GDD element works by advancing the phase of each comb line in proportion to the square of its detuning, so that the pairwise beat notes formed by comb lines separated by $\Omega_\text{q}$ add constructively rather than canceling. We show in Fig. \ref{fig:advantages}b (with calculations described in Supplementary Information Section 8.4) that drives composed of combinations of harmonics and sub-harmonics of $\Omega_\text{q}$ generate amplitude modulation comparable to or larger than that produced by a drive at $\Omega_\text{q}$ itself. For example, a drive at $\Omega_\text{q}/2$ and $\Omega_\text{q}/4$ produces more amplitude modulation at $\Omega_\text{q}$ than a drive at $\Omega_\text{q}$ itself, and a drive at $\Omega_\text{q}/3$ and $\Omega_\text{q}/6$ reaches $\SI{97}{\percent}$ of the direct-drive value. Intuitively, this is possible because the multi-harmonic drive can more flexibly configure the amplitudes and phases of the comb lines such that significant beat notes align constructively upon propagation through the GDD element.

The above analysis shows that driving at multiple sub-harmonics allows a device to address a qubit transition even with no mechanical resonance at its frequency, without sacrificing the Rabi frequency obtained per unit optical power delivered to the modulator. This is important because at higher frequencies mechanical modes have lower quality factors and strain profiles with faster spatial variations that could otherwise inhibit this platform's and other acousto-optic platforms' ability to address commonly utilized qubit species. The devices reported here are a direct example of this, as the $^{87}\text{Rb}$ hyperfine ground-state splitting is $\SI{6.835}{\giga\hertz}$, meaning the third and sixth sub-harmonics are $\SI{2.278}{\giga\hertz}$ and $\SI{1.139}{\giga\hertz}$, respectively. Both frequencies lie within a mechanical linewidth of $\Omega_2$ and $\Omega_1$ for several of the measured devices. Such a device, followed by an integrated GDD element such as a chirped Bragg grating \cite{cheng_communication-ready_2026, geng_meter-long_2026}, can generate Rabi frequencies at $\SI{97}{\percent}$ of the efficiency of a modulator driven directly at the transition frequency. 

Overall, this new class of devices adds qualitatively new capabilities to a platform that already demonstrably supports watt-level optical power at visible wavelengths \cite{freedman_gigahertz-frequency_2025, zhao_integrated_2025}, monolithic CMOS electronic integration \cite{zimmermann_monolithic_2026}, and several complementary components \cite{dong_high-speed_2022, stanfield_cmos-compatible_2019} that together make it a uniquely promising candidate for scalable quantum control of trapped ions, neutral atoms, and solid-state emitters.

\section*{Methods}

\subsection*{Generating programmed spectra}

Numerically identified target output spectra are realized with closed-loop optimization on the device. The drive state is specified by three parameters, namely the powers of the microwave signals at $\Omega$ and $2\Omega$ and their relative phase. For each candidate state, the output optical spectrum is measured with the characterization setup shown in Fig. \ref{fig:char}d, and a cost function is evaluated with respect to the target. Because the relative phase produced by the signal generator (Berkeley Nucleonics 855b) is scrambled when a microwave power is changed, the search has an inner loop in which the relative phase is optimized for a given pair of powers, and an outer loop in which the microwave powers are optimized. The inner phase loop is optimized via a coarse grid search from $0-\SI{360}{\degree}$ that is refined by adaptive-step descent. The outer loop is optimized by the Nelder-Mead method. 

\subsection*{Numerical optimizations}

We numerically optimize the modulation parameters $\{\beta_n\}$ and $\{\psi_n\}$ for a given spectrum by assuming phase modulation of the form
\begin{align}
    \phi(t) = \sum_{n=1}^N \beta_n \sin(n\Omega t + \psi_n),
\end{align}
which produces an output spectrum for which the $p$th sideband has field amplitude $a_p$. We then
choose an objective function based off the target and optimize the parameters with a suitable optimization algorithm. For optimizing the frequency-shifting conversion efficiency in Fig. \ref{fig:advantages}a, we simply minimize $\mathcal{L} = -\eta_1=-|a_1|^2$ while limiting each modulation depth to at most $\SI{3}{\radian}$. For single-sideband frequency shifting (Supplementary Information Section 8.2), we minimize $-\eta_1$ subject to suppressing a specified set $S$ of sideband orders with the objective $\mathcal{L}=-\eta_1+\lambda\sum_{p\in S}|a_p|^2$. In this case, we limit each modulation depth to at most $\SI{4}{\radian}$. Lastly, for optimizing the amplitude modulation produced by a GDD element (Fig. \ref{fig:advantages}b), we introduce an additional parameter $\alpha$, that fully encapsulates the order-dependent phase imparted by the GDD element $\theta_n=\alpha n^2$. We then maximize the amplitude modulation produced at a target frequency according to the expression given in Supplementary Information Section 8.4, and we limit each modulation depth to at most $\SI{6}{\radian}$ and $\text{GDD}\cdot \Omega_q^2$ to at most 15. This bound on $\text{GDD}$ ensures the required dispersion is physically realizable for typical qubit transition frequencies \cite{geng_meter-long_2026}.

\subsection*{Optical and mechanical simulations}
Simulations of the device’s optical modes and mechanical modes are
performed with COMSOL Multiphysics FEM software using the Wave
Optics, Solid Mechanics, and Piezoelectricity modules.

\section*{Funding} This work and all authors were supported by the U.S. Department of Energy, Office of Science, National Quantum Information Science Research Centers, Quantum Systems Accelerator. Sandia National Laboratories is a multimission laboratory managed and operated by National Technology \& Engineering Solutions of Sandia, LLC, a wholly owned subsidiary of Honeywell International, Inc., for the U.S. DOE's National Nuclear Security Administration under contract DE-NA-0003525. The views in the article do not necessarily represent the views of the U.S. Department of Energy or the United States Government.

\section*{Competing interests} M.E. is a founder and shareholder of LightlogiQ Inc., which develops technology relevant to the research presented in this study. N.O. is a founder and shareholder of Manzano Systems Inc., which develops technology relevant to the research presented in this study. The University of Colorado Boulder has filed a provisional patent application covering the devices and methods described in this work, with M.E. and J.F. listed as inventors. The application is pending and unpublished; the application number is withheld pending non-provisional filing. No other authors declare competing interests.

\section*{Data availability} Data underlying the results presented in this paper are available from the corresponding authors upon reasonable request.

\section*{Author contributions} M.E. and J.F. conceived the project, and M.E. supervised the work. M.E. and N.O. produced the qualitative device designs. J.F. and M.S. built device models and performed simulations of mechanical frequencies and optomechanical coupling. D.D. laid out the devices for fabrication, and A.L. coordinated the fabrication. J.F. built and conducted the experiments. J.F. analyzed the results and wrote the manuscript with guidance from M.E. and input from all authors.

\newpage

\bibliography{refs}

\begin{thebibliography}{52}%
\makeatletter
\providecommand \@ifxundefined [1]{%
 \@ifx{#1\undefined}
}%
\providecommand \@ifnum [1]{%
 \ifnum #1\expandafter \@firstoftwo
 \else \expandafter \@secondoftwo
 \fi
}%
\providecommand \@ifx [1]{%
 \ifx #1\expandafter \@firstoftwo
 \else \expandafter \@secondoftwo
 \fi
}%
\providecommand \natexlab [1]{#1}%
\providecommand \enquote  [1]{``#1''}%
\providecommand \bibnamefont  [1]{#1}%
\providecommand \bibfnamefont [1]{#1}%
\providecommand \citenamefont [1]{#1}%
\providecommand \href@noop [0]{\@secondoftwo}%
\providecommand \href [0]{\begingroup \@sanitize@url \@href}%
\providecommand \@href[1]{\@@startlink{#1}\@@href}%
\providecommand \@@href[1]{\endgroup#1\@@endlink}%
\providecommand \@sanitize@url [0]{\catcode `\\12\catcode `\$12\catcode `\&12\catcode `\#12\catcode `\^12\catcode `\_12\catcode `\%12\relax}%
\providecommand \@@startlink[1]{}%
\providecommand \@@endlink[0]{}%
\providecommand \url  [0]{\begingroup\@sanitize@url \@url }%
\providecommand \@url [1]{\endgroup\@href {#1}{\urlprefix }}%
\providecommand \urlprefix  [0]{URL }%
\providecommand \Eprint [0]{\href }%
\providecommand \doibase [0]{https://doi.org/}%
\providecommand \selectlanguage [0]{\@gobble}%
\providecommand \bibinfo  [0]{\@secondoftwo}%
\providecommand \bibfield  [0]{\@secondoftwo}%
\providecommand \translation [1]{[#1]}%
\providecommand \BibitemOpen [0]{}%
\providecommand \bibitemStop [0]{}%
\providecommand \bibitemNoStop [0]{.\EOS\space}%
\providecommand \EOS [0]{\spacefactor3000\relax}%
\providecommand \BibitemShut  [1]{\csname bibitem#1\endcsname}%
\let\auto@bib@innerbib\@empty
%</preamble>
\bibitem [{\citenamefont {Park}\ \emph {et~al.}(2024)\citenamefont {Park}, \citenamefont {Notaros}, \citenamefont {Mohanty}, \citenamefont {Kim}, \citenamefont {Notaros},\ and\ \citenamefont {Mouradian}}]{park_technologies_2024}%
  \BibitemOpen
  \bibfield  {author} {\bibinfo {author} {\bibfnamefont {S.}~\bibnamefont {Park}}, \bibinfo {author} {\bibfnamefont {M.}~\bibnamefont {Notaros}}, \bibinfo {author} {\bibfnamefont {A.}~\bibnamefont {Mohanty}}, \bibinfo {author} {\bibfnamefont {D.}~\bibnamefont {Kim}}, \bibinfo {author} {\bibfnamefont {J.}~\bibnamefont {Notaros}},\ and\ \bibinfo {author} {\bibfnamefont {S.}~\bibnamefont {Mouradian}},\ }\href {https://doi.org/10.1016/j.pquantelec.2024.100534} {\bibfield  {journal} {\bibinfo  {journal} {Progress in Quantum Electronics}\ }\textbf {\bibinfo {volume} {97}},\ \bibinfo {pages} {100534} (\bibinfo {year} {2024})}\BibitemShut {NoStop}%
\bibitem [{\citenamefont {Bruzewicz}\ \emph {et~al.}(2019)\citenamefont {Bruzewicz}, \citenamefont {Chiaverini}, \citenamefont {McConnell},\ and\ \citenamefont {Sage}}]{bruzewicz_trapped-ion_2019}%
  \BibitemOpen
  \bibfield  {author} {\bibinfo {author} {\bibfnamefont {C.~D.}\ \bibnamefont {Bruzewicz}}, \bibinfo {author} {\bibfnamefont {J.}~\bibnamefont {Chiaverini}}, \bibinfo {author} {\bibfnamefont {R.}~\bibnamefont {McConnell}},\ and\ \bibinfo {author} {\bibfnamefont {J.~M.}\ \bibnamefont {Sage}},\ }\href {https://doi.org/10.1063/1.5088164} {\bibfield  {journal} {\bibinfo  {journal} {Applied Physics Reviews}\ }\textbf {\bibinfo {volume} {6}},\ \bibinfo {pages} {021314} (\bibinfo {year} {2019})}\BibitemShut {NoStop}%
\bibitem [{\citenamefont {Menssen}\ \emph {et~al.}(2026)\citenamefont {Menssen} \emph {et~al.}}]{menssen_strategic_2026}%
  \BibitemOpen
  \bibfield  {author} {\bibinfo {author} {\bibfnamefont {A.~J.}\ \bibnamefont {Menssen}} \emph {et~al.},\ }\href {https://doi.org/10.48550/arXiv.2607.21554} {\bibinfo {title} {Strategic {Plan} for {Neutral} {Atom} {Quantum} {Computation}}} (\bibinfo {year} {2026}),\ \bibinfo {note} {arXiv:2607.21554 [quant-ph]}\BibitemShut {NoStop}%
\bibitem [{\citenamefont {Ransford}\ \emph {et~al.}(2026)\citenamefont {Ransford} \emph {et~al.}}]{ransford_98-qubit_2026}%
  \BibitemOpen
  \bibfield  {author} {\bibinfo {author} {\bibfnamefont {A.}~\bibnamefont {Ransford}} \emph {et~al.},\ }\href {https://doi.org/10.1038/s41586-026-10676-4} {\bibfield  {journal} {\bibinfo  {journal} {Nature}\ }\textbf {\bibinfo {volume} {655}},\ \bibinfo {pages} {81} (\bibinfo {year} {2026})}\BibitemShut {NoStop}%
\bibitem [{\citenamefont {Bluvstein}\ \emph {et~al.}(2026)\citenamefont {Bluvstein}, \citenamefont {Geim}, \citenamefont {Li}, \citenamefont {Evered}, \citenamefont {Bonilla~Ataides}, \citenamefont {Baranes}, \citenamefont {Gu}, \citenamefont {Manovitz}, \citenamefont {Xu}, \citenamefont {Kalinowski}, \citenamefont {Majidy}, \citenamefont {Kokail}, \citenamefont {Maskara}, \citenamefont {Trapp}, \citenamefont {Stewart}, \citenamefont {Hollerith}, \citenamefont {Zhou}, \citenamefont {Gullans}, \citenamefont {Yelin}, \citenamefont {Greiner}, \citenamefont {Vuletić}, \citenamefont {Cain},\ and\ \citenamefont {Lukin}}]{bluvstein_fault-tolerant_2026}%
  \BibitemOpen
  \bibfield  {author} {\bibinfo {author} {\bibfnamefont {D.}~\bibnamefont {Bluvstein}}, \bibinfo {author} {\bibfnamefont {A.~A.}\ \bibnamefont {Geim}}, \bibinfo {author} {\bibfnamefont {S.~H.}\ \bibnamefont {Li}}, \bibinfo {author} {\bibfnamefont {S.~J.}\ \bibnamefont {Evered}}, \bibinfo {author} {\bibfnamefont {J.~P.}\ \bibnamefont {Bonilla~Ataides}}, \bibinfo {author} {\bibfnamefont {G.}~\bibnamefont {Baranes}}, \bibinfo {author} {\bibfnamefont {A.}~\bibnamefont {Gu}}, \bibinfo {author} {\bibfnamefont {T.}~\bibnamefont {Manovitz}}, \bibinfo {author} {\bibfnamefont {M.}~\bibnamefont {Xu}}, \bibinfo {author} {\bibfnamefont {M.}~\bibnamefont {Kalinowski}}, \bibinfo {author} {\bibfnamefont {S.}~\bibnamefont {Majidy}}, \bibinfo {author} {\bibfnamefont {C.}~\bibnamefont {Kokail}}, \bibinfo {author} {\bibfnamefont {N.}~\bibnamefont {Maskara}}, \bibinfo {author} {\bibfnamefont {E.~C.}\ \bibnamefont {Trapp}}, \bibinfo {author} {\bibfnamefont {L.~M.}\ \bibnamefont {Stewart}}, \bibinfo {author} {\bibfnamefont
  {S.}~\bibnamefont {Hollerith}}, \bibinfo {author} {\bibfnamefont {H.}~\bibnamefont {Zhou}}, \bibinfo {author} {\bibfnamefont {M.~J.}\ \bibnamefont {Gullans}}, \bibinfo {author} {\bibfnamefont {S.~F.}\ \bibnamefont {Yelin}}, \bibinfo {author} {\bibfnamefont {M.}~\bibnamefont {Greiner}}, \bibinfo {author} {\bibfnamefont {V.}~\bibnamefont {Vuletić}}, \bibinfo {author} {\bibfnamefont {M.}~\bibnamefont {Cain}},\ and\ \bibinfo {author} {\bibfnamefont {M.~D.}\ \bibnamefont {Lukin}},\ }\href {https://doi.org/10.1038/s41586-025-09848-5} {\bibfield  {journal} {\bibinfo  {journal} {Nature}\ }\textbf {\bibinfo {volume} {649}},\ \bibinfo {pages} {39} (\bibinfo {year} {2026})}\BibitemShut {NoStop}%
\bibitem [{\citenamefont {Li}\ \emph {et~al.}(2024)\citenamefont {Li}, \citenamefont {Santis}, \citenamefont {Harris}, \citenamefont {Chen}, \citenamefont {Gao}, \citenamefont {Christen}, \citenamefont {Choi}, \citenamefont {Trusheim}, \citenamefont {Song}, \citenamefont {Errando-Herranz}, \citenamefont {Du}, \citenamefont {Hu}, \citenamefont {Clark}, \citenamefont {Ibrahim}, \citenamefont {Gilbert}, \citenamefont {Han},\ and\ \citenamefont {Englund}}]{li_heterogeneous_2024}%
  \BibitemOpen
  \bibfield  {author} {\bibinfo {author} {\bibfnamefont {L.}~\bibnamefont {Li}}, \bibinfo {author} {\bibfnamefont {L.~D.}\ \bibnamefont {Santis}}, \bibinfo {author} {\bibfnamefont {I.~B.~W.}\ \bibnamefont {Harris}}, \bibinfo {author} {\bibfnamefont {K.~C.}\ \bibnamefont {Chen}}, \bibinfo {author} {\bibfnamefont {Y.}~\bibnamefont {Gao}}, \bibinfo {author} {\bibfnamefont {I.}~\bibnamefont {Christen}}, \bibinfo {author} {\bibfnamefont {H.}~\bibnamefont {Choi}}, \bibinfo {author} {\bibfnamefont {M.}~\bibnamefont {Trusheim}}, \bibinfo {author} {\bibfnamefont {Y.}~\bibnamefont {Song}}, \bibinfo {author} {\bibfnamefont {C.}~\bibnamefont {Errando-Herranz}}, \bibinfo {author} {\bibfnamefont {J.}~\bibnamefont {Du}}, \bibinfo {author} {\bibfnamefont {Y.}~\bibnamefont {Hu}}, \bibinfo {author} {\bibfnamefont {G.}~\bibnamefont {Clark}}, \bibinfo {author} {\bibfnamefont {M.~I.}\ \bibnamefont {Ibrahim}}, \bibinfo {author} {\bibfnamefont {G.}~\bibnamefont {Gilbert}}, \bibinfo {author} {\bibfnamefont {R.}~\bibnamefont {Han}},\
  and\ \bibinfo {author} {\bibfnamefont {D.}~\bibnamefont {Englund}},\ }\href {https://doi.org/10.1038/s41586-024-07371-7} {\bibfield  {journal} {\bibinfo  {journal} {Nature}\ }\textbf {\bibinfo {volume} {630}},\ \bibinfo {pages} {70} (\bibinfo {year} {2024})}\BibitemShut {NoStop}%
\bibitem [{\citenamefont {Levine}\ \emph {et~al.}(2022)\citenamefont {Levine}, \citenamefont {Bluvstein}, \citenamefont {Keesling}, \citenamefont {Wang}, \citenamefont {Ebadi}, \citenamefont {Semeghini}, \citenamefont {Omran}, \citenamefont {Greiner}, \citenamefont {Vuletić},\ and\ \citenamefont {Lukin}}]{levine_dispersive_2022}%
  \BibitemOpen
  \bibfield  {author} {\bibinfo {author} {\bibfnamefont {H.}~\bibnamefont {Levine}}, \bibinfo {author} {\bibfnamefont {D.}~\bibnamefont {Bluvstein}}, \bibinfo {author} {\bibfnamefont {A.}~\bibnamefont {Keesling}}, \bibinfo {author} {\bibfnamefont {T.~T.}\ \bibnamefont {Wang}}, \bibinfo {author} {\bibfnamefont {S.}~\bibnamefont {Ebadi}}, \bibinfo {author} {\bibfnamefont {G.}~\bibnamefont {Semeghini}}, \bibinfo {author} {\bibfnamefont {A.}~\bibnamefont {Omran}}, \bibinfo {author} {\bibfnamefont {M.}~\bibnamefont {Greiner}}, \bibinfo {author} {\bibfnamefont {V.}~\bibnamefont {Vuletić}},\ and\ \bibinfo {author} {\bibfnamefont {M.~D.}\ \bibnamefont {Lukin}},\ }\href {https://doi.org/10.1103/PhysRevA.105.032618} {\bibfield  {journal} {\bibinfo  {journal} {Physical Review A}\ }\textbf {\bibinfo {volume} {105}},\ \bibinfo {pages} {032618} (\bibinfo {year} {2022})}\BibitemShut {NoStop}%
\bibitem [{\citenamefont {Kodigala}\ \emph {et~al.}(2024)\citenamefont {Kodigala}, \citenamefont {Gehl}, \citenamefont {Hoth}, \citenamefont {Lee}, \citenamefont {DeRose}, \citenamefont {Pomerene}, \citenamefont {Dallo}, \citenamefont {Trotter}, \citenamefont {Starbuck}, \citenamefont {Biedermann}, \citenamefont {Schwindt},\ and\ \citenamefont {Lentine}}]{doi:10.1126/sciadv.ade4454}%
  \BibitemOpen
  \bibfield  {author} {\bibinfo {author} {\bibfnamefont {A.}~\bibnamefont {Kodigala}}, \bibinfo {author} {\bibfnamefont {M.}~\bibnamefont {Gehl}}, \bibinfo {author} {\bibfnamefont {G.~W.}\ \bibnamefont {Hoth}}, \bibinfo {author} {\bibfnamefont {J.}~\bibnamefont {Lee}}, \bibinfo {author} {\bibfnamefont {C.~T.}\ \bibnamefont {DeRose}}, \bibinfo {author} {\bibfnamefont {A.}~\bibnamefont {Pomerene}}, \bibinfo {author} {\bibfnamefont {C.}~\bibnamefont {Dallo}}, \bibinfo {author} {\bibfnamefont {D.}~\bibnamefont {Trotter}}, \bibinfo {author} {\bibfnamefont {A.~L.}\ \bibnamefont {Starbuck}}, \bibinfo {author} {\bibfnamefont {G.}~\bibnamefont {Biedermann}}, \bibinfo {author} {\bibfnamefont {P.~D.~D.}\ \bibnamefont {Schwindt}},\ and\ \bibinfo {author} {\bibfnamefont {A.~L.}\ \bibnamefont {Lentine}},\ }\href {https://doi.org/10.1126/sciadv.ade4454} {\bibfield  {journal} {\bibinfo  {journal} {Science Advances}\ }\textbf {\bibinfo {volume} {10}},\ \bibinfo {pages} {eade4454} (\bibinfo {year} {2024})}\BibitemShut
  {NoStop}%
\bibitem [{\citenamefont {Manetsch}\ \emph {et~al.}(2025)\citenamefont {Manetsch}, \citenamefont {Nomura}, \citenamefont {Bataille}, \citenamefont {Lv}, \citenamefont {Leung},\ and\ \citenamefont {Endres}}]{manetsch_tweezer_2025}%
  \BibitemOpen
  \bibfield  {author} {\bibinfo {author} {\bibfnamefont {H.~J.}\ \bibnamefont {Manetsch}}, \bibinfo {author} {\bibfnamefont {G.}~\bibnamefont {Nomura}}, \bibinfo {author} {\bibfnamefont {E.}~\bibnamefont {Bataille}}, \bibinfo {author} {\bibfnamefont {X.}~\bibnamefont {Lv}}, \bibinfo {author} {\bibfnamefont {K.~H.}\ \bibnamefont {Leung}},\ and\ \bibinfo {author} {\bibfnamefont {M.}~\bibnamefont {Endres}},\ }\href {https://doi.org/10.1038/s41586-025-09641-4} {\bibfield  {journal} {\bibinfo  {journal} {Nature}\ }\textbf {\bibinfo {volume} {647}},\ \bibinfo {pages} {60} (\bibinfo {year} {2025})}\BibitemShut {NoStop}%
\bibitem [{\citenamefont {Chiu}\ \emph {et~al.}(2025)\citenamefont {Chiu}, \citenamefont {Trapp}, \citenamefont {Guo}, \citenamefont {Abobeih}, \citenamefont {Stewart}, \citenamefont {Hollerith}, \citenamefont {Stroganov}, \citenamefont {Kalinowski}, \citenamefont {Geim}, \citenamefont {Evered}, \citenamefont {Li}, \citenamefont {Lyu}, \citenamefont {Peters}, \citenamefont {Bluvstein}, \citenamefont {Wang}, \citenamefont {Greiner}, \citenamefont {Vuletić},\ and\ \citenamefont {Lukin}}]{chiu_continuous_2025}%
  \BibitemOpen
  \bibfield  {author} {\bibinfo {author} {\bibfnamefont {N.-C.}\ \bibnamefont {Chiu}}, \bibinfo {author} {\bibfnamefont {E.~C.}\ \bibnamefont {Trapp}}, \bibinfo {author} {\bibfnamefont {J.}~\bibnamefont {Guo}}, \bibinfo {author} {\bibfnamefont {M.~H.}\ \bibnamefont {Abobeih}}, \bibinfo {author} {\bibfnamefont {L.~M.}\ \bibnamefont {Stewart}}, \bibinfo {author} {\bibfnamefont {S.}~\bibnamefont {Hollerith}}, \bibinfo {author} {\bibfnamefont {P.~L.}\ \bibnamefont {Stroganov}}, \bibinfo {author} {\bibfnamefont {M.}~\bibnamefont {Kalinowski}}, \bibinfo {author} {\bibfnamefont {A.~A.}\ \bibnamefont {Geim}}, \bibinfo {author} {\bibfnamefont {S.~J.}\ \bibnamefont {Evered}}, \bibinfo {author} {\bibfnamefont {S.~H.}\ \bibnamefont {Li}}, \bibinfo {author} {\bibfnamefont {X.}~\bibnamefont {Lyu}}, \bibinfo {author} {\bibfnamefont {L.~M.}\ \bibnamefont {Peters}}, \bibinfo {author} {\bibfnamefont {D.}~\bibnamefont {Bluvstein}}, \bibinfo {author} {\bibfnamefont {T.~T.}\ \bibnamefont {Wang}}, \bibinfo {author} {\bibfnamefont
  {M.}~\bibnamefont {Greiner}}, \bibinfo {author} {\bibfnamefont {V.}~\bibnamefont {Vuletić}},\ and\ \bibinfo {author} {\bibfnamefont {M.~D.}\ \bibnamefont {Lukin}},\ }\href {https://doi.org/10.1038/s41586-025-09596-6} {\bibfield  {journal} {\bibinfo  {journal} {Nature}\ }\textbf {\bibinfo {volume} {646}},\ \bibinfo {pages} {1075} (\bibinfo {year} {2025})}\BibitemShut {NoStop}%
\bibitem [{\citenamefont {Holman}\ \emph {et~al.}(2026)\citenamefont {Holman}, \citenamefont {Xu}, \citenamefont {Sun}, \citenamefont {Wu}, \citenamefont {Wang}, \citenamefont {Zhu}, \citenamefont {Seo}, \citenamefont {Yu},\ and\ \citenamefont {Will}}]{holman_trapping_2026}%
  \BibitemOpen
  \bibfield  {author} {\bibinfo {author} {\bibfnamefont {A.}~\bibnamefont {Holman}}, \bibinfo {author} {\bibfnamefont {Y.}~\bibnamefont {Xu}}, \bibinfo {author} {\bibfnamefont {X.}~\bibnamefont {Sun}}, \bibinfo {author} {\bibfnamefont {J.}~\bibnamefont {Wu}}, \bibinfo {author} {\bibfnamefont {M.}~\bibnamefont {Wang}}, \bibinfo {author} {\bibfnamefont {Z.}~\bibnamefont {Zhu}}, \bibinfo {author} {\bibfnamefont {B.}~\bibnamefont {Seo}}, \bibinfo {author} {\bibfnamefont {N.}~\bibnamefont {Yu}},\ and\ \bibinfo {author} {\bibfnamefont {S.}~\bibnamefont {Will}},\ }\href {https://doi.org/10.1038/s41586-025-09961-5} {\bibfield  {journal} {\bibinfo  {journal} {Nature}\ }\textbf {\bibinfo {volume} {649}},\ \bibinfo {pages} {859} (\bibinfo {year} {2026})}\BibitemShut {NoStop}%
\bibitem [{\citenamefont {Hall}\ \emph {et~al.}(2025)\citenamefont {Hall}, \citenamefont {He}, \citenamefont {Chiu}, \citenamefont {Morrissey}, \citenamefont {Gradkowski}, \citenamefont {Jayaram}, \citenamefont {Cheng}, \citenamefont {Hoang}, \citenamefont {Bergman}, \citenamefont {O’Brien},\ and\ \citenamefont {Hosseini}}]{hall_innovations_2025}%
  \BibitemOpen
  \bibfield  {author} {\bibinfo {author} {\bibfnamefont {M.~L.}\ \bibnamefont {Hall}}, \bibinfo {author} {\bibfnamefont {X.}~\bibnamefont {He}}, \bibinfo {author} {\bibfnamefont {C.~P.}\ \bibnamefont {Chiu}}, \bibinfo {author} {\bibfnamefont {P.~E.}\ \bibnamefont {Morrissey}}, \bibinfo {author} {\bibfnamefont {K.}~\bibnamefont {Gradkowski}}, \bibinfo {author} {\bibfnamefont {V.}~\bibnamefont {Jayaram}}, \bibinfo {author} {\bibfnamefont {F.}~\bibnamefont {Cheng}}, \bibinfo {author} {\bibfnamefont {T.~T.}\ \bibnamefont {Hoang}}, \bibinfo {author} {\bibfnamefont {K.}~\bibnamefont {Bergman}}, \bibinfo {author} {\bibfnamefont {P.}~\bibnamefont {O’Brien}},\ and\ \bibinfo {author} {\bibfnamefont {K.}~\bibnamefont {Hosseini}},\ }\bibfield  {journal} {\bibinfo  {journal} {Optical Interconnects and Packaging 2025}\ }\bibinfo {series} {Proceedings of {SPIE} - {The} {International} {Society} for {Optical} {Engineering}},\ \href {https://doi.org/10.1117/12.3040471} {10.1117/12.3040471} (\bibinfo {year}
  {2025})\BibitemShut {NoStop}%
\bibitem [{\citenamefont {Blumenthal}\ \emph {et~al.}(2018)\citenamefont {Blumenthal}, \citenamefont {Heideman}, \citenamefont {Geuzebroek}, \citenamefont {Leinse},\ and\ \citenamefont {Roeloffzen}}]{blumenthal_silicon_2018}%
  \BibitemOpen
  \bibfield  {author} {\bibinfo {author} {\bibfnamefont {D.~J.}\ \bibnamefont {Blumenthal}}, \bibinfo {author} {\bibfnamefont {R.}~\bibnamefont {Heideman}}, \bibinfo {author} {\bibfnamefont {D.}~\bibnamefont {Geuzebroek}}, \bibinfo {author} {\bibfnamefont {A.}~\bibnamefont {Leinse}},\ and\ \bibinfo {author} {\bibfnamefont {C.}~\bibnamefont {Roeloffzen}},\ }\href {https://doi.org/10.1109/JPROC.2018.2861576} {\bibfield  {journal} {\bibinfo  {journal} {Proceedings of the IEEE}\ }\textbf {\bibinfo {volume} {106}},\ \bibinfo {pages} {2209} (\bibinfo {year} {2018})}\BibitemShut {NoStop}%
\bibitem [{\citenamefont {Freedman}\ \emph {et~al.}(2025)\citenamefont {Freedman}, \citenamefont {Storey}, \citenamefont {Dominguez}, \citenamefont {Leenheer}, \citenamefont {Magri}, \citenamefont {Otterstrom},\ and\ \citenamefont {Eichenfield}}]{freedman_gigahertz-frequency_2025}%
  \BibitemOpen
  \bibfield  {author} {\bibinfo {author} {\bibfnamefont {J.~M.}\ \bibnamefont {Freedman}}, \bibinfo {author} {\bibfnamefont {M.~J.}\ \bibnamefont {Storey}}, \bibinfo {author} {\bibfnamefont {D.}~\bibnamefont {Dominguez}}, \bibinfo {author} {\bibfnamefont {A.}~\bibnamefont {Leenheer}}, \bibinfo {author} {\bibfnamefont {S.}~\bibnamefont {Magri}}, \bibinfo {author} {\bibfnamefont {N.~T.}\ \bibnamefont {Otterstrom}},\ and\ \bibinfo {author} {\bibfnamefont {M.}~\bibnamefont {Eichenfield}},\ }\href {https://doi.org/10.1038/s41467-025-65937-z} {\bibfield  {journal} {\bibinfo  {journal} {Nature Communications}\ }\textbf {\bibinfo {volume} {16}},\ \bibinfo {pages} {10959} (\bibinfo {year} {2025})}\BibitemShut {NoStop}%
\bibitem [{\citenamefont {Zhao}\ \emph {et~al.}(2025)\citenamefont {Zhao}, \citenamefont {Singh}, \citenamefont {Singh}, \citenamefont {Thoreen}, \citenamefont {DeAngelo}, \citenamefont {Dominguez}, \citenamefont {Leenheer}, \citenamefont {Peyskens}, \citenamefont {Lukin}, \citenamefont {Englund}, \citenamefont {Eichenfield}, \citenamefont {Gemelke},\ and\ \citenamefont {Wan}}]{zhao_integrated_2025}%
  \BibitemOpen
  \bibfield  {author} {\bibinfo {author} {\bibfnamefont {M.}~\bibnamefont {Zhao}}, \bibinfo {author} {\bibfnamefont {M.}~\bibnamefont {Singh}}, \bibinfo {author} {\bibfnamefont {A.}~\bibnamefont {Singh}}, \bibinfo {author} {\bibfnamefont {H.}~\bibnamefont {Thoreen}}, \bibinfo {author} {\bibfnamefont {R.~J.}\ \bibnamefont {DeAngelo}}, \bibinfo {author} {\bibfnamefont {D.}~\bibnamefont {Dominguez}}, \bibinfo {author} {\bibfnamefont {A.}~\bibnamefont {Leenheer}}, \bibinfo {author} {\bibfnamefont {F.}~\bibnamefont {Peyskens}}, \bibinfo {author} {\bibfnamefont {A.}~\bibnamefont {Lukin}}, \bibinfo {author} {\bibfnamefont {D.}~\bibnamefont {Englund}}, \bibinfo {author} {\bibfnamefont {M.}~\bibnamefont {Eichenfield}}, \bibinfo {author} {\bibfnamefont {N.}~\bibnamefont {Gemelke}},\ and\ \bibinfo {author} {\bibfnamefont {N.~H.}\ \bibnamefont {Wan}},\ }\href {https://doi.org/10.48550/arXiv.2508.09920} {\bibinfo {title} {An integrated photonics platform for high-speed, ultrahigh-extinction, many-channel quantum control}}
  (\bibinfo {year} {2025}),\ \bibinfo {note} {arXiv:2508.09920 [quant-ph]}\BibitemShut {NoStop}%
\bibitem [{\citenamefont {Mehta}\ \emph {et~al.}(2020)\citenamefont {Mehta}, \citenamefont {Zhang}, \citenamefont {Malinowski}, \citenamefont {Nguyen}, \citenamefont {Stadler},\ and\ \citenamefont {Home}}]{mehta_integrated_2020}%
  \BibitemOpen
  \bibfield  {author} {\bibinfo {author} {\bibfnamefont {K.~K.}\ \bibnamefont {Mehta}}, \bibinfo {author} {\bibfnamefont {C.}~\bibnamefont {Zhang}}, \bibinfo {author} {\bibfnamefont {M.}~\bibnamefont {Malinowski}}, \bibinfo {author} {\bibfnamefont {T.-L.}\ \bibnamefont {Nguyen}}, \bibinfo {author} {\bibfnamefont {M.}~\bibnamefont {Stadler}},\ and\ \bibinfo {author} {\bibfnamefont {J.~P.}\ \bibnamefont {Home}},\ }\href {https://doi.org/10.1038/s41586-020-2823-6} {\bibfield  {journal} {\bibinfo  {journal} {Nature}\ }\textbf {\bibinfo {volume} {586}},\ \bibinfo {pages} {533} (\bibinfo {year} {2020})}\BibitemShut {NoStop}%
\bibitem [{\citenamefont {Zimmermann}\ \emph {et~al.}(2026)\citenamefont {Zimmermann}, \citenamefont {Zhai}, \citenamefont {Leenheer}, \citenamefont {Boyle}, \citenamefont {Mishra}, \citenamefont {Dominguez}, \citenamefont {Koppa}, \citenamefont {Jehle}, \citenamefont {Panuski}, \citenamefont {Dong}, \citenamefont {Gilbert}, \citenamefont {Englund},\ and\ \citenamefont {Eichenfield}}]{zimmermann_monolithic_2026}%
  \BibitemOpen
  \bibfield  {author} {\bibinfo {author} {\bibfnamefont {M.}~\bibnamefont {Zimmermann}}, \bibinfo {author} {\bibfnamefont {A.}~\bibnamefont {Zhai}}, \bibinfo {author} {\bibfnamefont {A.~J.}\ \bibnamefont {Leenheer}}, \bibinfo {author} {\bibfnamefont {J.}~\bibnamefont {Boyle}}, \bibinfo {author} {\bibfnamefont {M.}~\bibnamefont {Mishra}}, \bibinfo {author} {\bibfnamefont {D.}~\bibnamefont {Dominguez}}, \bibinfo {author} {\bibfnamefont {M.}~\bibnamefont {Koppa}}, \bibinfo {author} {\bibfnamefont {W.}~\bibnamefont {Jehle}}, \bibinfo {author} {\bibfnamefont {C.}~\bibnamefont {Panuski}}, \bibinfo {author} {\bibfnamefont {M.}~\bibnamefont {Dong}}, \bibinfo {author} {\bibfnamefont {G.}~\bibnamefont {Gilbert}}, \bibinfo {author} {\bibfnamefont {D.}~\bibnamefont {Englund}},\ and\ \bibinfo {author} {\bibfnamefont {M.}~\bibnamefont {Eichenfield}},\ }\href {https://doi.org/10.48550/arXiv.2607.01514} {\bibinfo {title} {Monolithic {Integration} of {Piezo}-{Optomechanical} {Photonics} and {CMOS} {Electronics}}} (\bibinfo
  {year} {2026}),\ \bibinfo {note} {arXiv:2607.01514 [physics.optics]}\BibitemShut {NoStop}%
\bibitem [{\citenamefont {Xiang}\ \emph {et~al.}(2022)\citenamefont {Xiang}, \citenamefont {Jin},\ and\ \citenamefont {Bowers}}]{xiang_silicon_2022}%
  \BibitemOpen
  \bibfield  {author} {\bibinfo {author} {\bibfnamefont {C.}~\bibnamefont {Xiang}}, \bibinfo {author} {\bibfnamefont {W.}~\bibnamefont {Jin}},\ and\ \bibinfo {author} {\bibfnamefont {J.~E.}\ \bibnamefont {Bowers}},\ }\href {https://doi.org/10.1364/PRJ.452936} {\bibfield  {journal} {\bibinfo  {journal} {Photonics Research}\ }\textbf {\bibinfo {volume} {10}},\ \bibinfo {pages} {A82} (\bibinfo {year} {2022})}\BibitemShut {NoStop}%
\bibitem [{\citenamefont {Shen}\ \emph {et~al.}(2020)\citenamefont {Shen}, \citenamefont {Chang}, \citenamefont {Liu}, \citenamefont {Wang}, \citenamefont {Yang}, \citenamefont {Xiang}, \citenamefont {Wang}, \citenamefont {He}, \citenamefont {Liu}, \citenamefont {Xie}, \citenamefont {Guo}, \citenamefont {Kinghorn}, \citenamefont {Wu}, \citenamefont {Ji}, \citenamefont {Kippenberg}, \citenamefont {Vahala},\ and\ \citenamefont {Bowers}}]{shen_integrated_2020}%
  \BibitemOpen
  \bibfield  {author} {\bibinfo {author} {\bibfnamefont {B.}~\bibnamefont {Shen}}, \bibinfo {author} {\bibfnamefont {L.}~\bibnamefont {Chang}}, \bibinfo {author} {\bibfnamefont {J.}~\bibnamefont {Liu}}, \bibinfo {author} {\bibfnamefont {H.}~\bibnamefont {Wang}}, \bibinfo {author} {\bibfnamefont {Q.-F.}\ \bibnamefont {Yang}}, \bibinfo {author} {\bibfnamefont {C.}~\bibnamefont {Xiang}}, \bibinfo {author} {\bibfnamefont {R.~N.}\ \bibnamefont {Wang}}, \bibinfo {author} {\bibfnamefont {J.}~\bibnamefont {He}}, \bibinfo {author} {\bibfnamefont {T.}~\bibnamefont {Liu}}, \bibinfo {author} {\bibfnamefont {W.}~\bibnamefont {Xie}}, \bibinfo {author} {\bibfnamefont {J.}~\bibnamefont {Guo}}, \bibinfo {author} {\bibfnamefont {D.}~\bibnamefont {Kinghorn}}, \bibinfo {author} {\bibfnamefont {L.}~\bibnamefont {Wu}}, \bibinfo {author} {\bibfnamefont {Q.-X.}\ \bibnamefont {Ji}}, \bibinfo {author} {\bibfnamefont {T.~J.}\ \bibnamefont {Kippenberg}}, \bibinfo {author} {\bibfnamefont {K.}~\bibnamefont {Vahala}},\ and\ \bibinfo
  {author} {\bibfnamefont {J.~E.}\ \bibnamefont {Bowers}},\ }\href {https://doi.org/10.1038/s41586-020-2358-x} {\bibfield  {journal} {\bibinfo  {journal} {Nature}\ }\textbf {\bibinfo {volume} {582}},\ \bibinfo {pages} {365} (\bibinfo {year} {2020})}\BibitemShut {NoStop}%
\bibitem [{\citenamefont {Chauhan}\ \emph {et~al.}(2021)\citenamefont {Chauhan}, \citenamefont {Isichenko}, \citenamefont {Liu}, \citenamefont {Wang}, \citenamefont {Zhao}, \citenamefont {Behunin}, \citenamefont {Rakich}, \citenamefont {Jayich}, \citenamefont {Fertig}, \citenamefont {Hoyt},\ and\ \citenamefont {Blumenthal}}]{chauhan_visible_2021}%
  \BibitemOpen
  \bibfield  {author} {\bibinfo {author} {\bibfnamefont {N.}~\bibnamefont {Chauhan}}, \bibinfo {author} {\bibfnamefont {A.}~\bibnamefont {Isichenko}}, \bibinfo {author} {\bibfnamefont {K.}~\bibnamefont {Liu}}, \bibinfo {author} {\bibfnamefont {J.}~\bibnamefont {Wang}}, \bibinfo {author} {\bibfnamefont {Q.}~\bibnamefont {Zhao}}, \bibinfo {author} {\bibfnamefont {R.~O.}\ \bibnamefont {Behunin}}, \bibinfo {author} {\bibfnamefont {P.~T.}\ \bibnamefont {Rakich}}, \bibinfo {author} {\bibfnamefont {A.~M.}\ \bibnamefont {Jayich}}, \bibinfo {author} {\bibfnamefont {C.}~\bibnamefont {Fertig}}, \bibinfo {author} {\bibfnamefont {C.~W.}\ \bibnamefont {Hoyt}},\ and\ \bibinfo {author} {\bibfnamefont {D.~J.}\ \bibnamefont {Blumenthal}},\ }\href {https://doi.org/10.1038/s41467-021-24926-8} {\bibfield  {journal} {\bibinfo  {journal} {Nature Communications}\ }\textbf {\bibinfo {volume} {12}},\ \bibinfo {pages} {4685} (\bibinfo {year} {2021})}\BibitemShut {NoStop}%
\bibitem [{\citenamefont {Klaver}\ \emph {et~al.}(2026)\citenamefont {Klaver}, \citenamefont {te~Morsche}, \citenamefont {Botter}, \citenamefont {Hashemi}, \citenamefont {Segat~Frare}, \citenamefont {Mishra}, \citenamefont {Ye}, \citenamefont {Mbonde}, \citenamefont {Torab~Ahmadi}, \citenamefont {Majidian~Taleghani}, \citenamefont {Jonker}, \citenamefont {Braamhaar}, \citenamefont {Selvaganapathy}, \citenamefont {Mascher}, \citenamefont {van~der Slot}, \citenamefont {Bradley},\ and\ \citenamefont {Marpaung}}]{klaver_surface_2026}%
  \BibitemOpen
  \bibfield  {author} {\bibinfo {author} {\bibfnamefont {Y.}~\bibnamefont {Klaver}}, \bibinfo {author} {\bibfnamefont {R.}~\bibnamefont {te~Morsche}}, \bibinfo {author} {\bibfnamefont {R.~A.}\ \bibnamefont {Botter}}, \bibinfo {author} {\bibfnamefont {B.}~\bibnamefont {Hashemi}}, \bibinfo {author} {\bibfnamefont {B.~L.}\ \bibnamefont {Segat~Frare}}, \bibinfo {author} {\bibfnamefont {A.}~\bibnamefont {Mishra}}, \bibinfo {author} {\bibfnamefont {K.}~\bibnamefont {Ye}}, \bibinfo {author} {\bibfnamefont {H.~M.}\ \bibnamefont {Mbonde}}, \bibinfo {author} {\bibfnamefont {P.}~\bibnamefont {Torab~Ahmadi}}, \bibinfo {author} {\bibfnamefont {N.}~\bibnamefont {Majidian~Taleghani}}, \bibinfo {author} {\bibfnamefont {E.}~\bibnamefont {Jonker}}, \bibinfo {author} {\bibfnamefont {R.~B.~G.}\ \bibnamefont {Braamhaar}}, \bibinfo {author} {\bibfnamefont {P.~R.}\ \bibnamefont {Selvaganapathy}}, \bibinfo {author} {\bibfnamefont {P.}~\bibnamefont {Mascher}}, \bibinfo {author} {\bibfnamefont {P.~J.~M.}\ \bibnamefont {van~der Slot}},
  \bibinfo {author} {\bibfnamefont {J.~D.~B.}\ \bibnamefont {Bradley}},\ and\ \bibinfo {author} {\bibfnamefont {D.}~\bibnamefont {Marpaung}},\ }\href {https://doi.org/10.1038/s41566-026-01873-8} {\bibfield  {journal} {\bibinfo  {journal} {Nature Photonics}\ }\textbf {\bibinfo {volume} {20}},\ \bibinfo {pages} {637} (\bibinfo {year} {2026})}\BibitemShut {NoStop}%
\bibitem [{\citenamefont {Gundavarapu}\ \emph {et~al.}(2019)\citenamefont {Gundavarapu}, \citenamefont {Brodnik}, \citenamefont {Puckett}, \citenamefont {Huffman}, \citenamefont {Bose}, \citenamefont {Behunin}, \citenamefont {Wu}, \citenamefont {Qiu}, \citenamefont {Pinho}, \citenamefont {Chauhan}, \citenamefont {Nohava}, \citenamefont {Rakich}, \citenamefont {Nelson}, \citenamefont {Salit},\ and\ \citenamefont {Blumenthal}}]{gundavarapu_sub-hertz_2019}%
  \BibitemOpen
  \bibfield  {author} {\bibinfo {author} {\bibfnamefont {S.}~\bibnamefont {Gundavarapu}}, \bibinfo {author} {\bibfnamefont {G.~M.}\ \bibnamefont {Brodnik}}, \bibinfo {author} {\bibfnamefont {M.}~\bibnamefont {Puckett}}, \bibinfo {author} {\bibfnamefont {T.}~\bibnamefont {Huffman}}, \bibinfo {author} {\bibfnamefont {D.}~\bibnamefont {Bose}}, \bibinfo {author} {\bibfnamefont {R.}~\bibnamefont {Behunin}}, \bibinfo {author} {\bibfnamefont {J.}~\bibnamefont {Wu}}, \bibinfo {author} {\bibfnamefont {T.}~\bibnamefont {Qiu}}, \bibinfo {author} {\bibfnamefont {C.}~\bibnamefont {Pinho}}, \bibinfo {author} {\bibfnamefont {N.}~\bibnamefont {Chauhan}}, \bibinfo {author} {\bibfnamefont {J.}~\bibnamefont {Nohava}}, \bibinfo {author} {\bibfnamefont {P.~T.}\ \bibnamefont {Rakich}}, \bibinfo {author} {\bibfnamefont {K.~D.}\ \bibnamefont {Nelson}}, \bibinfo {author} {\bibfnamefont {M.}~\bibnamefont {Salit}},\ and\ \bibinfo {author} {\bibfnamefont {D.~J.}\ \bibnamefont {Blumenthal}},\ }\href
  {https://doi.org/10.1038/s41566-018-0313-2} {\bibfield  {journal} {\bibinfo  {journal} {Nature Photonics}\ }\textbf {\bibinfo {volume} {13}},\ \bibinfo {pages} {60} (\bibinfo {year} {2019})}\BibitemShut {NoStop}%
\bibitem [{\citenamefont {Morin}\ \emph {et~al.}(2021)\citenamefont {Morin}, \citenamefont {Chang}, \citenamefont {Jin}, \citenamefont {Li}, \citenamefont {Guo}, \citenamefont {Park}, \citenamefont {Tran}, \citenamefont {Komljenovic},\ and\ \citenamefont {Bowers}}]{morin_cmos-foundry-based_2021}%
  \BibitemOpen
  \bibfield  {author} {\bibinfo {author} {\bibfnamefont {T.~J.}\ \bibnamefont {Morin}}, \bibinfo {author} {\bibfnamefont {L.}~\bibnamefont {Chang}}, \bibinfo {author} {\bibfnamefont {W.}~\bibnamefont {Jin}}, \bibinfo {author} {\bibfnamefont {C.}~\bibnamefont {Li}}, \bibinfo {author} {\bibfnamefont {J.}~\bibnamefont {Guo}}, \bibinfo {author} {\bibfnamefont {H.}~\bibnamefont {Park}}, \bibinfo {author} {\bibfnamefont {M.~A.}\ \bibnamefont {Tran}}, \bibinfo {author} {\bibfnamefont {T.}~\bibnamefont {Komljenovic}},\ and\ \bibinfo {author} {\bibfnamefont {J.~E.}\ \bibnamefont {Bowers}},\ }\href {https://doi.org/10.1364/OPTICA.426065} {\bibfield  {journal} {\bibinfo  {journal} {Optica}\ }\textbf {\bibinfo {volume} {8}},\ \bibinfo {pages} {755} (\bibinfo {year} {2021})}\BibitemShut {NoStop}%
\bibitem [{\citenamefont {Mishra}\ \emph {et~al.}(2026)\citenamefont {Mishra}, \citenamefont {Choi}, \citenamefont {He}, \citenamefont {Talcott}, \citenamefont {Kearney}, \citenamefont {Gehl}, \citenamefont {Leenheer}, \citenamefont {Dominguez}, \citenamefont {Otterstrom},\ and\ \citenamefont {Eichenfield}}]{mishra_ultra-low_2026}%
  \BibitemOpen
  \bibfield  {author} {\bibinfo {author} {\bibfnamefont {M.}~\bibnamefont {Mishra}}, \bibinfo {author} {\bibfnamefont {G.}~\bibnamefont {Choi}}, \bibinfo {author} {\bibfnamefont {W.}~\bibnamefont {He}}, \bibinfo {author} {\bibfnamefont {G.~M.}\ \bibnamefont {Talcott}}, \bibinfo {author} {\bibfnamefont {K.}~\bibnamefont {Kearney}}, \bibinfo {author} {\bibfnamefont {M.}~\bibnamefont {Gehl}}, \bibinfo {author} {\bibfnamefont {A.}~\bibnamefont {Leenheer}}, \bibinfo {author} {\bibfnamefont {D.}~\bibnamefont {Dominguez}}, \bibinfo {author} {\bibfnamefont {N.~T.}\ \bibnamefont {Otterstrom}},\ and\ \bibinfo {author} {\bibfnamefont {M.}~\bibnamefont {Eichenfield}},\ }\href {https://doi.org/10.48550/arXiv.2603.02584} {\bibinfo {title} {Ultra-low loss piezo-optomechanical low-confinement silicon nitride platform for visible wavelength quantum photonic circuits}} (\bibinfo {year} {2026}),\ \bibinfo {note} {arXiv:2603.02584 [physics.optics]}\BibitemShut {NoStop}%
\bibitem [{\citenamefont {Jin}\ \emph {et~al.}(2021)\citenamefont {Jin}, \citenamefont {Yang}, \citenamefont {Chang}, \citenamefont {Shen}, \citenamefont {Wang}, \citenamefont {Leal}, \citenamefont {Wu}, \citenamefont {Gao}, \citenamefont {Feshali}, \citenamefont {Paniccia}, \citenamefont {Vahala},\ and\ \citenamefont {Bowers}}]{jin_hertz-linewidth_2021}%
  \BibitemOpen
  \bibfield  {author} {\bibinfo {author} {\bibfnamefont {W.}~\bibnamefont {Jin}}, \bibinfo {author} {\bibfnamefont {Q.-F.}\ \bibnamefont {Yang}}, \bibinfo {author} {\bibfnamefont {L.}~\bibnamefont {Chang}}, \bibinfo {author} {\bibfnamefont {B.}~\bibnamefont {Shen}}, \bibinfo {author} {\bibfnamefont {H.}~\bibnamefont {Wang}}, \bibinfo {author} {\bibfnamefont {M.~A.}\ \bibnamefont {Leal}}, \bibinfo {author} {\bibfnamefont {L.}~\bibnamefont {Wu}}, \bibinfo {author} {\bibfnamefont {M.}~\bibnamefont {Gao}}, \bibinfo {author} {\bibfnamefont {A.}~\bibnamefont {Feshali}}, \bibinfo {author} {\bibfnamefont {M.}~\bibnamefont {Paniccia}}, \bibinfo {author} {\bibfnamefont {K.~J.}\ \bibnamefont {Vahala}},\ and\ \bibinfo {author} {\bibfnamefont {J.~E.}\ \bibnamefont {Bowers}},\ }\href {https://doi.org/10.1038/s41566-021-00761-7} {\bibfield  {journal} {\bibinfo  {journal} {Nature Photonics}\ }\textbf {\bibinfo {volume} {15}},\ \bibinfo {pages} {346} (\bibinfo {year} {2021})}\BibitemShut {NoStop}%
\bibitem [{\citenamefont {Niffenegger}\ \emph {et~al.}(2020)\citenamefont {Niffenegger}, \citenamefont {Stuart}, \citenamefont {Sorace-Agaskar}, \citenamefont {Kharas}, \citenamefont {Bramhavar}, \citenamefont {Bruzewicz}, \citenamefont {Loh}, \citenamefont {Maxson}, \citenamefont {McConnell}, \citenamefont {Reens}, \citenamefont {West}, \citenamefont {Sage},\ and\ \citenamefont {Chiaverini}}]{niffenegger_integrated_2020}%
  \BibitemOpen
  \bibfield  {author} {\bibinfo {author} {\bibfnamefont {R.~J.}\ \bibnamefont {Niffenegger}}, \bibinfo {author} {\bibfnamefont {J.}~\bibnamefont {Stuart}}, \bibinfo {author} {\bibfnamefont {C.}~\bibnamefont {Sorace-Agaskar}}, \bibinfo {author} {\bibfnamefont {D.}~\bibnamefont {Kharas}}, \bibinfo {author} {\bibfnamefont {S.}~\bibnamefont {Bramhavar}}, \bibinfo {author} {\bibfnamefont {C.~D.}\ \bibnamefont {Bruzewicz}}, \bibinfo {author} {\bibfnamefont {W.}~\bibnamefont {Loh}}, \bibinfo {author} {\bibfnamefont {R.~T.}\ \bibnamefont {Maxson}}, \bibinfo {author} {\bibfnamefont {R.}~\bibnamefont {McConnell}}, \bibinfo {author} {\bibfnamefont {D.}~\bibnamefont {Reens}}, \bibinfo {author} {\bibfnamefont {G.~N.}\ \bibnamefont {West}}, \bibinfo {author} {\bibfnamefont {J.~M.}\ \bibnamefont {Sage}},\ and\ \bibinfo {author} {\bibfnamefont {J.}~\bibnamefont {Chiaverini}},\ }\href {https://doi.org/10.1038/s41586-020-2811-x} {\bibfield  {journal} {\bibinfo  {journal} {Nature}\ }\textbf {\bibinfo {volume} {586}},\ \bibinfo
  {pages} {538} (\bibinfo {year} {2020})}\BibitemShut {NoStop}%
\bibitem [{\citenamefont {Tian}\ \emph {et~al.}(2024)\citenamefont {Tian}, \citenamefont {Liu}, \citenamefont {Attanasio}, \citenamefont {Siddharth}, \citenamefont {Blésin}, \citenamefont {Wang}, \citenamefont {Voloshin}, \citenamefont {Lihachev}, \citenamefont {Riemensberger}, \citenamefont {Kenning}, \citenamefont {Tian}, \citenamefont {Chang}, \citenamefont {Bancora}, \citenamefont {Snigirev}, \citenamefont {Shadymov}, \citenamefont {Kippenberg},\ and\ \citenamefont {Bhave}}]{tian_piezoelectric_2024}%
  \BibitemOpen
  \bibfield  {author} {\bibinfo {author} {\bibfnamefont {H.}~\bibnamefont {Tian}}, \bibinfo {author} {\bibfnamefont {J.}~\bibnamefont {Liu}}, \bibinfo {author} {\bibfnamefont {A.}~\bibnamefont {Attanasio}}, \bibinfo {author} {\bibfnamefont {A.}~\bibnamefont {Siddharth}}, \bibinfo {author} {\bibfnamefont {T.}~\bibnamefont {Blésin}}, \bibinfo {author} {\bibfnamefont {R.~N.}\ \bibnamefont {Wang}}, \bibinfo {author} {\bibfnamefont {A.}~\bibnamefont {Voloshin}}, \bibinfo {author} {\bibfnamefont {G.}~\bibnamefont {Lihachev}}, \bibinfo {author} {\bibfnamefont {J.}~\bibnamefont {Riemensberger}}, \bibinfo {author} {\bibfnamefont {S.~E.}\ \bibnamefont {Kenning}}, \bibinfo {author} {\bibfnamefont {Y.}~\bibnamefont {Tian}}, \bibinfo {author} {\bibfnamefont {T.~H.}\ \bibnamefont {Chang}}, \bibinfo {author} {\bibfnamefont {A.}~\bibnamefont {Bancora}}, \bibinfo {author} {\bibfnamefont {V.}~\bibnamefont {Snigirev}}, \bibinfo {author} {\bibfnamefont {V.}~\bibnamefont {Shadymov}}, \bibinfo {author} {\bibfnamefont {T.~J.}\
  \bibnamefont {Kippenberg}},\ and\ \bibinfo {author} {\bibfnamefont {S.~A.}\ \bibnamefont {Bhave}},\ }\href {https://doi.org/10.1364/AOP.529288} {\bibfield  {journal} {\bibinfo  {journal} {Advances in Optics and Photonics}\ }\textbf {\bibinfo {volume} {16}},\ \bibinfo {pages} {749} (\bibinfo {year} {2024})},\ \bibinfo {note} {publisher: Optica Publishing Group}\BibitemShut {NoStop}%
\bibitem [{\citenamefont {Hosseini}\ \emph {et~al.}(2015)\citenamefont {Hosseini}, \citenamefont {Dekker}, \citenamefont {Hoekman}, \citenamefont {Dekkers}, \citenamefont {Bos}, \citenamefont {Leinse},\ and\ \citenamefont {Heideman}}]{hosseini_stress-optic_2015}%
  \BibitemOpen
  \bibfield  {author} {\bibinfo {author} {\bibfnamefont {N.}~\bibnamefont {Hosseini}}, \bibinfo {author} {\bibfnamefont {R.}~\bibnamefont {Dekker}}, \bibinfo {author} {\bibfnamefont {M.}~\bibnamefont {Hoekman}}, \bibinfo {author} {\bibfnamefont {M.}~\bibnamefont {Dekkers}}, \bibinfo {author} {\bibfnamefont {J.}~\bibnamefont {Bos}}, \bibinfo {author} {\bibfnamefont {A.}~\bibnamefont {Leinse}},\ and\ \bibinfo {author} {\bibfnamefont {R.}~\bibnamefont {Heideman}},\ }\href {https://doi.org/10.1364/OE.23.014018} {\bibfield  {journal} {\bibinfo  {journal} {Optics Express}\ }\textbf {\bibinfo {volume} {23}},\ \bibinfo {pages} {14018} (\bibinfo {year} {2015})}\BibitemShut {NoStop}%
\bibitem [{\citenamefont {Erdil}\ \emph {et~al.}(2025)\citenamefont {Erdil}, \citenamefont {{Izhar}}, \citenamefont {Deng}, \citenamefont {Klein}, \citenamefont {Tang}, \citenamefont {Idjadi}, \citenamefont {Ashtiani}, \citenamefont {Aflatouni},\ and\ \citenamefont {Olsson}}]{erdil_wideband_2025}%
  \BibitemOpen
  \bibfield  {author} {\bibinfo {author} {\bibfnamefont {M.}~\bibnamefont {Erdil}}, \bibinfo {author} {\bibnamefont {{Izhar}}}, \bibinfo {author} {\bibfnamefont {Y.}~\bibnamefont {Deng}}, \bibinfo {author} {\bibfnamefont {E.}~\bibnamefont {Klein}}, \bibinfo {author} {\bibfnamefont {Z.}~\bibnamefont {Tang}}, \bibinfo {author} {\bibfnamefont {M.~H.}\ \bibnamefont {Idjadi}}, \bibinfo {author} {\bibfnamefont {F.}~\bibnamefont {Ashtiani}}, \bibinfo {author} {\bibfnamefont {F.}~\bibnamefont {Aflatouni}},\ and\ \bibinfo {author} {\bibfnamefont {R.~H.}\ \bibnamefont {Olsson}, \bibfnamefont {III}},\ }\href {https://doi.org/10.1021/acsphotonics.4c01520} {\bibfield  {journal} {\bibinfo  {journal} {ACS Photonics}\ }\textbf {\bibinfo {volume} {12}},\ \bibinfo {pages} {2356} (\bibinfo {year} {2025})}\BibitemShut {NoStop}%
\bibitem [{\citenamefont {Bian}\ \emph {et~al.}(2024)\citenamefont {Bian}, \citenamefont {Li}, \citenamefont {Liu}, \citenamefont {Xu}, \citenamefont {Zhao}, \citenamefont {Qiu}, \citenamefont {Dong}, \citenamefont {Zhong}, \citenamefont {Wu}, \citenamefont {Zheng},\ and\ \citenamefont {Hu}}]{bian_demonstration_2024}%
  \BibitemOpen
  \bibfield  {author} {\bibinfo {author} {\bibfnamefont {K.}~\bibnamefont {Bian}}, \bibinfo {author} {\bibfnamefont {Z.}~\bibnamefont {Li}}, \bibinfo {author} {\bibfnamefont {Y.}~\bibnamefont {Liu}}, \bibinfo {author} {\bibfnamefont {S.}~\bibnamefont {Xu}}, \bibinfo {author} {\bibfnamefont {X.}~\bibnamefont {Zhao}}, \bibinfo {author} {\bibfnamefont {Y.}~\bibnamefont {Qiu}}, \bibinfo {author} {\bibfnamefont {Y.}~\bibnamefont {Dong}}, \bibinfo {author} {\bibfnamefont {Q.}~\bibnamefont {Zhong}}, \bibinfo {author} {\bibfnamefont {T.}~\bibnamefont {Wu}}, \bibinfo {author} {\bibfnamefont {S.}~\bibnamefont {Zheng}},\ and\ \bibinfo {author} {\bibfnamefont {T.}~\bibnamefont {Hu}},\ }\href {https://doi.org/10.1364/PRJ.517719} {\bibfield  {journal} {\bibinfo  {journal} {Photonics Research}\ }\textbf {\bibinfo {volume} {12}},\ \bibinfo {pages} {1138} (\bibinfo {year} {2024})}\BibitemShut {NoStop}%
\bibitem [{\citenamefont {Stanfield}\ \emph {et~al.}(2019)\citenamefont {Stanfield}, \citenamefont {Leenheer}, \citenamefont {Michael}, \citenamefont {Sims},\ and\ \citenamefont {Eichenfield}}]{stanfield_cmos-compatible_2019}%
  \BibitemOpen
  \bibfield  {author} {\bibinfo {author} {\bibfnamefont {P.~R.}\ \bibnamefont {Stanfield}}, \bibinfo {author} {\bibfnamefont {A.~J.}\ \bibnamefont {Leenheer}}, \bibinfo {author} {\bibfnamefont {C.~P.}\ \bibnamefont {Michael}}, \bibinfo {author} {\bibfnamefont {R.}~\bibnamefont {Sims}},\ and\ \bibinfo {author} {\bibfnamefont {M.}~\bibnamefont {Eichenfield}},\ }\href {https://doi.org/10.1364/OE.27.028588} {\bibfield  {journal} {\bibinfo  {journal} {Optics Express}\ }\textbf {\bibinfo {volume} {27}},\ \bibinfo {pages} {28588} (\bibinfo {year} {2019})}\BibitemShut {NoStop}%
\bibitem [{\citenamefont {Liu}\ \emph {et~al.}(2020)\citenamefont {Liu}, \citenamefont {Tian}, \citenamefont {Lucas}, \citenamefont {Raja}, \citenamefont {Lihachev}, \citenamefont {Wang}, \citenamefont {He}, \citenamefont {Liu}, \citenamefont {Anderson}, \citenamefont {Weng}, \citenamefont {Bhave},\ and\ \citenamefont {Kippenberg}}]{liu_monolithic_2020}%
  \BibitemOpen
  \bibfield  {author} {\bibinfo {author} {\bibfnamefont {J.}~\bibnamefont {Liu}}, \bibinfo {author} {\bibfnamefont {H.}~\bibnamefont {Tian}}, \bibinfo {author} {\bibfnamefont {E.}~\bibnamefont {Lucas}}, \bibinfo {author} {\bibfnamefont {A.~S.}\ \bibnamefont {Raja}}, \bibinfo {author} {\bibfnamefont {G.}~\bibnamefont {Lihachev}}, \bibinfo {author} {\bibfnamefont {R.~N.}\ \bibnamefont {Wang}}, \bibinfo {author} {\bibfnamefont {J.}~\bibnamefont {He}}, \bibinfo {author} {\bibfnamefont {T.}~\bibnamefont {Liu}}, \bibinfo {author} {\bibfnamefont {M.~H.}\ \bibnamefont {Anderson}}, \bibinfo {author} {\bibfnamefont {W.}~\bibnamefont {Weng}}, \bibinfo {author} {\bibfnamefont {S.~A.}\ \bibnamefont {Bhave}},\ and\ \bibinfo {author} {\bibfnamefont {T.~J.}\ \bibnamefont {Kippenberg}},\ }\href {https://doi.org/10.1038/s41586-020-2465-8} {\bibfield  {journal} {\bibinfo  {journal} {Nature}\ }\textbf {\bibinfo {volume} {583}},\ \bibinfo {pages} {385} (\bibinfo {year} {2020})}\BibitemShut {NoStop}%
\bibitem [{\citenamefont {Jin}\ \emph {et~al.}(2018)\citenamefont {Jin}, \citenamefont {Polcawich}, \citenamefont {Morton},\ and\ \citenamefont {Bowers}}]{jin_piezoelectrically_2018}%
  \BibitemOpen
  \bibfield  {author} {\bibinfo {author} {\bibfnamefont {W.}~\bibnamefont {Jin}}, \bibinfo {author} {\bibfnamefont {R.~G.}\ \bibnamefont {Polcawich}}, \bibinfo {author} {\bibfnamefont {P.~A.}\ \bibnamefont {Morton}},\ and\ \bibinfo {author} {\bibfnamefont {J.~E.}\ \bibnamefont {Bowers}},\ }\href {https://doi.org/10.1364/OE.26.003174} {\bibfield  {journal} {\bibinfo  {journal} {Optics Express}\ }\textbf {\bibinfo {volume} {26}},\ \bibinfo {pages} {3174} (\bibinfo {year} {2018})}\BibitemShut {NoStop}%
\bibitem [{\citenamefont {Menssen}\ \emph {et~al.}(2023)\citenamefont {Menssen}, \citenamefont {Hermans}, \citenamefont {Christen}, \citenamefont {Propson}, \citenamefont {Li}, \citenamefont {Leenheer}, \citenamefont {Zimmermann}, \citenamefont {Dong}, \citenamefont {Larocque}, \citenamefont {Raniwala}, \citenamefont {Gilbert}, \citenamefont {Eichenfield},\ and\ \citenamefont {Englund}}]{menssen_scalable_2023}%
  \BibitemOpen
  \bibfield  {author} {\bibinfo {author} {\bibfnamefont {A.~J.}\ \bibnamefont {Menssen}}, \bibinfo {author} {\bibfnamefont {A.}~\bibnamefont {Hermans}}, \bibinfo {author} {\bibfnamefont {I.}~\bibnamefont {Christen}}, \bibinfo {author} {\bibfnamefont {T.}~\bibnamefont {Propson}}, \bibinfo {author} {\bibfnamefont {C.}~\bibnamefont {Li}}, \bibinfo {author} {\bibfnamefont {A.~J.}\ \bibnamefont {Leenheer}}, \bibinfo {author} {\bibfnamefont {M.}~\bibnamefont {Zimmermann}}, \bibinfo {author} {\bibfnamefont {M.}~\bibnamefont {Dong}}, \bibinfo {author} {\bibfnamefont {H.}~\bibnamefont {Larocque}}, \bibinfo {author} {\bibfnamefont {H.}~\bibnamefont {Raniwala}}, \bibinfo {author} {\bibfnamefont {G.}~\bibnamefont {Gilbert}}, \bibinfo {author} {\bibfnamefont {M.}~\bibnamefont {Eichenfield}},\ and\ \bibinfo {author} {\bibfnamefont {D.~R.}\ \bibnamefont {Englund}},\ }\href {https://doi.org/10.1364/OPTICA.489504} {\bibfield  {journal} {\bibinfo  {journal} {Optica}\ }\textbf {\bibinfo {volume} {10}},\ \bibinfo {pages} {1366}
  (\bibinfo {year} {2023})}\BibitemShut {NoStop}%
\bibitem [{\citenamefont {Dong}\ \emph {et~al.}(2022)\citenamefont {Dong}, \citenamefont {Clark}, \citenamefont {Leenheer}, \citenamefont {Zimmermann}, \citenamefont {Dominguez}, \citenamefont {Menssen}, \citenamefont {Heim}, \citenamefont {Gilbert}, \citenamefont {Englund},\ and\ \citenamefont {Eichenfield}}]{dong_high-speed_2022}%
  \BibitemOpen
  \bibfield  {author} {\bibinfo {author} {\bibfnamefont {M.}~\bibnamefont {Dong}}, \bibinfo {author} {\bibfnamefont {G.}~\bibnamefont {Clark}}, \bibinfo {author} {\bibfnamefont {A.~J.}\ \bibnamefont {Leenheer}}, \bibinfo {author} {\bibfnamefont {M.}~\bibnamefont {Zimmermann}}, \bibinfo {author} {\bibfnamefont {D.}~\bibnamefont {Dominguez}}, \bibinfo {author} {\bibfnamefont {A.~J.}\ \bibnamefont {Menssen}}, \bibinfo {author} {\bibfnamefont {D.}~\bibnamefont {Heim}}, \bibinfo {author} {\bibfnamefont {G.}~\bibnamefont {Gilbert}}, \bibinfo {author} {\bibfnamefont {D.}~\bibnamefont {Englund}},\ and\ \bibinfo {author} {\bibfnamefont {M.}~\bibnamefont {Eichenfield}},\ }\href {https://doi.org/10.1038/s41566-021-00903-x} {\bibfield  {journal} {\bibinfo  {journal} {Nature Photonics}\ }\textbf {\bibinfo {volume} {16}},\ \bibinfo {pages} {59} (\bibinfo {year} {2022})}\BibitemShut {NoStop}%
\bibitem [{\citenamefont {Saha}\ \emph {et~al.}(2026)\citenamefont {Saha}, \citenamefont {Wen}, \citenamefont {Greenspon}, \citenamefont {Zimmermann}, \citenamefont {Palm}, \citenamefont {Witte}, \citenamefont {Goh}, \citenamefont {Li}, \citenamefont {Bumstead}, \citenamefont {Schädler}, \citenamefont {Fortin}, \citenamefont {Dong}, \citenamefont {Leenheer}, \citenamefont {Clark}, \citenamefont {Gilbert}, \citenamefont {Eichenfield},\ and\ \citenamefont {Englund}}]{saha_nanophotonic_2026}%
  \BibitemOpen
  \bibfield  {author} {\bibinfo {author} {\bibfnamefont {M.}~\bibnamefont {Saha}}, \bibinfo {author} {\bibfnamefont {Y.~H.}\ \bibnamefont {Wen}}, \bibinfo {author} {\bibfnamefont {A.~S.}\ \bibnamefont {Greenspon}}, \bibinfo {author} {\bibfnamefont {M.}~\bibnamefont {Zimmermann}}, \bibinfo {author} {\bibfnamefont {K.~J.}\ \bibnamefont {Palm}}, \bibinfo {author} {\bibfnamefont {A.}~\bibnamefont {Witte}}, \bibinfo {author} {\bibfnamefont {Y.~M.}\ \bibnamefont {Goh}}, \bibinfo {author} {\bibfnamefont {C.}~\bibnamefont {Li}}, \bibinfo {author} {\bibfnamefont {J.}~\bibnamefont {Bumstead}}, \bibinfo {author} {\bibfnamefont {K.}~\bibnamefont {Schädler}}, \bibinfo {author} {\bibfnamefont {R.}~\bibnamefont {Fortin}}, \bibinfo {author} {\bibfnamefont {M.}~\bibnamefont {Dong}}, \bibinfo {author} {\bibfnamefont {A.~J.}\ \bibnamefont {Leenheer}}, \bibinfo {author} {\bibfnamefont {G.}~\bibnamefont {Clark}}, \bibinfo {author} {\bibfnamefont {G.}~\bibnamefont {Gilbert}}, \bibinfo {author} {\bibfnamefont {M.}~\bibnamefont
  {Eichenfield}},\ and\ \bibinfo {author} {\bibfnamefont {D.}~\bibnamefont {Englund}},\ }\href {https://doi.org/10.1038/s41586-025-10038-6} {\bibfield  {journal} {\bibinfo  {journal} {Nature}\ }\textbf {\bibinfo {volume} {651}},\ \bibinfo {pages} {356} (\bibinfo {year} {2026})}\BibitemShut {NoStop}%
\bibitem [{\citenamefont {Tian}\ \emph {et~al.}(2021)\citenamefont {Tian}, \citenamefont {Liu}, \citenamefont {Siddharth}, \citenamefont {Wang}, \citenamefont {Blésin}, \citenamefont {He}, \citenamefont {Kippenberg},\ and\ \citenamefont {Bhave}}]{tian_magnetic-free_2021}%
  \BibitemOpen
  \bibfield  {author} {\bibinfo {author} {\bibfnamefont {H.}~\bibnamefont {Tian}}, \bibinfo {author} {\bibfnamefont {J.}~\bibnamefont {Liu}}, \bibinfo {author} {\bibfnamefont {A.}~\bibnamefont {Siddharth}}, \bibinfo {author} {\bibfnamefont {R.~N.}\ \bibnamefont {Wang}}, \bibinfo {author} {\bibfnamefont {T.}~\bibnamefont {Blésin}}, \bibinfo {author} {\bibfnamefont {J.}~\bibnamefont {He}}, \bibinfo {author} {\bibfnamefont {T.~J.}\ \bibnamefont {Kippenberg}},\ and\ \bibinfo {author} {\bibfnamefont {S.~A.}\ \bibnamefont {Bhave}},\ }\href {https://doi.org/10.1038/s41566-021-00882-z} {\bibfield  {journal} {\bibinfo  {journal} {Nature Photonics}\ }\textbf {\bibinfo {volume} {15}},\ \bibinfo {pages} {828} (\bibinfo {year} {2021})}\BibitemShut {NoStop}%
\bibitem [{\citenamefont {Tian}\ \emph {et~al.}(2020)\citenamefont {Tian}, \citenamefont {Liu}, \citenamefont {Dong}, \citenamefont {Skehan}, \citenamefont {Zervas}, \citenamefont {Kippenberg},\ and\ \citenamefont {Bhave}}]{tian_hybrid_2020}%
  \BibitemOpen
  \bibfield  {author} {\bibinfo {author} {\bibfnamefont {H.}~\bibnamefont {Tian}}, \bibinfo {author} {\bibfnamefont {J.}~\bibnamefont {Liu}}, \bibinfo {author} {\bibfnamefont {B.}~\bibnamefont {Dong}}, \bibinfo {author} {\bibfnamefont {J.~C.}\ \bibnamefont {Skehan}}, \bibinfo {author} {\bibfnamefont {M.}~\bibnamefont {Zervas}}, \bibinfo {author} {\bibfnamefont {T.~J.}\ \bibnamefont {Kippenberg}},\ and\ \bibinfo {author} {\bibfnamefont {S.~A.}\ \bibnamefont {Bhave}},\ }\href {https://doi.org/10.1038/s41467-020-16812-6} {\bibfield  {journal} {\bibinfo  {journal} {Nature Communications}\ }\textbf {\bibinfo {volume} {11}},\ \bibinfo {pages} {3073} (\bibinfo {year} {2020})}\BibitemShut {NoStop}%
\bibitem [{\citenamefont {Kenning}\ \emph {et~al.}(2025)\citenamefont {Kenning}, \citenamefont {Chang}, \citenamefont {Attanasio}, \citenamefont {Jin}, \citenamefont {Feshali}, \citenamefont {Tian}, \citenamefont {Paniccia},\ and\ \citenamefont {Bhave}}]{kenning_broadband_2025}%
  \BibitemOpen
  \bibfield  {author} {\bibinfo {author} {\bibfnamefont {S.~E.}\ \bibnamefont {Kenning}}, \bibinfo {author} {\bibfnamefont {T.-H.}\ \bibnamefont {Chang}}, \bibinfo {author} {\bibfnamefont {A.~G.}\ \bibnamefont {Attanasio}}, \bibinfo {author} {\bibfnamefont {W.}~\bibnamefont {Jin}}, \bibinfo {author} {\bibfnamefont {A.}~\bibnamefont {Feshali}}, \bibinfo {author} {\bibfnamefont {Y.}~\bibnamefont {Tian}}, \bibinfo {author} {\bibfnamefont {M.}~\bibnamefont {Paniccia}},\ and\ \bibinfo {author} {\bibfnamefont {S.~A.}\ \bibnamefont {Bhave}},\ }\href {https://doi.org/10.1038/s41467-025-67618-3} {\bibfield  {journal} {\bibinfo  {journal} {Nature Communications}\ }\textbf {\bibinfo {volume} {17}},\ \bibinfo {pages} {897} (\bibinfo {year} {2025})}\BibitemShut {NoStop}%
\bibitem [{\citenamefont {Lukens}\ and\ \citenamefont {Lougovski}(2017)}]{lukens_frequency-encoded_2017}%
  \BibitemOpen
  \bibfield  {author} {\bibinfo {author} {\bibfnamefont {J.~M.}\ \bibnamefont {Lukens}}\ and\ \bibinfo {author} {\bibfnamefont {P.}~\bibnamefont {Lougovski}},\ }\href {https://doi.org/10.1364/OPTICA.4.000008} {\bibfield  {journal} {\bibinfo  {journal} {Optica}\ }\textbf {\bibinfo {volume} {4}},\ \bibinfo {pages} {8} (\bibinfo {year} {2017})}\BibitemShut {NoStop}%
\bibitem [{\citenamefont {Lu}\ \emph {et~al.}(2018)\citenamefont {Lu}, \citenamefont {Lukens}, \citenamefont {Peters}, \citenamefont {Odele}, \citenamefont {Leaird}, \citenamefont {Weiner},\ and\ \citenamefont {Lougovski}}]{lu_electro-optic_2018}%
  \BibitemOpen
  \bibfield  {author} {\bibinfo {author} {\bibfnamefont {H.-H.}\ \bibnamefont {Lu}}, \bibinfo {author} {\bibfnamefont {J.~M.}\ \bibnamefont {Lukens}}, \bibinfo {author} {\bibfnamefont {N.~A.}\ \bibnamefont {Peters}}, \bibinfo {author} {\bibfnamefont {O.~D.}\ \bibnamefont {Odele}}, \bibinfo {author} {\bibfnamefont {D.~E.}\ \bibnamefont {Leaird}}, \bibinfo {author} {\bibfnamefont {A.~M.}\ \bibnamefont {Weiner}},\ and\ \bibinfo {author} {\bibfnamefont {P.}~\bibnamefont {Lougovski}},\ }\href {https://doi.org/10.1103/PhysRevLett.120.030502} {\bibfield  {journal} {\bibinfo  {journal} {Physical Review Letters}\ }\textbf {\bibinfo {volume} {120}},\ \bibinfo {pages} {030502} (\bibinfo {year} {2018})}\BibitemShut {NoStop}%
\bibitem [{\citenamefont {Cohen}\ \emph {et~al.}(2024)\citenamefont {Cohen}, \citenamefont {Wu}, \citenamefont {Myilswamy}, \citenamefont {Fatema}, \citenamefont {Lingaraju},\ and\ \citenamefont {Weiner}}]{cohen_silicon_2024}%
  \BibitemOpen
  \bibfield  {author} {\bibinfo {author} {\bibfnamefont {L.~M.}\ \bibnamefont {Cohen}}, \bibinfo {author} {\bibfnamefont {K.}~\bibnamefont {Wu}}, \bibinfo {author} {\bibfnamefont {K.~V.}\ \bibnamefont {Myilswamy}}, \bibinfo {author} {\bibfnamefont {S.}~\bibnamefont {Fatema}}, \bibinfo {author} {\bibfnamefont {N.~B.}\ \bibnamefont {Lingaraju}},\ and\ \bibinfo {author} {\bibfnamefont {A.~M.}\ \bibnamefont {Weiner}},\ }\href {https://doi.org/10.1038/s41467-024-52051-9} {\bibfield  {journal} {\bibinfo  {journal} {Nature Communications}\ }\textbf {\bibinfo {volume} {15}},\ \bibinfo {pages} {7878} (\bibinfo {year} {2024})}\BibitemShut {NoStop}%
\bibitem [{\citenamefont {Antonio}\ \emph {et~al.}(2012)\citenamefont {Antonio}, \citenamefont {Zanette},\ and\ \citenamefont {López}}]{antonio_frequency_2012}%
  \BibitemOpen
  \bibfield  {author} {\bibinfo {author} {\bibfnamefont {D.}~\bibnamefont {Antonio}}, \bibinfo {author} {\bibfnamefont {D.~H.}\ \bibnamefont {Zanette}},\ and\ \bibinfo {author} {\bibfnamefont {D.}~\bibnamefont {López}},\ }\href {https://doi.org/10.1038/ncomms1813} {\bibfield  {journal} {\bibinfo  {journal} {Nature Communications}\ }\textbf {\bibinfo {volume} {3}},\ \bibinfo {pages} {806} (\bibinfo {year} {2012})}\BibitemShut {NoStop}%
\bibitem [{\citenamefont {Chen}\ \emph {et~al.}(2017)\citenamefont {Chen}, \citenamefont {Zanette}, \citenamefont {Czaplewski}, \citenamefont {Shaw},\ and\ \citenamefont {López}}]{chen_direct_2017}%
  \BibitemOpen
  \bibfield  {author} {\bibinfo {author} {\bibfnamefont {C.}~\bibnamefont {Chen}}, \bibinfo {author} {\bibfnamefont {D.~H.}\ \bibnamefont {Zanette}}, \bibinfo {author} {\bibfnamefont {D.~A.}\ \bibnamefont {Czaplewski}}, \bibinfo {author} {\bibfnamefont {S.}~\bibnamefont {Shaw}},\ and\ \bibinfo {author} {\bibfnamefont {D.}~\bibnamefont {López}},\ }\href {https://doi.org/10.1038/ncomms15523} {\bibfield  {journal} {\bibinfo  {journal} {Nature Communications}\ }\textbf {\bibinfo {volume} {8}},\ \bibinfo {pages} {15523} (\bibinfo {year} {2017})}\BibitemShut {NoStop}%
\bibitem [{\citenamefont {Shoshani}\ and\ \citenamefont {Shaw}(2021)}]{shoshani_resonant_2021}%
  \BibitemOpen
  \bibfield  {author} {\bibinfo {author} {\bibfnamefont {O.}~\bibnamefont {Shoshani}}\ and\ \bibinfo {author} {\bibfnamefont {S.~W.}\ \bibnamefont {Shaw}},\ }\href {https://doi.org/10.1007/s11071-021-06405-3} {\bibfield  {journal} {\bibinfo  {journal} {Nonlinear Dynamics}\ }\textbf {\bibinfo {volume} {104}},\ \bibinfo {pages} {1801} (\bibinfo {year} {2021})}\BibitemShut {NoStop}%
\bibitem [{\citenamefont {Ganesan}\ \emph {et~al.}(2017)\citenamefont {Ganesan}, \citenamefont {Do},\ and\ \citenamefont {Seshia}}]{ganesan_phononic_2017}%
  \BibitemOpen
  \bibfield  {author} {\bibinfo {author} {\bibfnamefont {A.}~\bibnamefont {Ganesan}}, \bibinfo {author} {\bibfnamefont {C.}~\bibnamefont {Do}},\ and\ \bibinfo {author} {\bibfnamefont {A.}~\bibnamefont {Seshia}},\ }\href {https://doi.org/10.1103/PhysRevLett.118.033903} {\bibfield  {journal} {\bibinfo  {journal} {Physical Review Letters}\ }\textbf {\bibinfo {volume} {118}},\ \bibinfo {pages} {033903} (\bibinfo {year} {2017})}\BibitemShut {NoStop}%
\bibitem [{\citenamefont {Graham}\ \emph {et~al.}(2022)\citenamefont {Graham}, \citenamefont {Song}, \citenamefont {Scott}, \citenamefont {Poole}, \citenamefont {Phuttitarn}, \citenamefont {Jooya}, \citenamefont {Eichler}, \citenamefont {Jiang}, \citenamefont {Marra}, \citenamefont {Grinkemeyer}, \citenamefont {Kwon}, \citenamefont {Ebert}, \citenamefont {Cherek}, \citenamefont {Lichtman}, \citenamefont {Gillette}, \citenamefont {Gilbert}, \citenamefont {Bowman}, \citenamefont {Ballance}, \citenamefont {Campbell}, \citenamefont {Dahl}, \citenamefont {Crawford}, \citenamefont {Blunt}, \citenamefont {Rogers}, \citenamefont {Noel},\ and\ \citenamefont {Saffman}}]{graham_multi-qubit_2022}%
  \BibitemOpen
  \bibfield  {author} {\bibinfo {author} {\bibfnamefont {T.~M.}\ \bibnamefont {Graham}}, \bibinfo {author} {\bibfnamefont {Y.}~\bibnamefont {Song}}, \bibinfo {author} {\bibfnamefont {J.}~\bibnamefont {Scott}}, \bibinfo {author} {\bibfnamefont {C.}~\bibnamefont {Poole}}, \bibinfo {author} {\bibfnamefont {L.}~\bibnamefont {Phuttitarn}}, \bibinfo {author} {\bibfnamefont {K.}~\bibnamefont {Jooya}}, \bibinfo {author} {\bibfnamefont {P.}~\bibnamefont {Eichler}}, \bibinfo {author} {\bibfnamefont {X.}~\bibnamefont {Jiang}}, \bibinfo {author} {\bibfnamefont {A.}~\bibnamefont {Marra}}, \bibinfo {author} {\bibfnamefont {B.}~\bibnamefont {Grinkemeyer}}, \bibinfo {author} {\bibfnamefont {M.}~\bibnamefont {Kwon}}, \bibinfo {author} {\bibfnamefont {M.}~\bibnamefont {Ebert}}, \bibinfo {author} {\bibfnamefont {J.}~\bibnamefont {Cherek}}, \bibinfo {author} {\bibfnamefont {M.~T.}\ \bibnamefont {Lichtman}}, \bibinfo {author} {\bibfnamefont {M.}~\bibnamefont {Gillette}}, \bibinfo {author} {\bibfnamefont {J.}~\bibnamefont
  {Gilbert}}, \bibinfo {author} {\bibfnamefont {D.}~\bibnamefont {Bowman}}, \bibinfo {author} {\bibfnamefont {T.}~\bibnamefont {Ballance}}, \bibinfo {author} {\bibfnamefont {C.}~\bibnamefont {Campbell}}, \bibinfo {author} {\bibfnamefont {E.~D.}\ \bibnamefont {Dahl}}, \bibinfo {author} {\bibfnamefont {O.}~\bibnamefont {Crawford}}, \bibinfo {author} {\bibfnamefont {N.~S.}\ \bibnamefont {Blunt}}, \bibinfo {author} {\bibfnamefont {B.}~\bibnamefont {Rogers}}, \bibinfo {author} {\bibfnamefont {T.}~\bibnamefont {Noel}},\ and\ \bibinfo {author} {\bibfnamefont {M.}~\bibnamefont {Saffman}},\ }\href {https://doi.org/10.1038/s41586-022-04603-6} {\bibfield  {journal} {\bibinfo  {journal} {Nature}\ }\textbf {\bibinfo {volume} {604}},\ \bibinfo {pages} {457} (\bibinfo {year} {2022})}\BibitemShut {NoStop}%
\bibitem [{\citenamefont {Moses}\ \emph {et~al.}(2023)\citenamefont {Moses} \emph {et~al.}}]{moses_race-track_2023}%
  \BibitemOpen
  \bibfield  {author} {\bibinfo {author} {\bibfnamefont {S.}~\bibnamefont {Moses}} \emph {et~al.},\ }\href {https://doi.org/10.1103/PhysRevX.13.041052} {\bibfield  {journal} {\bibinfo  {journal} {Physical Review X}\ }\textbf {\bibinfo {volume} {13}},\ \bibinfo {pages} {041052} (\bibinfo {year} {2023})}\BibitemShut {NoStop}%
\bibitem [{\citenamefont {Wang}\ \emph {et~al.}(2018)\citenamefont {Wang}, \citenamefont {Zhang}, \citenamefont {Chen}, \citenamefont {Bertrand}, \citenamefont {Shams-Ansari}, \citenamefont {Chandrasekhar}, \citenamefont {Winzer},\ and\ \citenamefont {Lončar}}]{wang_integrated_2018}%
  \BibitemOpen
  \bibfield  {author} {\bibinfo {author} {\bibfnamefont {C.}~\bibnamefont {Wang}}, \bibinfo {author} {\bibfnamefont {M.}~\bibnamefont {Zhang}}, \bibinfo {author} {\bibfnamefont {X.}~\bibnamefont {Chen}}, \bibinfo {author} {\bibfnamefont {M.}~\bibnamefont {Bertrand}}, \bibinfo {author} {\bibfnamefont {A.}~\bibnamefont {Shams-Ansari}}, \bibinfo {author} {\bibfnamefont {S.}~\bibnamefont {Chandrasekhar}}, \bibinfo {author} {\bibfnamefont {P.}~\bibnamefont {Winzer}},\ and\ \bibinfo {author} {\bibfnamefont {M.}~\bibnamefont {Lončar}},\ }\href {https://doi.org/10.1038/s41586-018-0551-y} {\bibfield  {journal} {\bibinfo  {journal} {Nature}\ }\textbf {\bibinfo {volume} {562}},\ \bibinfo {pages} {101} (\bibinfo {year} {2018})}\BibitemShut {NoStop}%
\bibitem [{\citenamefont {Zhu}\ \emph {et~al.}(2026)\citenamefont {Zhu}, \citenamefont {Ren}, \citenamefont {Liu}, \citenamefont {Yang}, \citenamefont {Xu},\ and\ \citenamefont {Zhu}}]{zhu_generation_2026}%
  \BibitemOpen
  \bibfield  {author} {\bibinfo {author} {\bibfnamefont {X.}~\bibnamefont {Zhu}}, \bibinfo {author} {\bibfnamefont {Z.}~\bibnamefont {Ren}}, \bibinfo {author} {\bibfnamefont {X.}~\bibnamefont {Liu}}, \bibinfo {author} {\bibfnamefont {S.}~\bibnamefont {Yang}}, \bibinfo {author} {\bibfnamefont {K.}~\bibnamefont {Xu}},\ and\ \bibinfo {author} {\bibfnamefont {Y.}~\bibnamefont {Zhu}},\ }\href {https://doi.org/10.1016/j.measurement.2026.120766} {\bibfield  {journal} {\bibinfo  {journal} {Measurement}\ }\textbf {\bibinfo {volume} {269}},\ \bibinfo {pages} {120766} (\bibinfo {year} {2026})}\BibitemShut {NoStop}%
\bibitem [{\citenamefont {Cheng}\ \emph {et~al.}(2026)\citenamefont {Cheng}, \citenamefont {Xie}, \citenamefont {Wang}, \citenamefont {Nie}, \citenamefont {Jin}, \citenamefont {Luo}, \citenamefont {Wang}, \citenamefont {Zhou}, \citenamefont {Gong}, \citenamefont {Chang}, \citenamefont {Hu},\ and\ \citenamefont {Yang}}]{cheng_communication-ready_2026}%
  \BibitemOpen
  \bibfield  {author} {\bibinfo {author} {\bibfnamefont {Y.}~\bibnamefont {Cheng}}, \bibinfo {author} {\bibfnamefont {Z.}~\bibnamefont {Xie}}, \bibinfo {author} {\bibfnamefont {Y.}~\bibnamefont {Wang}}, \bibinfo {author} {\bibfnamefont {B.}~\bibnamefont {Nie}}, \bibinfo {author} {\bibfnamefont {X.}~\bibnamefont {Jin}}, \bibinfo {author} {\bibfnamefont {H.}~\bibnamefont {Luo}}, \bibinfo {author} {\bibfnamefont {J.}~\bibnamefont {Wang}}, \bibinfo {author} {\bibfnamefont {Z.}~\bibnamefont {Zhou}}, \bibinfo {author} {\bibfnamefont {Q.}~\bibnamefont {Gong}}, \bibinfo {author} {\bibfnamefont {L.}~\bibnamefont {Chang}}, \bibinfo {author} {\bibfnamefont {Y.}~\bibnamefont {Hu}},\ and\ \bibinfo {author} {\bibfnamefont {Q.-F.}\ \bibnamefont {Yang}},\ }\href {https://doi.org/10.1364/OPTICA.586478} {\bibfield  {journal} {\bibinfo  {journal} {Optica}\ }\textbf {\bibinfo {volume} {13}},\ \bibinfo {pages} {519} (\bibinfo {year} {2026})}\BibitemShut {NoStop}%
\bibitem [{\citenamefont {Geng}\ \emph {et~al.}(2026)\citenamefont {Geng}, \citenamefont {Tong}, \citenamefont {Zhang}, \citenamefont {Tang}, \citenamefont {Du}, \citenamefont {Xia}, \citenamefont {Liu}, \citenamefont {Yu}, \citenamefont {Huang}, \citenamefont {Huang}, \citenamefont {Li}, \citenamefont {Dai}, \citenamefont {Wong}, \citenamefont {Chen},\ and\ \citenamefont {Xiang}}]{geng_meter-long_2026}%
  \BibitemOpen
  \bibfield  {author} {\bibinfo {author} {\bibfnamefont {Z.}~\bibnamefont {Geng}}, \bibinfo {author} {\bibfnamefont {Y.}~\bibnamefont {Tong}}, \bibinfo {author} {\bibfnamefont {C.}~\bibnamefont {Zhang}}, \bibinfo {author} {\bibfnamefont {H.}~\bibnamefont {Tang}}, \bibinfo {author} {\bibfnamefont {Z.}~\bibnamefont {Du}}, \bibinfo {author} {\bibfnamefont {Y.}~\bibnamefont {Xia}}, \bibinfo {author} {\bibfnamefont {M.}~\bibnamefont {Liu}}, \bibinfo {author} {\bibfnamefont {D.}~\bibnamefont {Yu}}, \bibinfo {author} {\bibfnamefont {Y.}~\bibnamefont {Huang}}, \bibinfo {author} {\bibfnamefont {Y.}~\bibnamefont {Huang}}, \bibinfo {author} {\bibfnamefont {Z.}~\bibnamefont {Li}}, \bibinfo {author} {\bibfnamefont {T.}~\bibnamefont {Dai}}, \bibinfo {author} {\bibfnamefont {K.~K.-Y.}\ \bibnamefont {Wong}}, \bibinfo {author} {\bibfnamefont {H.}~\bibnamefont {Chen}},\ and\ \bibinfo {author} {\bibfnamefont {C.}~\bibnamefont {Xiang}},\ }\href {https://doi.org/10.48550/arXiv.2604.12564} {\bibinfo {title} {Meter-long broadband
  chirped {Bragg} gratings for on-chip dispersion control and pulse shaping}} (\bibinfo {year} {2026}),\ \bibinfo {note} {arXiv:2604.12564 [physics.optics]}\BibitemShut {NoStop}%
\end{thebibliography}%


\begin{thebibliography}{10}

\bibitem{doi:10.1126/sciadv.ade4454}
Ashok Kodigala, Michael Gehl, Gregory~W. Hoth, Jongmin Lee, Christopher~T. DeRose, Andrew Pomerene, Christina Dallo, Douglas Trotter, Andrew~L. Starbuck, Grant Biedermann, Peter D.~D. Schwindt, and Anthony~L. Lentine.
\newblock High-performance silicon photonic single-sideband modulators for cold-atom interferometry.
\newblock {\em Science Advances}, 10(28):eade4454, 2024.

\bibitem{kittlaus_electrically_2021}
Eric~A. Kittlaus, William~M. Jones, Peter~T. Rakich, Nils~T. Otterstrom, Richard~E. Muller, and Mina Rais-Zadeh.
\newblock Electrically driven acousto-optics and broadband non-reciprocity in silicon photonics.
\newblock {\em Nature Photonics}, 15(1):43--52, January 2021.

\bibitem{hu_-chip_2021}
Yaowen Hu, Mengjie Yu, Di~Zhu, Neil Sinclair, Amirhassan Shams-Ansari, Linbo Shao, Jeffrey Holzgrafe, Eric Puma, Mian Zhang, and Marko Lončar.
\newblock On-chip electro-optic frequency shifters and beam splitters.
\newblock {\em Nature}, 599(7886):587--593, November 2021.

\bibitem{assumpcao_thin_2024}
Daniel Assumpcao, Dylan Renaud, Aida Baradari, Beibei Zeng, Chawina De-Eknamkul, C.~J. Xin, Amirhassan Shams-Ansari, David Barton, Bartholomeus Machielse, and Marko Loncar.
\newblock A thin film lithium niobate near-infrared platform for multiplexing quantum nodes.
\newblock {\em Nature Communications}, 15(1):10459, December 2024.

\bibitem{shao_integrated_2020}
Linbo Shao, Neil Sinclair, James Leatham, Yaowen Hu, Mengjie Yu, Terry Turpin, Devon Crowe, and Marko Lončar.
\newblock Integrated microwave acousto-optic frequency shifter on thin-film lithium niobate.
\newblock {\em Optics Express}, 28(16):23728--23738, August 2020.

\bibitem{yu_gigahertz_2021}
Zejie Yu and Xiankai Sun.
\newblock Gigahertz {Acousto}-{Optic} {Modulation} and {Frequency} {Shifting} on {Etchless} {Lithium} {Niobate} {Integrated} {Platform}.
\newblock {\em ACS Photonics}, 8(3):798--803, March 2021.

\bibitem{chen_mono-drive_2025}
Yikun Chen, Hanke Feng, Zhenzheng Wang, Ke~Zhang, Xiangzhi Xie, Yuansong Zeng, Yujie Ren, and Cheng Wang.
\newblock Mono-drive single-sideband modulation via optical delay lines on thin-film lithium niobate.
\newblock {\em Optica}, 12(5):666--673, May 2025.

\bibitem{cui_nonreciprocal_2026}
Zekun Cui, Tianyi Li, Xujia Zhang, Jinwei Su, Liangjun Lu, Linjie Zhou, Jianping Chen, and Kan Wu.
\newblock Nonreciprocal and {High}-{Efficiency} {Monolithic} {Intermodal} {Conversion} {Driven} by {Surface} {Acoustic} {Waves} on {Thin}-{Film} {Lithium} {Niobate}.
\newblock {\em Laser \& Photonics Reviews}, 20(15):e71167, 2026.

\bibitem{zhang_integrated-waveguide-based_2024}
Liang Zhang, Chaohan Cui, Pao-Kang Chen, and Linran Fan.
\newblock Integrated-waveguide-based acousto-optic modulation with complete optical conversion.
\newblock {\em Optica}, 11(2):184--189, February 2024.

\bibitem{freedman_gigahertz-frequency_2025}
Jacob~M. Freedman, Matthew~J. Storey, Daniel Dominguez, Andrew Leenheer, Sebastian Magri, Nils~T. Otterstrom, and Matt Eichenfield.
\newblock Gigahertz-frequency acousto-optic phase modulation of visible light in a {CMOS}-fabricated photonic circuit.
\newblock {\em Nature Communications}, 16(1):10959, December 2025.

\bibitem{mishra_ultra-low_2026}
Mayank Mishra, Gwangho Choi, Wenhua He, Gina~M. Talcott, Katherine Kearney, Michael Gehl, Andrew Leenheer, Daniel Dominguez, Nils~T. Otterstrom, and Matt Eichenfield.
\newblock Ultra-low loss piezo-optomechanical low-confinement silicon nitride platform for visible wavelength quantum photonic circuits, March 2026.
\newblock arXiv:2603.02584 [physics.optics].

\bibitem{thomas_third-order_1968}
J.~F. Thomas.
\newblock Third-{Order} {Elastic} {Constants} of {Aluminum}.
\newblock {\em Physical Review}, 175(3):955--962, November 1968.

\bibitem{bogardus_thirdorder_1965}
E.~H. Bogardus.
\newblock Third‐{Order} {Elastic} {Constants} of {Ge}, {MgO}, and {Fused} {SiO2}.
\newblock {\em Journal of Applied Physics}, 36(8):2504--2513, August 1965.

\bibitem{lepkowski_third-order_2021}
S~P Łepkowski and Abdur-Rehman Anwar.
\newblock Third-order elastic constants and biaxial relaxation coefficient in wurtzite group-{III} nitrides by hybrid-density functional theory calculations.
\newblock {\em Journal of Physics: Condensed Matter}, 33(35):355402, July 2021.

\bibitem{pal_second-order_2011}
Joydeep Pal, Geoffrey Tse, Vesel Haxha, Max~A. Migliorato, and Stanko Tomić.
\newblock Second-order piezoelectricity in wurtzite {III}-{N} semiconductors.
\newblock {\em Physical Review B}, 84(8):085211, August 2011.

\bibitem{levine_dispersive_2022}
Harry Levine, Dolev Bluvstein, Alexander Keesling, Tout~T. Wang, Sepehr Ebadi, Giulia Semeghini, Ahmed Omran, Markus Greiner, Vladan Vuletić, and Mikhail~D. Lukin.
\newblock Dispersive optical systems for scalable {Raman} driving of hyperfine qubits.
\newblock {\em Physical Review A}, 105(3):032618, March 2022.

\end{thebibliography}

\end{document}

\begin{center}
    {\bf \large Supplementary Information for \say{Resonant multi-harmonic acousto-optics for programmable frequency control of visible light in a CMOS platform}}
\end{center}
\vspace{1em}

\tableofcontents
\newpage

\section{Comparison to other integrated frequency-shifting technologies}

Below we present a comparison table to other integrated optical frequency-shifting technologies. Dual-parallel Mach-Zehnder interferometer is abbreviated DP-MZM, Bragg AO refers to acousto-optic devices based on Bragg reflection, and intermodal AO refers to acousto-optic devices based on intermodal scattering between two distinct waveguide modes. The acronym TFLN refers to thin-film lithium niobate.

\begin{table*}[htbp]
\centering
\small
\setlength{\tabcolsep}{3.5pt}
\caption{Integrated optical frequency shifters. CE: conversion efficiency, CS/SS: carrier and unwanted-sideband suppression as reported.}
\vspace{3mm}
\label{tab:ssb_comparison}
\begin{tabular}{llllrrrrr}
\toprule
Reference & Material & Mechanism & $\lambda$ [nm] & Shift [GHz] & CE [\%] & $P_\text{RF}$ [dBm] & CS [dB] & SS [dB] \\
\midrule
This work & SiN--AlN & multi-harm.\ AO & 730 & 1.14 & 40 & 20.2 & 60 & 53 \\
\cite{doi:10.1126/sciadv.ade4454} Kodigala et al. & Si & DP-MZM EO & 1560 & 1 & 20.7 & 21 & 30 & 48 \\
\cite{kittlaus_electrically_2021} Kittlaus et al. & Si--AlN & intermodal AO & 1604 & 3.11 & $<$ 1 & 10 & --- & 30 \\
\cite{hu_-chip_2021} Hu et al. & TFLN & coupled-ring EO & 1600 & 12.5 & 80 & 20 & 33 & $<$ 30 \\
\cite{assumpcao_thin_2024} Assumpcao et al. & TFLN & serrodyne EO & 730 & $\le$15 & 50--90 & 43.9 & --- & --- \\
\cite{shao_integrated_2020} Shao et al. & TFLN & Bragg AO & 1550 & 2.89 & 3.5 & 30 & 35 & $>$ 40 \\
\cite{yu_gigahertz_2021} Yu \& Sun & TFLN & Bragg AO & 1550 & 3 & $2.4$ & $30$ & 27 & 51 \\
\cite{chen_mono-drive_2025} Chen et al. & TFLN & DP-MZM EO & 1550 & 50 & --- & 10 & 16.9 & 22.5 \\
\cite{cui_nonreciprocal_2026} Cui et al. & TFLN & intermodal AO & 1550 & 0.89 & 21.6 & 28 & --- & 25.4 \\
\cite{zhang_integrated-waveguide-based_2024} Zhang et al. & GaN & intermodal AO & 1550 & 1 & 82 & 27.5 & 13 & --- \\
\bottomrule
\end{tabular}
\end{table*}

\section{Modeling of mechanical and optical modes}

\begin{figure}[H]
    \centering
    \includegraphics[width=1\linewidth]{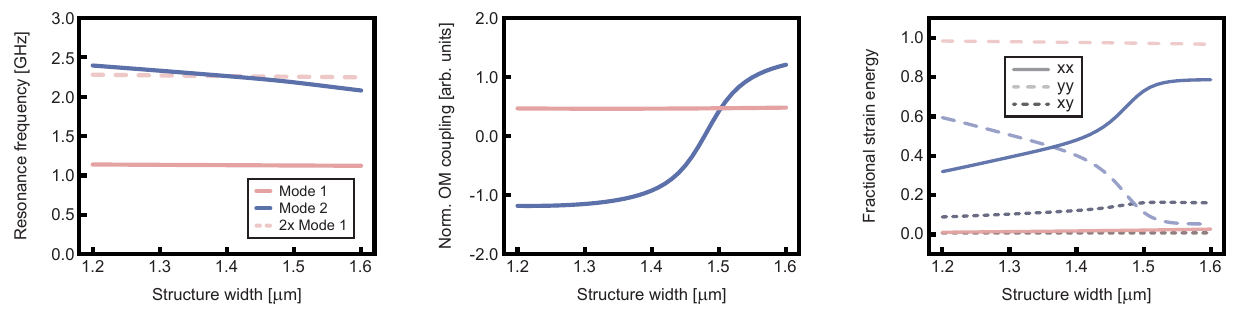}
    \caption{{\bf Optomechanical modeling.} {\bf a} Mechanical resonance frequency versus structure width for Mode 1 and Mode 2. The resonances are harmonically spaced at approximately $w_\text{struct}=\SI{1.4}{\micro\metre}$. {\bf b} Normalized optomechanical coupling versus structure width for Mode 1 and Mode 2. {\bf c} Fractional strain energy in the $xx$, $yy$, and $xy$ components as a function of structure width for Mode 1 and Mode 2. Mode 2's dominant strain term transitions from $yy$ to $xx$ over this range of structure widths.}
    \label{fig:modeling}
\end{figure}

\subsection{Width-dependence of mechanical resonance frequencies}
To model the resonance frequencies of the device's mechanical modes as a function of the structure width, we simulate the device using finite-element-method software that accounts for the linear elastic mechanics, electrostatics, and piezoelectric coupling between them. We use a sidewall angle of $\SI{3}{\degree}$ that is suggested from cross-sectional scanning-electron-microscope images from previous work \cite{freedman_gigahertz-frequency_2025} and the nominal layer thicknesses listed in Table \ref{tab:layer_thicknesses}. 
\begin{table}[t]
\centering
\caption{Layer thicknesses used in the finite-element simulations. Layers are listed from top to bottom.}
\label{tab:layer_thicknesses}
\begin{tabular}{l l r}
\toprule
Layer & Thickness (\si{\nano\metre}) \\
\midrule
SiO$_2$ cladding above SiN      &  690 \\
SiN core                        &  300 \\
SiO$_2$ cladding below SiN      &  710 \\
Top electrode                   &  200 \\
AlN                             &  450 \\
Bottom electrode                &  150 \\
SiO$_2$ below bottom electrode  &  100 \\
\bottomrule
\end{tabular}
\end{table}
The resonance frequencies of Mode 1 and Mode 2 are shown in Supplementary Fig. \ref{fig:modeling}a, where we see that they are harmonically spaced at approximately $w_\text{struct} = \SI{1.4}{\micro\metre}$. For ease of presentation, the displayed simulation results use a constant core width of $w_\text{core}=\SI{1.1}{\micro\metre}$. This parameter can also be swept, however it has a much smaller effect on the mechanical resonance frequencies than the structure width.

\subsection{Optomechanical coupling}
\label{subsec:omc}
To model the optomechanical coupling between a given breathing-mode mechanical resonance and guided optical mode of the structure, we calculate the photoelastic and moving-boundary overlap integrals \cite{kittlaus_electrically_2021}:
\begin{align}
    g_\text{pe}\propto\frac{\iint E^*_i \Delta\epsilon_{ij}E_j \text{ }dA}{\sqrt{U_\text{mech}}\iint E_i^*\epsilon_{ij}E_j \text{ }dA}, \text{ and } g_\text{mb} \propto \frac{\oint u_in_i\left(\Delta\epsilon_{12}E_{||,i}^*E_{||,i} - \Delta\epsilon_{12}^{-1}D_{\perp,i}^*D_{\perp,i}\right)dl}{\sqrt{U_\text{mech}}\iint E_i^*\epsilon_{ij}E_j \text{ }dA}.
\end{align}
In the above, $E_i$ is the electric field profile of the optical mode, $E_{||,i}$ is the component of the electric field parallel to the core-cladding boundary, $D_{\perp,i}$ is the component of the electric displacement field perpendicular to core-cladding boundary, $\Delta\epsilon_{12}$ is the difference in permittivity between the core and cladding, $\Delta\epsilon^{-1}_{12}$ is the difference in reciprocals of the permittivity between the core and cladding, $\Delta\epsilon_{ij} = -\epsilon_{ik}p_{klmn}S_{mn}\epsilon_{lj}$ is the photoelastic perturbation to the permittivity tensor, $\epsilon_{ij}$ is the unperturbed permittivity tensor, $u_i$ is the displacement profile of the mechanical mode, $n_i$ is the normal vector to the core-cladding boundary, and $U_\text{mech}$ is the energy per unit length of the mechanical mode. In these calculations, we use following Voigt-contracted photoelastic constants for silicon nitride: $p_{11} = -0.125$, $p_{12}=0.047$, and $p_{44} = (p_{11}-p_{12})/2 = -0.086$. For silicon dioxide we use: $p_{11}=0.121$, $p_{12}=0.270$, and $p_{44}=(p_{11}-p_{12})/2=-0.0745$ \cite{mishra_ultra-low_2026}. 

The amount by which the optical propagation constant changes is proportional to the sum of these normalized overlap integrals. We normalize the optomechanical coupling by the square root of the mechanical energy so that the modulation efficiency of two mechanical modes can be directly compared (provided they have the same electromechanical coupling and quality factor). Using these integrals and the results of electromagnetic and mechanical simulations of the device cross-section as a function of $w_\text{struct}$, we calculate the optomechanical coupling of Mode 1 and Mode 2 to the fundamental transverse-electric optical mode and display the results in Supplementary Fig. \ref{fig:modeling}b. We note that overall, Mode 1 and Mode 2 are both substantially optomechanically coupled, but Mode 2 undergoes a transition from negative coupling (meaning the change to the effective refractive index is out of phase with the mechanical amplitude) to positive coupling. At approximately $w_\text{struct}=\SI{1.5}{\micro\metre}$, the coupling passes through zero. This change from negative to positive coupling occurs because the mode transitions from being dominated by $yy$ strain to $xx$ strain (Supplementary Fig. \ref{fig:modeling}c). These two strain components influence the permittivity perturbation through photoelastic coefficients with different signs, and eventually as the fractional $xx$ energy grows for Mode 2, its contribution overtakes that of the $yy$ component. The optomechanical coupling strength does not account for differences in their electromechanical coupling, quality factor, and the device's frequency-dependent impedance, meaning the observed $V_\pi$ is determined by a confluence of factors and these simulations alone cannot predict it. However, it still serves as a crucial tool to identify which mechanical modes will produce usable phase modulation.

\section{Two-harmonic phase modulation}
\label{sec:multi-harmonic theory}

\subsection{Two-harmonic optical spectrum}
\label{subsec:two-harmonic}

We consider an input harmonic optical signal, $U(t) = U_0e^{-i\omega t}$, with angular frequency $\omega$. This signal then undergoes sinusoidal phase modulation at the angular frequency, $\Omega$, and the second harmonic, $2\Omega$, to produce a signal of the form
\begin{align}
    U(t) = U_0 \exp\left\{i\left[-\omega t + \beta_1 \sin\left(\Omega t\right) + \beta_2 \sin\left(2\Omega t + \psi\right)\right]\right\},
\end{align}
where $\beta_1$ is the modulation depth at $\Omega$, $\beta_2$ is the modulation depth at $2\Omega$, and $\psi$ is the phase of the modulation at $2\Omega$ relative to that at $\Omega$ (see Section \ref{subsec:psi1_gone}). We can rewrite this according to the Jacobi-Anger expansion as
\begin{align}
    U(t) &= U_0 e^{-i\omega t} \sum_{n=-\infty}^{\infty} J_n(\beta_1) e^{in\Omega t} \sum_{m = -\infty}^{\infty} J_m(\beta_2)e^{im(2\Omega t+\psi)}
    \\[5mm]
        &= U_0 e^{-i\omega t} \sum_{n=-\infty}^{\infty} \sum_{m = -\infty}^{\infty} J_n(\beta_1)J_m(\beta_2)e^{im\psi} e^{i(n+2m)\Omega t}.
\end{align}
From this, we can see the two-tone modulation generates optical signals at the angular frequencies $\omega-(n+2m)\Omega$, which are weighted by a product of associated Bessel functions with a phase, $m\psi$. So, the total optical signal at one of these individual frequencies (comb lines), $\tilde{U}_{p}$, is given by the phasor sum of the relevant complex amplitudes, which can be written as
\begin{align}
    \tilde{U}_p = U_0\sum_{k=-\infty}^\infty J_{2k+p}(\beta_1)J_{-k}(\beta_2)e^{-ik\psi}.
    \label{eq:up}
\end{align}
Thus, by optimizing $\beta_1$, $\beta_2$, and $\psi$, the amplitudes of each combline can be optimized for some desired outcome.

\subsection{Conditions for resonant two-harmonic modulation}

In order to precisely define the conditions under which dual-harmonic resonant phase modulation is possible ({\it i.e.} when harmonic alignment is possible), one must specify the minimum required modulation depth at each frequency and the maximum voltage that can be applied to the device at each frequency. Otherwise, one could achieve the modulation by driving the device \say{off resonance} with arbitrarily small modulation depth or drive with an arbitrarily large voltage to obtain a desired modulation depth. In practice, modulation depths of order $\pi$ are required to produce spectra of interest, and voltages below around $\SI{3}{\volt}$ are desired for compatibility with CMOS driving electronics. We model each mechanical mode as a damped harmonic oscillator with angular resonance frequency $\Omega_n$ and energy damping rate $\gamma_n$, so the energy spectra have a FWHM of $\gamma_n/2\pi$ in units of frequency. 

With this, the half-wave voltages $V_{\pi,n}(\Omega)$ are reciprocals of the square root of a Lorentzian function because the modulation depths are proportional to the mechanical amplitude (not energy). These take the form
\begin{align}
    V_{\pi,n}(\Omega) = V_{\pi,n}(\Omega_n)\sqrt{1+4\left(\frac{\Omega-\Omega_n}{\gamma_n}\right)^2}.
    \label{eq:vpin}
\end{align}

We now let the minimum modulation depth required by Mode 1 be $\beta_1$ and that required by Mode 2 be $\beta_2$. In general, these values differ from one another; for example, to produce maximum conversion to the first-order sideband the required modulation depth at $\Omega$ is $\SI{1.94}{\radian}$, and the required modulation depth at $2\Omega$ is $\SI{0.88}{\radian}$. We also assume both modes have the same maximum voltage $V_\text{max}$, which is the case, for example, if compatibility with CMOS driving electronics determines this constraint. We now define the dimensionless ratio
\begin{align}
    r_n = \frac{\pi V_\text{max}}{\beta_n V_{\pi,n}(\Omega_n)},
\end{align}
which is the maximum on-resonance modulation depth divided by the required modulation depth. The range of frequencies over which each mode can produce the minimum modulation depth is then given by
\begin{align}
    r_n^2 &\geq 1+4\left(\frac{\Omega-\Omega_n}{\gamma_n}\right)^2
    \\[3mm]
    |\Omega-\Omega_n| &\leq \frac{\gamma_n\sqrt{r_n^2-1}}{2}.
\end{align}

Harmonic alignment is achieved when twice the range of frequencies over which Mode 1 can produce $\beta_1$ has a nonempty intersection with the range of frequencies over which Mode 2 can produce $\beta_2$. This intersection occurs when
\begin{align}
    |2\Omega_1 - \Omega_2| &\leq \gamma_1\sqrt{r_1^2-1}+\frac{\gamma_2}{2}\sqrt{r_2^2-1}.
    \\[5mm]
    |2\Omega_1 - \Omega_2| &\leq \gamma_1\sqrt{\left(\frac{\pi V_\text{max}}{\beta_1 V_{\pi,1}(\Omega_1)}\right)^2-1}+\frac{\gamma_2}{2}\sqrt{\left(\frac{\pi V_\text{max}}{\beta_2 V_{\pi,2}(\Omega_2)}\right)^2-1}.
\end{align}

We note here that for mechanical responses that are not well described by a single damped harmonic oscillator, the same principle applies, except the two frequency ranges around Mode 1 and Mode 2 will not in general be writable as a closed-form algebraic expression.

\subsection{Elimination of fundamental modulation phase}
\label{subsec:psi1_gone}

Explicitly assuming a phase $\psi_1$ for the modulation at $\Omega$ gives
\begin{align}
    \phi(t) = \beta_1\sin(\Omega t+\psi_1) + \beta_2\sin(2\Omega t + \psi_2).
\end{align}
Translating the time origin by $\tau = -\psi_1/\Omega$ so that $t\rightarrow t+\tau$ yields
\begin{align}
    \phi(t) = \beta_1\sin(\Omega t) + \beta_2\sin(2\Omega t+\psi_2 -2\psi_1).
\end{align}
The carrier acquires a global phase $\omega\tau$, which has no effect on any measured output spectra. Two-harmonic phase modulation is therefore characterized by the single relative phase $\psi=\psi_2-2\psi_1$.

\section{Measurement setup}
\label{sec:signal_analysis}

We design our measurement system to convert each on-chip optical comb line power to a proportionate power at a unique microwave frequency that can be measured with an RFSA. Below, we mathematically describe the measured signals, including the non-idealities present throughout the measurement. These non-idealities ultimately cause the constant of proportionality between the on-chip optical power and its corresponding microwave power to have a weak but non-negligible frequency dependence.

\begin{figure}[H]
    \centering
    \includegraphics[width=1\linewidth]{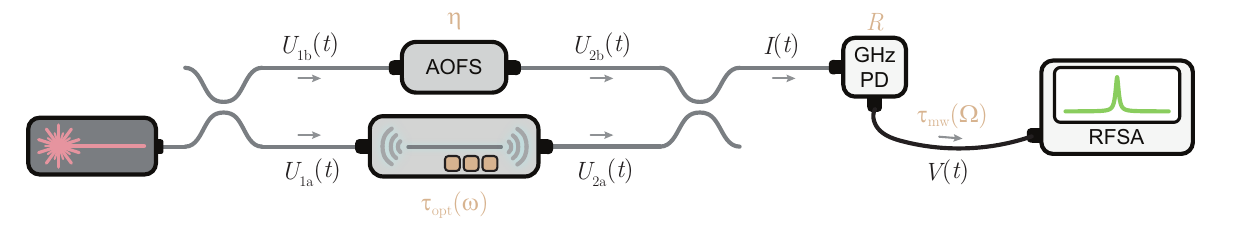}
    \caption{{\bf Experimental diagram with signal propagation labeled.}}
    \label{fig:signal_propagation}
\end{figure}

Following Supplementary Fig. \ref{fig:signal_propagation}, we consider an on-chip device producing pure sinusoidal phase modulation and mathematically trace the propagation of the optical signal $U(t)$ through our measurement and its conversion to the microwave domain. A continuous-wave laser with angular frequency $\omega_0$ propagates through an initial beam splitter with transmission and reflection coefficients of $t$ and $r$, which produces
\begin{align}
    U_{1\text{a}}(t)&=tU_0e^{-i\omega_0 t}, \text{ and}
    \\[4mm]
    U_{1\text{b}}(t)&=rU_0e^{-i\omega_0 t}.
\end{align}
The signal in arm b of the interferometer then passes through the acousto-optic frequency shifter, which shifts the signal by an angular frequency $\Delta$ with real-valued amplitude efficiency $\eta$. Meanwhile, the signal in arm a is grating-coupled to and from the chip, in which it is phase modulated by some periodic modulation function $\phi(t)$ with period $2\pi/\Omega_0$. This yields
\begin{align}
    U_{2\text{a}}(t) &= tU_0e^{-i\omega_0 t} \sum_{n=-\infty}^\infty \tau_\text{opt}(\omega_0-n\Omega_0)a_ne^{in\Omega_0t}, \text{ and}
    \\[4mm]
    U_{2\text{b}}(t) &= \eta rU_0e^{-i(\omega_0 + \Delta) t},
\end{align}
where $\tau_\text{opt}(\omega)$ is a frequency-dependent transmission coefficient through the fiber-to-chip-back-to-fiber system and $a_n$ is the $n$th coefficient of the Fourier series for $e^{i\phi(t)}$. In the case of sinusoidal phase modulation, the modulation function is $\phi(t)=\beta\sin(\Omega_0t)$, and the Fourier coefficients are $a_n = J_n(\beta)$. The frequency-dependent transmission arises because of Fabry-Perot-like cavities formed between the facet of the input fiber and the input grating coupler, between the two grating couplers, between the output grating coupler and output fiber facet, {\it etc}. After this, the signals in each of the arms are interferometrically combined with a second beam splitter to produce an intensity signal given by
\begin{align}
    I(t) &= \left(tU_\text{2b} + rU_\text{2a}\right)\left(tU_\text{2b} + rU_\text{2a}\right)^*
    \\[3mm]
        &= |t|^2|\eta|^2|U_0|^2 + r^*tU^*_\text{2a}U_\text{2b} + rt^*U_\text{2a}U^*_\text{2b} + |r|^2|U_0|^2
    \\[3mm]
        &= \left(|\eta|^2|t|^2 + |r|^2\right)|U_0|^2 + |U_0|^2\left\{ r^*t\left[\eta e^{-i(\omega_0 + \Delta)t} r^*e^{i\omega_0 t}\sum_{n=-\infty}^\infty \tau_{n,\text{opt}}^* a_n^* e^{-in\Omega_0 t}\right] + \text{c.c.} \right\}
    \\[3mm]
        &= \left(|\eta|^2|t|^2 + |r|^2\right)|U_0|^2 + 2\eta|t^*r|^2|U_0|^2\sum_{n=-\infty}^\infty |\tau_{n,\text{opt}}| |a_n| \cos\left[(n\Omega_0 +\Delta)t + \theta_n \right],
\end{align}
where $\theta_n = 2\arg(t^*r) + \arg(\tau_{n,\text{opt}}) + \arg(a_n)$. This intensity signal is the sum of a constant term and terms oscillating at $n\Omega_0 + \Delta$ which correspond to the comb lines generated from the on-chip device. The signal is converted to a voltage $V(t)$ with an AC-coupled high-speed photodetector with frequency-dependent responsivity $R(\Omega)$. This microwave signal then propagates along a cable connecting the photodetector to the spectrum analyzer (RFSA), which introduces both frequency-dependent propagation loss and a microwave cavity formed between the cable terminations together described by a single transmission coefficient $\tau_\text{mw}(\Omega)$. Ignoring the suppressed constant-intensity term, the voltage thus produced is given by
\begin{align}
    V(t) = 2\eta|t^*r|^2\sum_{n=-\infty}^\infty |\tau_{n,\text{mw}}|R_n|\tau_{n,\text{opt}}| a_n \cos\left[(n\Omega_0 +\Delta)t + \theta_n \right],
\end{align}
where $R_n = R(n\Omega_0 + \Delta)$, $\tau_{n,\text{mw}} = \tau_\text{mw}(n\Omega_0 + \Delta)$, and $\theta_n$ is assumed to be updated according to the phase of the microwave transmission coefficient. Finally, the power spectrum $P_n$ of the voltage signal incident on the RFSA is directly measured and is of the form
\begin{align}
    P_n &= \left(2\eta|t^*r|^2\right)^2 \left(|\tau_{n,\text{mw}}|R_n|\tau_{n,\text{opt}}|\right)^2 a_n^2
    \\[3mm]
        &= H_n\left(2\eta|t^*r|^2\right)^2 a_n^2,
\end{align}
where $P_n$ is the power of the microwave signal oscillating at $n\Omega_0 + \Delta$, and $H_n$ is the lumped frequency-dependent transmission. In the absence of $H_n$, each $P_n$ is proportional to the on-chip power of each comb line produced by the pure-phase-modulation process with the same constant of proportionality across all comb lines.

\section{Device statistics}

We measured 36 devices with identical cross-sectional design ($w_\text{struct}=\SI{1.4}{\micro\metre}$ and $w_\text{core}=\SI{1.1}{\micro\metre}$) from three different wafers (12 from each wafer). The devices have different support structures for their release and also different lengths (varying from $\SI{500}{\micro\metre}$ to $\SI{2}{\milli\metre}$). In Supplementary Fig. \ref{fig:yield} the half-wave power as a function of driving frequency is shown for Mode 2 (solid curve) and Mode 1 (dashed curve) with a doubled frequency axis for Mode 1.

From this data, we extract the following statistics on resonance frequency and bandwidth:
\begin{table}[htbp]
\centering
\vspace{1em}
\label{tab:mode_stats}
\begin{tabular}{lcccc}
\toprule
Mode & $\Omega/2\pi$ [GHz] & $\text{SD}(\Omega/2\pi)$ [MHz]& $\text{BW}$ [MHz] & $\text{SD}(\text{BW})$ [MHz] \\
\midrule
1 & $1.149$ & $12.8$ & $16.5$ & $7.8$ \\
2 & $2.298$ & $27.1$ & $24.4$ & $11.7$ \\
\bottomrule
\end{tabular}
\caption{Measured resonance frequencies and modulation bandwidths for the two optomechanically coupled modes, taken across $n=36$ devices with identical cross-sections drawn from three wafers.}
\end{table}
Here, $\SI{3}{\decibel}$ bandwidth refers to the range of frequencies over which $V_\pi$ remains within a factor of two of its minimum value. For calculation of equation \ref{eq:vpin}, these bandwidths equal $\sqrt{3}\gamma_n$, so $\gamma_1/2\pi = \SI{9.5}{\mega\hertz}$ and $\gamma_2/2\pi = \SI{14.1}{\mega\hertz}$. We attribute the spread in resonance frequency to thickness variations in the layers from the different wafers. As can be seen in Supplementary Fig. \ref{fig:yield}, within a group of devices from the same wafer (same colored fill), the spread in resonance frequency is much smaller. 

\begin{figure}[H]
    \centering
    \includegraphics[width=1\linewidth]{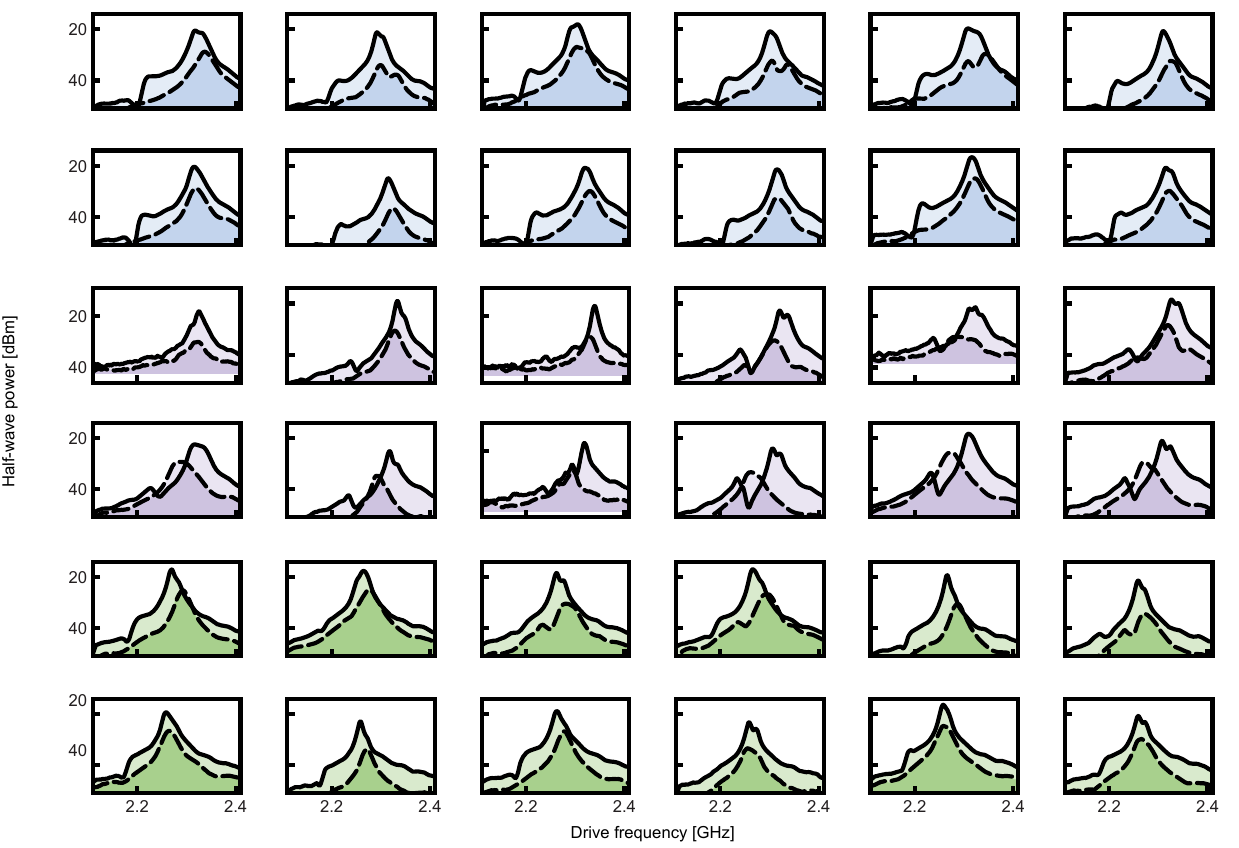}
    \caption{{\bf Half-wave power for 36 measured devices.} Same-colored plots represent devices from the same wafer. The solid curves are from Mode 2, and the dashed curves are from Mode 1 with its frequency axis doubled.}
    \label{fig:yield}
\end{figure}

\section{Quality factors and time-domain response}
\subsection{Fitting modulation responses to driven damped harmonic oscillators}
\label{subsec:sho_fits}

\begin{figure}[H]
    \centering
    \includegraphics[width=1\linewidth]{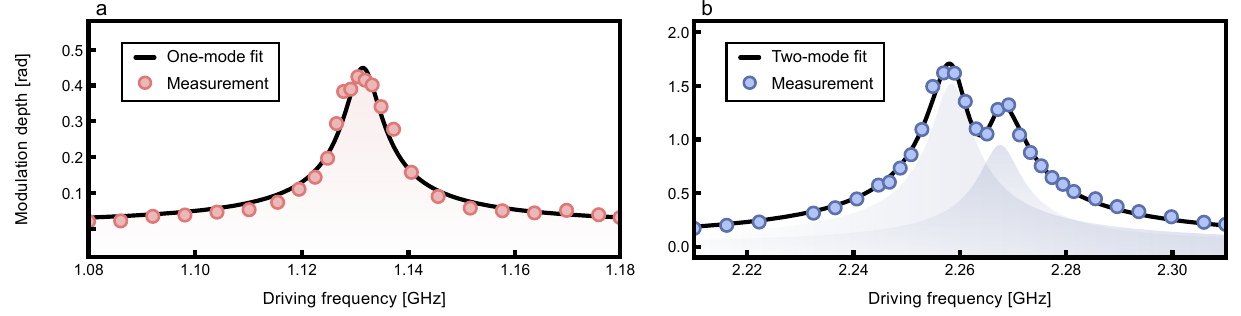}
    \caption{{\bf Fit of modulation response for Mode 1 and Mode 2.} {\bf a} Modulation depth versus driving frequency around Mode 1 with a single fit mechanical mode. {\bf b} Modulation depth versus driving frequency around Mode 2 with a doublet pair of mechanical modes fit. Individual fits are shown as semi-transparent fills.}
    \label{fig:quality factor}
\end{figure}

Around most mechanical modes, we observe a closely spaced pair of peaks in the modulation response (Supplementary Fig. \ref{fig:quality factor}). Throughout the main text, Mode 1 and Mode 2 refer to these pairs as a whole, but here we resolve Mode 2 into two constituent split resonances and leave Mode 1 as a single resonance. This splitting observed in Mode 2 is not captured by the two-dimensional model of Section~2, which assumes translational invariance along the device and a free lower boundary. As in previous work, we attribute the splitting to the phononic crystal formed by the structure's periodic supports \cite{freedman_gigahertz-frequency_2025}. Since the modes have different cross-sectional displacement profiles, they interact differently with the support structure, and this can explain that Mode 2 is split while Mode 1 is not. However, more detailed modeling of the splitting is warranted in future work.

With this, we model each constituent resonance as a damped, driven simple harmonic oscillator. We ignore the nonlinear coupling treated in Section~\ref{subsec:mech_nl} because the data in this section were taken at drive powers of $5$ dBm (for Mode 1) and $10$ dBm (for Mode 2), at which the nonlinearly generated modulation depth extrapolated from the data in Supplementary Figure \ref{fig:nonlinear_summ} is below $\SI{0.02}{\radian}$. The amplitude $x_n$ of the $n$th mechanical mode then obeys
\begin{align}
    \ddot{x}_n + \gamma_n\dot{x}_n + \Omega_n^2 x_n = f_n(t), \label{eq:sho}
\end{align}
where $\Omega_n$ is the angular resonance frequency and $f_n(t)$ is the piezoelectric driving force projected onto mode $n$, per unit modal mass. The rate $\gamma_n$ is the energy decay rate of the mode; {\it i.e.} with the drive removed, the energy stored in the mode decays as $e^{-\gamma_n t}$ and its amplitude as $e^{-\gamma_n t/2}$. The quality factor is defined as $Q_n = \Omega_n/\gamma_n$.

For a drive voltage $V(t) = V_0\cos(\Omega t)$ applied to the actuator, the force per unit modal mass is $f_n(t) = (\eta_n/m_n)V_0\cos(\Omega t)$, where $m_n$ is the modal mass and $\eta_n$ is the electromechanical coupling coefficient of mode $n$. Because the piezoelectrically induced stress overlaps differently with each mode shape, $\eta_n$ differs between modes. Writing the steady state as $x_n(t) = \text{Re}\left[\tilde{x}_n(\Omega)\,e^{i\Omega t}\right]$, the complex amplitude of mode $n$ is
\begin{align}
    \tilde{x}_n(\Omega) = \frac{\eta_n V_0/m_n}{\Omega_n^2 - \Omega^2 + i\gamma_n\Omega}. \label{eq:complex_amp}
\end{align}

The optical phase shift is linear in the mechanical displacement field (see Section \ref{subsec:omc}), so the modulation produced by several modes driven at the same frequency is the coherent sum of their contributions. Each mode $n$ contributes a modulation depth $\beta_n = G_n\tilde{x}_n(\Omega)$, where $G_n = d\beta_n/d\tilde{x}_n$. The total modulation is then a single sinusoid at $\Omega$ with complex amplitude
\begin{align}
    \beta(\Omega) = \sum_n \beta_n(\Omega) = \sum_n \frac{a_n}{\Omega_n^2 - \Omega^2 + i\gamma_n\Omega}, \qquad a_n = \frac{G_n\eta_n}{m_n}\,V_0. \label{eq:beta_coherent}
\end{align}
We measure the magnitude of modulation depth $|\beta(\Omega)|$ by fitting it to the ratio of sideband powers extracted from the measurement described in Section \ref{sec:signal_analysis}. We then fit $|\beta(\Omega)|$ using Eq.~\eqref{eq:beta_coherent} exactly by bounded nonlinear least squares with three free parameters per constituent ($a_n$, $\Omega_n$, $Q_n$). We note that $\beta_n(\Omega)$ is the square root of a Lorentzian in the limit that $\gamma_n \ll \Omega_n$ in the vicinity of the resonance. The fits are shown in Supplementary Fig. \ref{fig:quality factor} and the parameters are listed in Table~\ref{tab:sho_fits}.

\begin{table}[htbp]
\centering
\caption{Fitted parameters of Eq.~\eqref{eq:beta_coherent} for Mode 1 and Mode 2. Mode 1 is described by a single damped oscillator, and Mode 2 is described by two. The column $|a_n|Q_n/\Omega_n^2$ is the peak modulation depth each constituent would produce in the absence of the other.}
\vspace{3mm}
\label{tab:sho_fits}
\begin{tabular}{lcccrc}
\toprule
Constituent & $\Omega_n/2\pi$ [GHz] & $Q_n$ & $\gamma_n/2\pi$ [MHz] & $a_n$ [rad\,GHz$^2$] & $|a_n|Q_n/\Omega_n^2$ [rad] \\
\midrule
Mode 1     & 1.13144 & 164.5 & 6.88 & $+0.00350$ & 0.449 \\
\addlinespace
Mode 2a    & 2.25844 & 290.7 & 7.77 & $+0.02664$ & 1.518 \\
Mode 2b    & 2.26757 & 294.2 & 7.71 & $+0.01659$ & 0.949 \\
\bottomrule
\end{tabular}
\end{table}

The fitted amplitude $a_n$ constrains only the combination $G_n\eta_n/m_n$, so the electromechanical coupling $\eta_n$ and the acousto-optic conversion coefficient $G_n$ cannot be separated by the modulation-depth data alone. 

\subsection{Inferred switching time from measured frequency response}

\begin{figure}[H]
    \centering
    \includegraphics[width=1\linewidth]{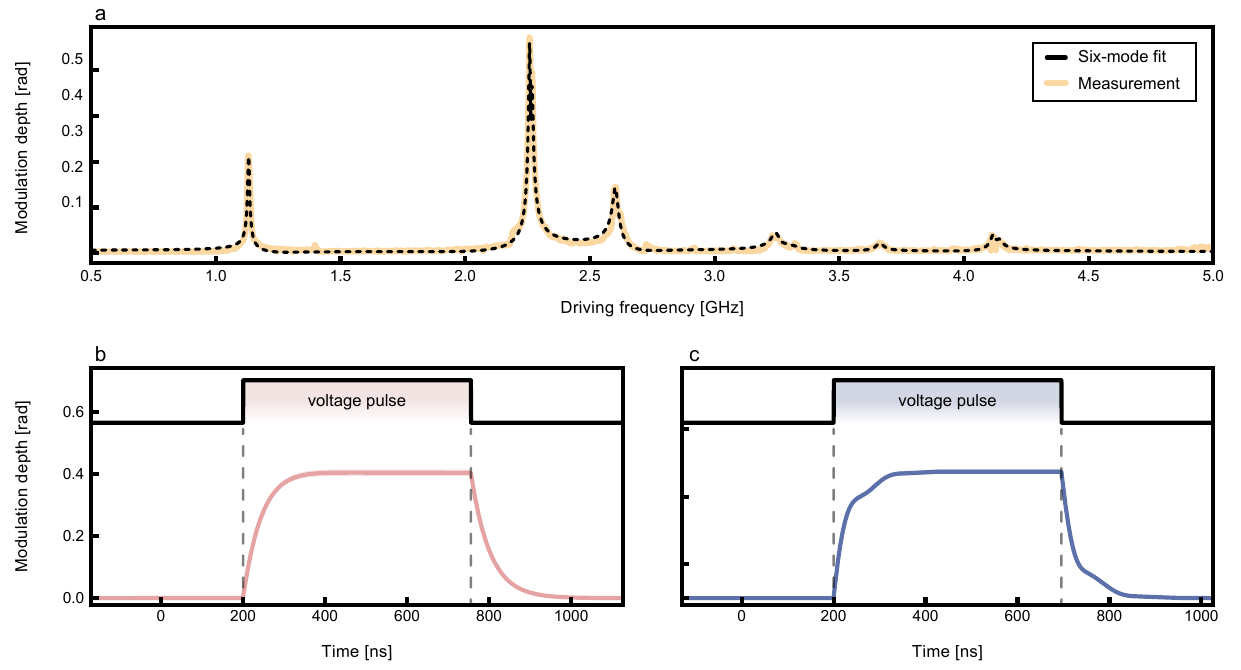}
    \caption{{\bf Inferred impulse response from measured frequency response.} {\bf a} Modulation depth versus driving frequency over a wide range of frequencies with the six-mode fit shown as a dashed line. {\bf b} Inferred modulation response at Mode 1 to a rectangular voltage pulse. {\bf c} Inferred modulation response at Mode 2 to a rectangular voltage pulse.}
    \label{fig:impulse}
\end{figure}

We measure $|\beta(\Omega)|$ over a frequency range from $\SI{500}{\mega\hertz}$ to $\SI{5}{\giga\hertz}$, outside of which we do not expect to observe a significant modulation response (Supplementary Fig. \ref{fig:impulse}a). Modeling the transformation of a voltage signal $V(t)$ to an optical phase modulation signal $\phi(t)$ as a linear system, we must know $\beta(\Omega)$ to calculate an impulse response function and therefore also a switching time. In general there are infinitely many possible phase profiles $e^{i\psi(\Omega)}$ consistent with $|\beta(\Omega)|$, however in our system this is heavily constrained by the physics. As in Section \ref{subsec:sho_fits}, we assume that the modulation response is a sum of simple harmonic oscillators that yield
\begin{align}
     \beta(\Omega) = \sum_n \frac{a_n}{\Omega_n^2 - \Omega^2 + i\gamma_n\Omega}.
\end{align}
With this, the phase profile of the frequency response is set by the relative signs of the real-valued $a_n$ coefficients. Using the $|\beta(\Omega)|$ data we fit six mechanical modes to either single or double simple harmonic oscillators. To bound the effect of this ambiguity in relative sign, we calculate an inferred impulse response function for each of the possible 32 relative sign configurations using
\begin{align}
    h(t) = \frac{1}{\pi V_0}\,\mathrm{Re}\int_{0}^{\infty} \beta(\Omega)\,e^{i\Omega t}\,d\Omega.
    \label{eq:impulse}
\end{align}
The phase modulation produced by any driving voltage is then given by
\begin{align}
    \phi(t) = \int_{0}^{\infty} h(t')\,V(t-t')\,dt'.
\end{align}
From here, we calculate the time-domain response of a rectangular voltage pulse with a carrier frequency at the resonance of Mode 1 and Mode 2 using each of these impulse responses and display the average in Supplementary Figs. \ref{fig:impulse}b and c. The total range of $10$-$\SI{90}{\percent}$ rise times is $\SI{101.4}{\nano\second}$ to $\SI{102.0}{\nano\second}$ for Mode 1 and $\SI{94.1}{\nano\second}$ to $\SI{94.5}{\nano\second}$ for Mode 2. We report the worst-case values in the main text.

\section{Mechanical nonlinearity}
\subsection{Resonantly enhanced mechanical second-harmonic generation}
\label{subsec:mech_nl}

By design, the devices have nearly harmonically spaced mechanical modes (Mode 1 and Mode 2) at the angular frequencies $\Omega_1$ and $\Omega_2$, respectively. Thus, when mechanical energy is present in Mode 1, it can couple to Mode 2 via a second-order mechanical nonlinearity \cite{thomas_third-order_1968, bogardus_thirdorder_1965, lepkowski_third-order_2021}. We model this system with a set of differential equations representing damped driven nonlinear harmonic oscillators:
\begin{align}
    &\ddot{x}_1 + \gamma_1\dot{x}_1 + \Omega_1^2x_1 + \alpha x^2=f(t) \label{nl1}
    \\[3mm]
    &\ddot{x}_2 + \gamma_2\dot{x}_2 + \Omega_2^2x_2 + \alpha x^2=f(t), \label{n12}
\end{align}
where $x_i$ is the amplitude of mode $i$, $\gamma_i$ is the damping rate of mode $i$, $\Omega_i$ is resonance angular frequency of mode $i$, $\alpha$ is the strength of the second-order nonlinearity, $x = x_1+x_2$ is the total mechanical amplitude, and $f(t)$ is the external drive of the system. We consider a driving term of the form $f(t) = f_0\cos(\Omega t)$ where $\Omega$ is assumed to be nearby $\Omega_1$. We then ignore the nonlinear term $\alpha x^2 = \alpha(x_1^2 + 2x_1x_2 + x_2^2)$ in equation \eqref{nl1} because we assume the $x_1^2$ and $x_2^2$ terms have no component oscillating at $\Omega$ and thus do not make significant contributions to $x_1$ (rotating-wave approximation). We further assume that the back-action term, $2x_1x_2$, is small enough to be ignored. We also note that the nonlinearity strength $\alpha$ is assumed to include the effect of nonlinear piezoelectricity \cite{pal_second-order_2011}, and we omit explicit piezoelectric coupling for simplicity. 

With this, we proceed by solving for the steady state of the linear damped driven harmonic oscillator for $x_1$ and then use this solution to calculate the nonlinearly driven steady state for $x_2$.

Assuming a steady state of the form $x_1(t)=A_1\cos(\Omega t+\theta_1)$ with real-valued $A_1$ and $\theta_1$, one finds that
\begin{align}
    A_1=\frac{f_0}{\sqrt{(\Omega_1^2-\Omega^2)^2+(\gamma_1\Omega)^2}}, \text{and } \theta_1=\arctan\left(\frac{\gamma_1\Omega}{\Omega_1^2-\Omega^2}\right) . \label{eq:mode1_amplitude}
\end{align}
From here, we solve for steady state of $x_2$ accounting for an effective nonlinear driving force nearby its resonance at $\Omega_2$. We ignore the direct driving force $f(t)$ because it is far detuned from $\Omega_2$. The nonlinear driving force is produced by the $\alpha x_1^2$ term and is given by $f_\text{NL}(t)=-(\alpha A_1^2/2)\cos(2\Omega t + 2\theta_1)$. With this, we solve for the steady state of $x_2$ assuming a solution of the form $x_2(t)=A_2\cos(2\Omega t + \theta_2)$, where $A_2$ and $\theta_2$ are once again real-valued. We find that
\begin{align}
    A_2 = \frac{\alpha A_1^2}{2\sqrt{\left[\left(\Omega_2^2-4\Omega^2\right)^2 + \left(2\Omega\gamma_2\right)^2\right]}} \text{, and } \theta_2 = \pi + 2\theta_1 - \arctan\left(\frac{2\Omega\gamma_2}{\Omega_2^2-4\Omega^2}\right). \label{eq:mode2_amplitude}
\end{align}

In Supplementary Fig. \ref{fig:analytic_vs_numerical}, we show these analytic steady-state solutions alongside solutions obtained from a full numerical solve that makes no approximations on equations \ref{nl1} and \ref{n12}. The analytically derived linear and parabolic dependence of the mode amplitudes on the driving strength are in good agreement with the numerical solutions until drive powers are high enough that substantial amplitude is transferred from Mode 1 to Mode 2. In the actual experiments, we expect our devices to exist in this undepleted pump regime, in which the analytic approximation is valid.

\begin{figure}[H]
    \centering
    \includegraphics[width=1\linewidth]{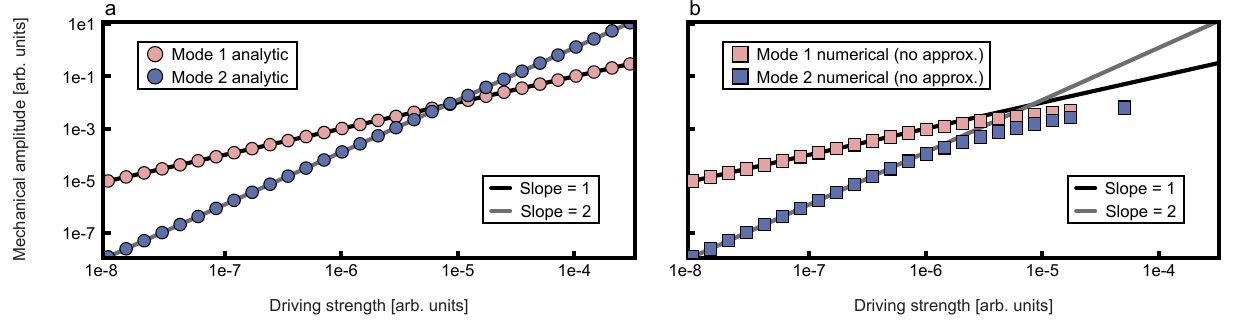}
    \caption{{\bf Resonant mechanical second-harmonic generation model.} {\bf a} Analytic solutions for the amplitudes of Mode 1 and Mode 2 with approximations based on equations \ref{eq:mode1_amplitude} and \ref{eq:mode2_amplitude}. {\bf b} Numerical solutions for the amplitudes of Mode 1 and Mode 2 with no approximations based on equations \ref{nl1} and \ref{n12}.}
    \label{fig:analytic_vs_numerical}
\end{figure}

\subsection{Nonlinear acousto-optic phase modulation}

Once the steady states for $x_1$ and $x_2$ are found, the effect of the nonlinearity on the observed phase-modulation comb can be determined knowing the proportionality constant between the modulation depth and strain amplitude for each mode, $d\beta_i/dx_i$. With these two constants, which are scaled versions of the optomechanical coupling rates, we have that the optical signal is of the form
\begin{align}
    U(t) &= U_0 \exp\left\{i\left[-\omega t + \frac{d\beta_1}{dx_1}A_1\cos\left(\Omega t + \psi_1\right) +\frac{d\beta_2}{dx_2}A_2\cos\left(2\Omega t+\psi_2\right)\right]\right\}
    \\[5mm]
    U(t) &= U_0e^{-i\omega t} \sum_{n=-\infty}^\infty \sum_{m=-\infty}^\infty i^{n+m} J_n\left(\frac{d\beta}{dx_1}A_1\right) J_m\left(\frac{d\beta}{dx_2}A_2\right) e^{i(n\psi_1 + m\psi_2)}e^{i(n+2m)\Omega t}.
\end{align}

In the dual-harmonic experiments that are the subject of the main text, we simultaneously drive the device at the frequency $\Omega$ (near the resonance frequency of mode 1) and $2\Omega$ (near the resonance frequency of mode 2). Thus, by the above analysis, we expect to effectively observe three different sinusoidal modulations: the directly driven modulation at $\Omega$ with depth $\beta_1$, the directly driven modulation at $2\Omega$ with depth $\beta_{2,\text{lin}}$, and the nonlinearly generated modulation at $2\Omega$ with depth $\beta_{2,\text{nl}}$. The optical signal is then of the form
\begin{align}
    U(t) = U_0 \exp \left\{i\left[-\omega t + \beta_1\sin(\Omega t) + \beta_{2,\text{lin}} \sin(2\Omega t + \psi_\text{lin}) + \beta_{2,\text{nl}}\sin(2\Omega t + \psi_\text{nl})\right]\right\}, \label{eq:nl_modulation}
\end{align}
where $\psi_\text{lin}$ is the programmed phase difference between the modulation at mode 1 and mode 2, and $\psi_\text{nl}$ is the phase of the nonlinear drive at mode 2 determined by the physics described in the above sections.

The linear (direct) and nonlinear modulation at $2\Omega$ combine into a single modulation term of the form $\beta_2\sin(2\Omega t + \psi)$, where
\begin{align}
    \beta_2 &= \sqrt{\beta_{2,\text{lin}}^2 + \beta_{2,\text{nl}}^2 + 2\beta_{2,\text{lin}}\beta_{2,\text{nl}}\cos(\psi_\text{lin} - \psi_\text{nl})}, \text{ and} \label{eq:beta2}
    \\[3mm]
    \psi &= \text{arctan2}\left[\beta_{2,\text{lin}}\sin(\psi_\text{lin}) + \beta_{2,\text{nl}}\sin(\psi_\text{nl}),\beta_{2,\text{lin}}\cos(\psi_\text{lin}) + \beta_{2,\text{nl}}\cos(\psi_\text{nl})\right].
\end{align}

This means the phase the signal generator applies ({\it i.e.} $\psi_\text{lin}$) is not actually the phase of the modulation signal at $2\Omega$. However, these dynamics do not modify which spectra can or cannot be generated.

\subsection{Fit to measured phase modulation}

In Fig. 3b-e of the main text, we fix the drive powers to the device at $\Omega$ and $2\Omega$ and sweep the relative phase between these two signals. In the absence of the above-mentioned nonlinearity, the conversion efficiency observed to the positive and negative $p$th sidebands ($\text{CE}_p$ and $\text{CE}_{-p}$) should be symmetric as a function of this relative phase. This is because the field amplitude in the $p$th sideband obeys the following condition upon phase translations of $\pi$:
\begin{align}
    \tilde{U}_p(\beta_1, \beta_2, \psi+\pi) &= U_0 \sum_{k=-\infty}^\infty J_{2k+p}(\beta_1)J_{-k}(\beta_2)(-1)^k e^{-ik\psi}
    \\[1.5mm]
    \tilde{U}_p(\beta_1, \beta_2, \psi+\pi) &= U_0 \sum_{k=-\infty}^\infty (-1)^{2k+p}J_{-2k-p}(\beta_1)(-1)^kJ_{k}(\beta_2)(-1)^k e^{-ik\psi}
    \\[1.5mm]
    \tilde{U}_p(\beta_1, \beta_2, \psi+\pi) &= (-1)^p U_0 \sum_{k=-\infty}^\infty J_{2k-p}(\beta_1)J_{-k}(\beta_2) e^{ik\psi}
    \\[1.5mm]
    \tilde{U}_p(\beta_1, \beta_2, \psi+\pi) &= (-1)^p \left[\tilde{U}_{-p}(\beta_1, \beta_2, \psi)\right]^*\,
\end{align}
where we have started with equation \eqref{eq:up}, made use of the property of Bessel functions that $J_p(\beta) = (-1)^pJ_{-p}(\beta)$, and re-indexed $k$. From here, it is clear that $\text{CE}_p(\beta_1, \beta_2, \psi+\pi) = \text{CE}_{-p}(\beta_1, \beta_2, \psi)$.

The presence of the nonlinearity does not change this fact. However, from the perspective of the measurement, the phase that is ostensibly programmed is actually just $\psi_\text{lin}$; and so the modulation depth at $2\Omega$ varies as a function of the applied phase according to equation \eqref{eq:beta2}. For this reason, the two curves within each plot in Fig. 3c-e of the main text are not mirror images of each other.

To verify that the cause of this lack of symmetry was indeed because of such a nonlinearity, we perform several measurements in which we fix the power applied at $\Omega$ and $2\Omega$ and sweep the relative phase between these driving signals through a full $2\pi$-cycle. Between each of these phase-sweep measurements we increase the power applied at $\Omega$. We then fit the observed sideband conversion efficiencies within each of these several phase sweeps to a model that assumes the signal shown in equation \eqref{eq:nl_modulation}, where the fitting parameters are $\beta_1$, $\beta_{2,\text{lin}}$, $\beta_{2,\text{nl}}$, and $\psi_\text{nl}$. The results of this are displayed in Supplementary Fig. \ref{fig:fit_grid}. As described in Section \ref{sec:signal_analysis}, there are frequency-dependent transmission coefficients through the setup, for which the fit also accounts. These transmission coefficients can account for a difference in overall scale factor between the pairs of curves in the main text's Fig. 3c-e after reflection, but cannot account for a difference in shape. This difference in shape is especially apparent in main text Fig. 3e and is caused by mechanical second harmonic generation. For this reason, we also fit a scale factor to each order for each phase sweep and report the statistics in Supplementary Table \ref{tab:nlfit}. The relatively small standard deviations are evidence the frequency-dependent transmission coefficients do not vary strongly over the course of several measurements.

\begin{figure}[H]
    \centering
    \includegraphics[width=1\linewidth]{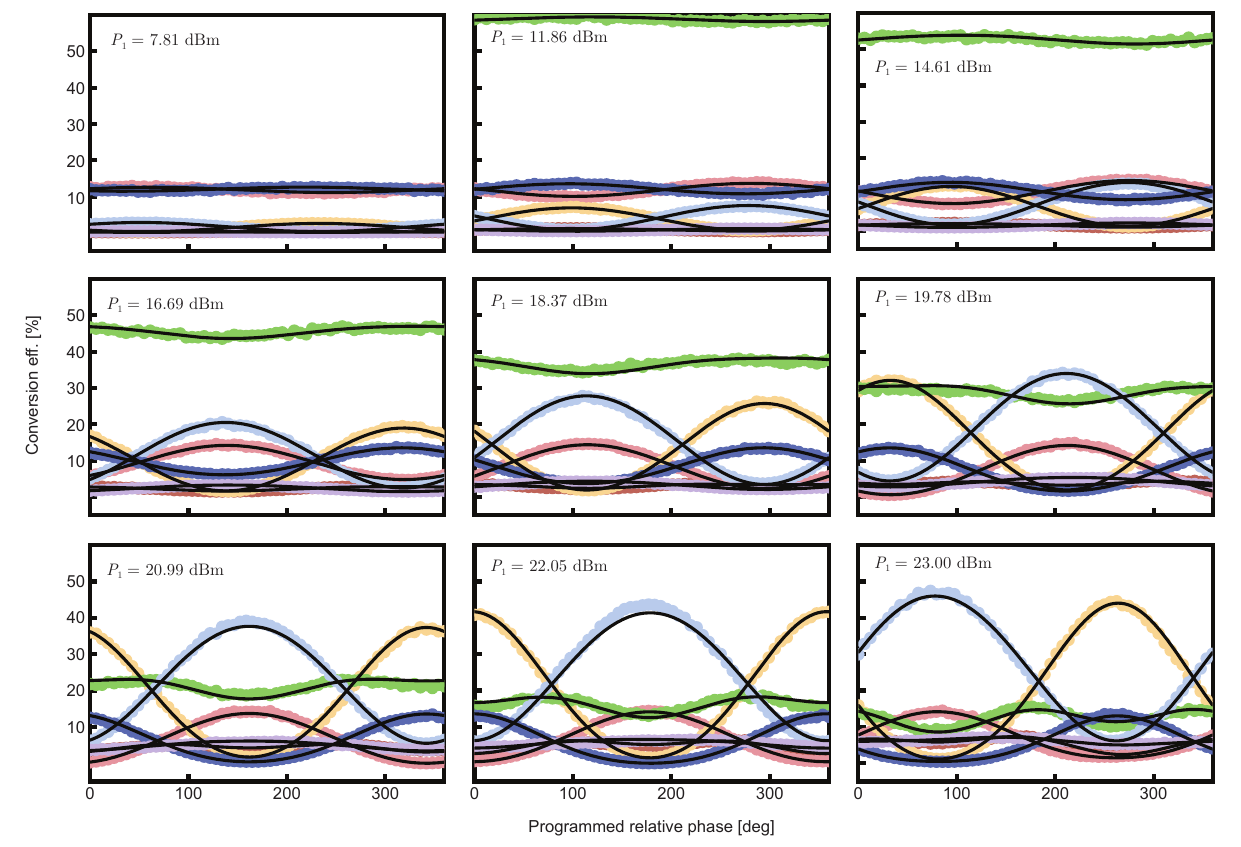}
    \caption{{\bf Dependence of conversion efficiency on phase with fitted nonlinear model.} Colored circles are measured values whose colors correspond to the same sideband as those in Fig. 3 of the main text, and the black solid curves are from the fitted theory that includes the nonlinearity.}
    \label{fig:fit_grid}
\end{figure}

\begin{table}[htbp]
\centering
\caption{Transmission scale factors $H_p$ of the characterization chain at sideband order $p$, pooled across the nine phase-sweep fits of the nonlinear model. Because $H_p$ describes the frequency-dependent transmission of the measurement setup rather than the device, it is expected to be independent of the drive power at $\Omega$.}
\vspace{3mm}
\label{tab:nlfit}
\begin{tabular}{lccccccc}
\toprule
 & $H_{-3}$ & $H_{-2}$ & $H_{-1}$ & $H_{0}$ & $H_{1}$ & $H_{2}$ & $H_{3}$ \\
\midrule
Mean & 0.971 & 0.805 & 0.842 & 0.979 & 0.901 & 0.863 & 0.934 \\
SD   & 0.081 & 0.036 & 0.019 & 0.017 & 0.040 & 0.025 & 0.073 \\
\bottomrule
\end{tabular}
\end{table}

From these fits, we obtain the functional dependence of the various fitting parameters upon the applied voltage at $\Omega$. This applied voltage acts as a proxy for the strain amplitude of mode 1 because as we can see from equation \eqref{eq:mode1_amplitude}, this amplitude is proportional to the applied force, which for a piezoelectrically driven system is proportional to the applied voltage. The fitted parameters $\beta_1$, $\beta_{2,\text{lin}}$, and $\beta_{2,\text{nl}}$ respectively act as proxies for the strain amplitude of mode 1 directly generated by the drive at $\Omega$, the strain amplitude of mode 2 directly generated by the drive at $2\Omega$, and the strain amplitude of mode 2 that is nonlinearly generated. These quantities are good proxies (proportional quantities) because the acousto-optically induced modulation depth caused by the presence of a mechanical mode is proportional to its strain amplitude.

Based on equations \eqref{eq:mode1_amplitude} and \eqref{eq:mode2_amplitude} for the strain amplitudes of mode 1 and mode 2, we expect that $\beta_1$ should vary linearly with the applied voltage at $\Omega$, $\beta_{2,\text{lin}}$ should not vary at all, and $\beta_{2,\text{nl}}$ should vary quadratically (assuming $\beta_1$ does indeed vary linearly). This is exactly what we observe in the results displayed in Supplementary Fig. \ref{fig:nonlinear_summ}. We also note that we characterized the harmonic suppression of the signal generator itself and found that the modulation depth it produces at $2\Omega$ when driven solely at $\Omega$ is not insignificant but is roughly five times smaller than the above-described effect at the maximum applied voltage of $\SI{3}{\volt}$.

\begin{figure}[H]
    \centering
    \includegraphics[width=1\linewidth]{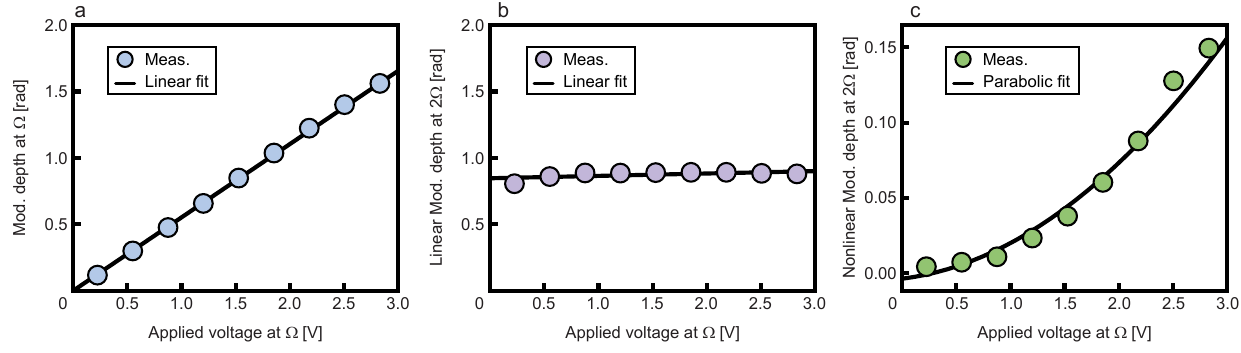}
    \caption{{\bf Summary of nonlinear fit.} {\bf a} Plot of fitted modulation depth, $\beta_1$, versus applied voltage at $\Omega$. {\bf b} Plot of fitted linear modulation depth, $\beta_{2,\text{lin}}$, versus applied voltage at $\Omega$. {\bf c} Plot of fitted nonlinear modulation depth, $\beta_{2,\text{nl}}$, versus applied voltage at $\Omega$.}
    \label{fig:nonlinear_summ}
\end{figure}

\section{Operation with more drive harmonics}
\subsection{Comb amplitudes with $N$ harmonics}
\label{sec:n_harm_amps}

Let there be $N$ drive harmonics for which the $n$th harmonic produces phase modulation with modulation depth $\beta_n$ at phase $\psi_n$. This produces an optical signal of the form
\begin{align}
    U(t) &= U_0\exp(-i\omega t) \exp \left[i\sum_{n=1}^N\beta_n\sin(n\Omega t + \psi_n)\right]
    \\[3mm]
        &= U_0\exp(-i\omega t) \prod_{n=1}^N\exp\left[i\beta_n \sin(n\Omega t+\psi_n)\right]
    \\[3mm]
        &= U_0e^{-i\omega t} \prod_{n=1}^N\left[\sum_{k_n=-\infty}^\infty J_{k_n}(\beta_n)e^{i{k_n}n\Omega t}e^{ik_n\psi_n}\right]. \label{eq:prod_of_sums}
\end{align}
Equation \ref{eq:prod_of_sums} says that the optical field is the product of the individually Jacobi-Anger-expanded phase modulation produced by each harmonic. Each term in the resulting product has picked up one frequency shift from each expansion, which shifts the optical frequency by $k_nn
\Omega$. This means to find the amplitude at the optical angular frequency $\omega-p\Omega$, we must sum all those terms produced by the product of equation \ref{eq:prod_of_sums} which satisfy the following condition:
\begin{align}
    \sum_{n=1}^Nk_nn=p.
\end{align}
Now, to succinctly write the expression for the $p$th field amplitude ({\it i.e.} the Fourier coefficient of the optical field at the angular frequency $\omega-p\Omega$), we define a function $f: \mathbb{Z}^N\rightarrow\mathbb{Z}$ as follows:
\begin{align}
    f\left[(k_1, k_2,... ,k_N)\right] = \sum_{n=1}^N k_nn.
\end{align}
With this, the $p$th field amplitude $\tilde{U}_p$ is
\begin{align}
    \tilde{U}_p = U_0\sum_{\substack{\mathbf{k}\in\mathbb{Z}^N \\[1mm] f(\mathbf{k})=p}} \left[\prod_{n=1}^N J_{k_n}(\beta_n)e^{ik_n\psi_n}\right].
    \label{eq:sum_of_paths}
\end{align}
Equation \ref{eq:prod_of_sums} can be interpreted visually with Supplementary Fig. \ref{fig:amp_graph}. In Supplementary Fig. \ref{fig:amp_graph}, the multi-tone phase modulation is shown as individual phase modulation occurring at each harmonic in series for the case $N=5$. The connection from a node in layer $n$ to a node $n+1$ is produced by the Bessel-weight (and phase) that the modulation at $(n+1)\Omega$ generates. With this, the summation in equation \ref{eq:sum_of_paths} is taken over all of those paths which reach the $p$th node in the final layer from the input continuous-wave signal in the root layer. There are of course infinite such paths when $p$ is not arbitrarily truncated as in Supplementary Fig. \ref{fig:amp_graph}, but in practice, the modulation depth allowed for each modulation limits the spectral extent that must be numerically considered.
\begin{figure}[H]
    \centering
    \includegraphics[width=1\linewidth]{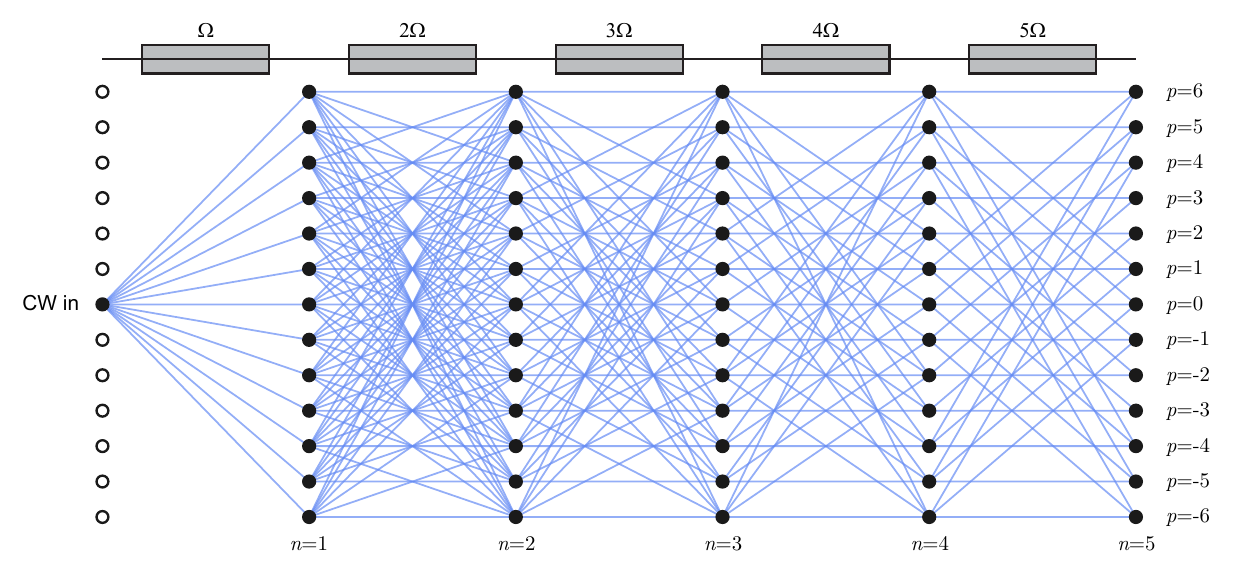}
    \caption{{\bf Graph representation of $N$-harmonic phase modulation.} Nodes represent the amplitude at the $p$th sideband and connections between nodes are the weight the phase modulation produces at the stage. For example, the weight between CW in and the $p=2$ node is $J_2(\beta_1)$.}
    \label{fig:amp_graph}
\end{figure}

Once these paths are identified, the amplitude each one contributes is determined by the product of the Bessel-function and phase weights of the path's constituent segments.

\subsection{Suppressing more sidebands using more drive harmonics}

\begin{figure}[H]
    \centering
    \includegraphics[width=1\linewidth]{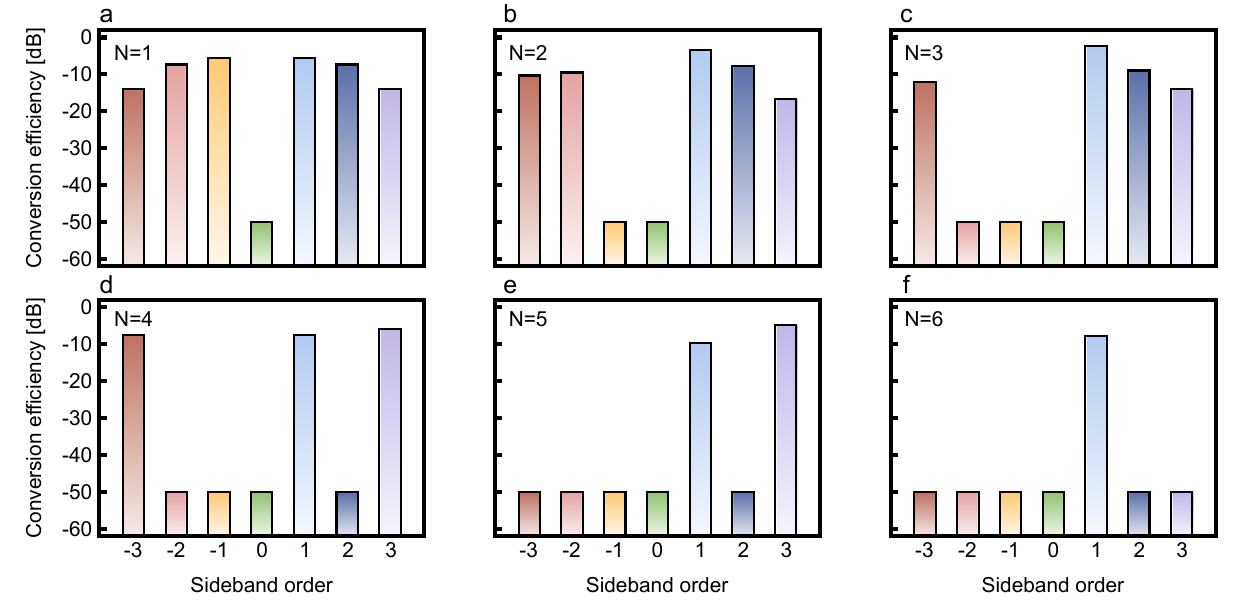}
    \caption{{\bf Suppressing more sidebands in single-sideband frequency shifts.} {\bf a-f} Optimized spectra for drives with a number of harmonics ranging from $N=1$ to $N=6$.}
    \label{fig:nulling}
\end{figure}

In the main text, we identify and generate a spectrum with highly suppressed carrier and $-1$ sidebands while converting a measured $\SI{40}{\percent}$ power to the $+1$ sideband. Here, we investigate the effect of adding more drive harmonics on the ability to suppress an increasing number of sidebands while maintaining relatively high conversion efficiency to the $+1$ sideband. Using equation \ref{eq:prod_of_sums}, we optimize for the modulation depths and phases of each constituent harmonic in the phase modulation under the constraints that as many nearby sidebands as possible are suppressed by $\SI{50}{\decibel}$ while maintaining a conversion efficiency of at least $\SI{10}{\percent}$ to the $+1$ sideband. We also limit the modulation depth available at each harmonic to $\SI{4}{\radian}$, which is below the maximum that has been experimentally achieved using these devices before failure occurs \cite{freedman_gigahertz-frequency_2025}. The order in which sidebands are progressively suppressed follows $(0, -1, -2, +2, -3, +3)$. Supplementary Figure \ref{fig:nulling}, shows these optimized spectra as a function of the number of drive harmonics and Supplementary Table \ref{tab:multiharmonic} contains the parameters that produced them. We empirically find that the number of orders that can be simultaneously suppressed more than $\SI{50}{\decibel}$ is the number of driving harmonics.

\begin{table}[htbp]
\centering
\caption{Optimal multi-harmonic drive parameters for single-sideband frequency shifting with progressively more suppressed orders.}
\label{tab:multiharmonic}
\vspace{3mm}
\begin{tabular}{clcrrrrrrr}
\toprule
$N$ & Suppressed orders & $\eta_{+1}$ [\%] & & $n{=}1$ & $n{=}2$ & $n{=}3$ & $n{=}4$ & $n{=}5$ & $n{=}6$ \\
\midrule
\multirow{2}{*}{2} & \multirow{2}{*}{$0,\,-1$} & \multirow{2}{*}{46.09}
  & $\beta_n$ [rad]  & 2.447 & 1.100 & --- & --- & --- & --- \\
  & & & $\psi_n$ [deg] & 0 & 180 & --- & --- & --- & --- \\
\addlinespace
\multirow{2}{*}{3} & \multirow{2}{*}{$0,\,-1,\,-2$} & \multirow{2}{*}{57.70}
  & $\beta_n$ [rad]  & 2.382 & 1.207 & 0.698 & --- & --- & --- \\
  & & & $\psi_n$ [deg] & 0 & 180 & 0 & --- & --- & --- \\
\addlinespace
\multirow{2}{*}{4} & \multirow{2}{*}{$0,\,-1,\,\pm 2$} & \multirow{2}{*}{13.69}
  & $\beta_n$ [rad]  & 2.278 & 1.578 & 1.624 & 2.553 & --- & --- \\
  & & & $\psi_n$ [deg] & 0 & 180 & 0 & 0 & --- & --- \\
\addlinespace
\multirow{2}{*}{5} & \multirow{2}{*}{$0,\,-1,\,\pm 2,\,-3$} & \multirow{2}{*}{12.93}
  & $\beta_n$ [rad]  & 2.549 & 1.000 & 3.116 & 1.249 & 1.294 & --- \\
  & & & $\psi_n$ [deg] & 0 & 180 & 0 & 0 & 180 & --- \\
\addlinespace
\multirow{2}{*}{6} & \multirow{2}{*}{$0,\,-1,\,\pm 2,\,\pm 3$} & \multirow{2}{*}{15.21}
  & $\beta_n$ [rad]  & 2.551 & 0.247 & 3.048 & 0.661 & 1.417 & 0.450 \\
  & & & $\psi_n$ [deg] & 0 & 170.6 & 0 & 0.1 & 179.4 & $-2.9$ \\
\bottomrule
\end{tabular}
\end{table}

\subsection{Improved conversion efficiency with more drive harmonics}
Using equation \ref{eq:prod_of_sums} and once again optimizing over the set of available modulation depths and phases, we numerically optimize the conversion efficiency to the $+1$ sideband as a function of the number of drive harmonics. The results of this optimization are conveyed in Supplementary Table \ref{tab:max_ce}. For this specific target spectrum, the weights and phases of the modulation harmonics converge to a solution that generates a sawtooth waveform in the time domain (serrodyne frequency shifting). In this limit, the optimal parameters approach $\beta_n = 2/n$ with alternating sign induced by $\psi_n$.

\begin{table}[htbp]
\centering
\caption{Maximum achievable single-sideband conversion efficiency versus the number of available harmonics $N$. For each $N$, the modulation depths $\beta_n$ and relative phases $\psi_n$ are optimized to maximize conversion to the $+1$ sideband with no constraint on the remaining orders. The phase origin is fixed by setting $\psi_1 = 0$; the optimum alternates $\psi_n = 0$ and $180^\circ$ for all $N$.}
\label{tab:max_ce}
\vspace{3mm}
\begin{tabular}{ccclrrrrrr}
\toprule
$N$ & $\eta_{+1}$ [\%] & $\eta_{+1}$ [dB] & & $n{=}1$ & $n{=}2$ & $n{=}3$ & $n{=}4$ & $n{=}5$ & $n{=}6$ \\
\midrule
\multirow{2}{*}{1} & \multirow{2}{*}{33.86} & \multirow{2}{*}{$-4.70$}
  & $\beta_n$ [rad]  & 1.841 & --- & --- & --- & --- & --- \\
  & & & $\psi_n$ [deg] & 0 & --- & --- & --- & --- & --- \\
\addlinespace
\multirow{2}{*}{2} & \multirow{2}{*}{55.00} & \multirow{2}{*}{$-2.60$}
  & $\beta_n$ [rad]  & 1.940 & 0.882 & --- & --- & --- & --- \\
  & & & $\psi_n$ [deg] & 0 & 180 & --- & --- & --- & --- \\
\addlinespace
\multirow{2}{*}{3} & \multirow{2}{*}{66.23} & \multirow{2}{*}{$-1.79$}
  & $\beta_n$ [rad]  & 1.969 & 0.939 & 0.575 & --- & --- & --- \\
  & & & $\psi_n$ [deg] & 0 & 180 & 0 & --- & --- & --- \\
\addlinespace
\multirow{2}{*}{4} & \multirow{2}{*}{73.04} & \multirow{2}{*}{$-1.36$}
  & $\beta_n$ [rad]  & 1.981 & 0.963 & 0.611 & 0.426 & --- & --- \\
  & & & $\psi_n$ [deg] & 0 & 180 & 0 & 180 & --- & --- \\
\addlinespace
\multirow{2}{*}{5} & \multirow{2}{*}{77.59} & \multirow{2}{*}{$-1.10$}
  & $\beta_n$ [rad]  & 1.987 & 0.975 & 0.629 & 0.450 & 0.338 & --- \\
  & & & $\psi_n$ [deg] & 0 & 180 & 0 & 180 & 0 & --- \\
\addlinespace
\multirow{2}{*}{6} & \multirow{2}{*}{80.83} & \multirow{2}{*}{$-0.92$}
  & $\beta_n$ [rad]  & 1.991 & 0.982 & 0.640 & 0.464 & 0.355 & 0.280 \\
  & & & $\psi_n$ [deg] & 0 & 180 & 0 & 180 & 0 & 180 \\
\bottomrule
\end{tabular}
\end{table}

\subsection{Multi- and sub-harmonic single-qubit gate generation}

Consider an input optical signal $U_\text{in}(t)=e^{i\phi(t)}$ where the phase modulation function $\phi(t)$ is a sum of harmonic sinusoids:
\begin{align}
    \phi(t) = \sum_{m=1}^M \beta_m \sin(m\Omega_0t + \psi_m). \label{eq:phi}
\end{align}
We are ultimately interested in calculating the intensity of this signal, so we omit the explicit carrier-frequency term ($e^{-i\omega t}$) and the overall field amplitude ($U_0$) for simplicity. Because the optical signal is periodic with period $2\pi/\Omega_0$, we can write it as
\begin{align}
    U_\text{in}(t) = \sum_{n=-\infty}^\infty c_ne^{in\Omega_0 t}.
\end{align}
Clearly, $U_\text{in}(t)$ produces a constant intensity. In the frequency domain, this is the statement that
\begin{align}
    D_k = \sum_{n=-\infty}^\infty c_n^*c_{n+k} = 0
\end{align}
for all nonzero $k$, where $D_k$ is Fourier coefficient of the intensity modulation signal at the frequency $k\Omega_0$. In other words, the contribution of each pairwise beatnote formed by optical comb lines spaced by $k\Omega_0$ sums to precisely zero for a purely phase-modulated optical signal. To convert this to intensity modulation in order to drive single-qubit gates, group-delay-dispersion elements are commonly utilized \cite{levine_dispersive_2022}. Such an element imparts a frequency-dependent phase shift to the optical signal to produce an output optical field of the form
\begin{align}
    U_\text{out}(t) = \sum_{n=-\infty}^\infty c_n e^{in\Omega_0 t}e^{i\alpha n^2},
\end{align}
where $\alpha = \frac{1}{2}\text{GDD}\cdot\Omega_0^2$ and $\text{GDD}$ is the group delay dispersion. The intensity modulation at $k\Omega_0$ is then given by
\begin{align}
    D_k &= \sum_{n=-\infty}^\infty c_n^*c_{n+k}e^{-i\alpha n^2}e^{i\alpha(n+k)^2}
    \\[3mm]
        &= e^{i\alpha k^2}\sum_{n=-\infty}^\infty c_n^*c_{n+k}e^{i2\alpha nk}. \label{eq:Dk}
\end{align}
One could stop here and use the result from equation \ref{eq:sum_of_paths} to perform an optimization on $D_k$ over the modulation depths and phases of the drive harmonics. However, by noticing the similarity of the above expression with that of a time-delayed interferometer, we can obtain a more descriptive expression.

The intensity signal $G(t)$ produced interfering $U_\text{in}(t)$ with a time-delayed copy of itself is
\begin{align}
    G(t) &= U_\text{in}^*(t-\tau)U_\text{in}(t)
    \\[3mm]
        &= \sum_{n=-\infty}^\infty c_n^* e^{-in\Omega_0(t-\tau)}\sum_{m=-\infty}^\infty c_me^{im\Omega_0t}
    \\[3mm]
        &= \sum_{n=-\infty}^\infty \sum_{m=-\infty}^\infty c_n^*c_me^{i(m-n)\Omega_0t}e^{in\Omega_0\tau}
\end{align}
where $\tau$ is the time delay. With this, the Fourier component of $G(t)$ at $k\Omega_0$ is
\begin{align}
    G_k = \sum_{n=-\infty}^\infty c_n^* c_{n+k} e^{in\Omega_0\tau}.
\end{align}
This has precisely the same form as equation \ref{eq:Dk} up to a global phase if the time delay is $\tau = 2k\alpha/\Omega_0 = k\cdot\text{GDD}\cdot\Omega_0$. Thus, the $k$th Fourier coefficient of the intensity modulation produced by $U_\text{out}(t)$ can be expressed in the following way:
\begin{align}
    D_k &= \left\langle U^*_\text{in}(t-\tau)U_\text{in}(t) \exp(-ik\Omega_0 t)\right\rangle_t
    \\[3mm]
        &= \left\langle \exp\left\{i\left[\phi(t)-\phi(t-\tau)\right]\right\}\exp(-ik\Omega_0 t)\right\rangle_t, \label{eq:t_avg}
\end{align}
where $\langle.\rangle_t$ denotes a time average over the period $2\pi/\Omega_0$. We now consider $\Delta\phi(t)=\phi(t)-\phi(t-\tau)$. Using equation \ref{eq:phi}, we have that
\begin{align}
    \Delta\phi(t) &= \sum_{m=1}^M \beta_m\left\{\sin(m\Omega_0t + \psi_m) - \sin\left[m\Omega_0(t-\tau) + \psi_m\right] \right\}
    \\[3mm]
                &= \sum_{m=1}^M 2\beta_m\sin(m\Omega_0\tau/2)\cos(m\Omega_0t + \psi_m - m\Omega_0\tau/2)
    \\[3mm]
                &= \sum_{m=1}^M b_m \cos(m\Omega_0t^\prime + \psi_m),
\end{align}
where $b_m = 2\beta_m\sin(mk\alpha)$ and $t^\prime = t-\tau/2$. In similar fashion to the derivation of Section \ref{sec:n_harm_amps}, we can write the term within the time average of equation \ref{eq:t_avg} as
\begin{align}
    \exp\left[i\Delta\phi(t)\right] = \prod_{m=1}^M\left[\sum_{p_m=-\infty}^\infty i^{p_m} J_{p_m}(b_m)e^{ip_m m \Omega_0t}e^{ip_m\psi_m}\right], \label{eq:exp_dphi}
\end{align}
where we have dropped the prime on $t^\prime$ because the time shift's only effect is to globally shift the phase of $D_k$. Now, to evaluate $D_k$ according to equation \ref{eq:t_avg}, we must identify those terms in equation \ref{eq:exp_dphi} which oscillate at $k\Omega_0$ so they do not get washed out by the time average. These terms are those for which
\begin{align}
    \sum_{m=1}^M p_mm = k.
\end{align}
With this, $D_k$ can be expressed as the sum of those coefficients of the terms in equation \ref{eq:exp_dphi} oscillating at $k\Omega_0$:
\begin{align}
    D_k = \sum_{\substack{\mathbf{p}\in\mathbb{Z}^M \\[1mm] f(\mathbf{p})=k}} \left[\prod_{m=1}^M i^{p_m} J_{p_m}(b_m)e^{ip_m\psi_m}\right], \label{eq:final_Dk}
\end{align}
where the function $f:\mathbb{Z}^M \rightarrow \mathbb{Z}$ is defined similarly to in Section \ref{sec:n_harm_amps}. The magnitude of equation \ref{eq:final_Dk} gives the efficiency with which the phase modulation is converted to amplitude modulation at the frequency $k\Omega_0$. This reformulation of the problem in terms of a time-delay interferometer has the nice interpretation that the $k$th efficiency is given by the $k$th Fourier coefficient of a signal phase-modulated by $\Delta\phi(t)$. 

In the simple case that there is single-tone phase modulation at $\Omega_0$ with modulation depth $\beta$ and phase $\psi$, this reduces to
\begin{align}
    D_k = i^k J_k\left[2\beta\sin(k\alpha)\right]e^{ik\psi}. \label{eq:m_is_1}
\end{align}
Since $U_\text{in}$ has no explicit field amplitude, $|D_k|$ can be considered the efficiency with which phase modulation is converted to amplitude modulation, and we observe our result in equation \ref{eq:m_is_1} is in agreement with previous work \cite{levine_dispersive_2022}.

Using equation \ref{eq:final_Dk}, we can numerically find optimal $\beta_m$ and $\psi_m$ that convert phase modulation at $m\Omega_0$ to amplitude modulation at some qubit transition frequency $\Omega_\text{q}=k\Omega_0$.

\begin{table}[htbp]
\centering
\caption{Calculated phase-to-amplitude conversion through a group-delay-dispersion element for drives composed of harmonics and subharmonics of the target frequency $\Omega_\text{q}$. For each drive, the modulation depths $\beta_n$ and the dispersion were optimized to maximize the amplitude-modulation depth $\eta$ generated at $\Omega_\text{q}$.}
\vspace{3mm}
\label{tab:gdd}
\begin{tabular}{lccrrrl}
\toprule
Drive tones $[\Omega_\text{q}]$ & $k$ & $\eta$ & $\eta/\eta_1$ & $\mathrm{GDD}\cdot\Omega_\text{q}^2$ & $\beta_n$ [rad] \\
\midrule
$1,\,2$                          & 1 & 0.7416 & 1.275 &  2.26 & 1.07,\ 0.57 \\
$1/6,\,1/3,\,1/2$                & 6 & 0.6792 & 1.167 & 14.65 & 5.79,\ 3.08,\ $1.40$ \\
$1/4,\,1/2$                      & 4 & 0.6052 & 1.040 &  9.67 & 3.65,\ 1.49 \\
$1$                              & 1 & 0.5819 & 1.000 &  3.14 & 0.92 \\
$1/6,\,1/3$                      & 6 & 0.5667 & 0.974 & 14.71 & 5.29,\ 2.04 \\
$1/2$                            & 2 & 0.4865 & 0.836 &  6.28 & 1.53 \\
$1/3$                            & 3 & 0.4344 & 0.746 &  9.42 & 2.10 \\
\bottomrule
\end{tabular}
\end{table}

\newpage

\bibliographystyle{unsrt}  % You can change "plain" to "abbrv", "ieeetr", etc.
\bibliography{sup_references.bib}